\documentclass[twocolumn]{emulateapj}
\usepackage{graphicx}
\usepackage[dvips]{color}
\usepackage{amsmath,amsthm,amssymb}
\usepackage{mathtools}

\newcommand{\be}{\begin{equation}}
\newcommand{\ee}{\end{equation}}
\newcommand{\ba}{\begin{eqnarray}}
\newcommand{\ea}{\end{eqnarray}}

\def\lsim{\raise0.3ex\hbox{$\;<$\kern-0.75em\raise-1.1ex\hbox{$\sim\;$}}}
\def\gsim{\raise0.3ex\hbox{$\;>$\kern-0.75em\raise-1.1ex\hbox{$\sim\;$}}}
\def\eps{\varepsilon}
\def\theta{\vartheta}

\shortauthors{Giacinti \& Kirk}
\shorttitle{Cosmic-Ray Anisotropy and Interstellar Turbulence}

\begin{document}

\author{Gwenael~Giacinti and John~G.~Kirk}
\affil{Max-Planck-Institut f\"ur Kernphysik, Postfach 103980, 69029 Heidelberg, Germany}

\title{Large-Scale Cosmic-Ray Anisotropy as a Probe of Interstellar Turbulence}

\begin{abstract}
  We calculate the {\em large-scale} cosmic-ray (CR) anisotropies predicted
  for a range of Goldreich-Sridhar (GS) and isotropic models of
  interstellar turbulence, and compare them with IceTop data. In
  general, the predicted CR anisotropy is not a pure dipole; the
  cold spots reported at $400$\,TeV and $2\,$PeV are
  consistent with a GS model that contains a smooth deficit of parallel-propagating waves and a
  broad resonance function, though some other possibilities cannot, as
  yet, be ruled out. In particular, isotropic fast magnetosonic wave
  turbulence can match the observations at high energy, but
  cannot accommodate an energy dependence in the shape
  of the CR anisotropy. Our findings suggest that improved data on the
  large-scale CR anisotropy could provide a valuable probe of the properties --- 
  notably the power-spectrum --- of the interstellar turbulence within a few 
  tens of parsecs from Earth.
\end{abstract}

\keywords{(ISM:) cosmic rays ---  ISM: magnetic fields}

\maketitle

\section{Introduction}
\label{Introduction}

In this paper, we focus on the possible role of turbulence in the local
interstellar medium, within about a cosmic-ray (CR) mean free path
from Earth, in shaping the {\em large-scale} or {\em global}
anisotropy of CRs of energy between 100 TeV and a few PeV, i.e., on
features in their angular distribution that are larger than several
tens of degrees in size.  Aside from a small distortion attributable
to the draping of magnetic field lines around the heliosphere,
\citet{Schwadron2014} find that the direction of the global anisotropy
of TeV~CR is compatible with that of the magnetic field as deduced
both from the IBEX ribbon, and from polarization of optical starlight
from stars within a few tens of
parsecs~\citep{Frisch:2012zj,Frisch:2015hfa}. In principle, this
could be due to a coincidental alignment of the CR density gradient
and the field direction, but we interpret it as supporting the
view that CRs diffuse mainly along field lines,
rather than across them, in which case the density gradient is not
strongly constrained. Although explanations have been advanced for
small-scale anisotropies that can arise from the configuration of
the interstellar magnetic
field~\citep{Giacinti:2011mz,Battaner2015,Lopez-Barquero:2015qpa},
both the amplitude and the shape of the global anisotropy remain
unexplained. Until now, the amplitude has attracted most of the attention
\citep{Blasi:2011fm,Pohl:2012xs,Kumar:2014dma,Mertsch:2014cua}. Here
we are primarily concerned with the shape and its potential to provide
information on the turbulent component of the local field.

Anisotropies in the arrival directions of TeV--PeV CRs 
have been detected in both hemispheres and at
multiple angular scales by several observatories 
\citep[a review of experimental results is presented in][]{DiSciascio:2014jwa}. 
However, the most useful data sets for our purposes
are those published by the IceCube/IceTop collaboration
\citep{Aartsen:2012ma,Aartsen:2016ivj} for high-energy 
($\gtrsim 100$\,TeV) CRs, because these data are free from features 
caused by heliospheric fields, which demand a different explanation 
\citep{Desiati:2011xg,Drury:2013uka}.

A dipole anisotropy, such as would be expected 
from the Compton-Getting effect if the Solar System 
were drifting through an isotropic distribution of CRs,
is clearly ruled out
by the data \citep{Aartsen:2012ma,Aartsen:2016ivj}, at least for CRs of 
energy $\gtrsim 100$\,TeV. 
Instead, IceTop has detected a relatively
small cold spot in its field of view, whereas the CR flux in the rest of
the observed sky is much more isotropic. The typical size of the cold
spot is $\approx 30^{\circ}$ for the data set with 400\,TeV median
energy, and $\approx 40^{\circ}$ at 2\,PeV~\citep{Aartsen:2012ma}. 
A dipole anisotropy would produce a larger cold spot, as well as 
an undetected maximum in the CR flux in the observed part of the sky 
at the opposite value of the cold
spot's right ascension. 
%
%

Although the partial sky coverage of IceTop/IceCube can give rise to
artefacts \citep[e.g.][]{Ahlers:2016njl}, we take these data at face
value and address the question of how they can be used to constrain
the nature of the turbulence underlying CR transport.  Although
attempts to go beyond the pure dipole description have been made ---
e.g., by \citet{Zhang:2014dsu}, who represented the large-scale
anisotropy phenomenologically as a sum of Legendre polynomials --- a
quantitative approach to the question that specifically relates it to
turbulence models has, so far, not been attempted.  

In this paper, we compute the anisotropies that arise from an
anisotropic spectrum of Alfv\'en waves of Goldreich-Sridhar
type~\citep{SridharGoldreich1994,Goldreich:1994zz}, as well as from an
isotropic distribution of fast magnetosonic modes with a
power-spectrum compatible with that found in the MHD simulations
of~\cite{Cho:2002qi}. The pitch-angle scattering rates for these
models have been calculated previously by, for example,
\citet{Chandran:2000hp} and \citet{Yan:2002qm,Yan:2004aq,Yan:2007uc},
but these authors did not use them to predict the CR anisotropy.  Each
hypothesis necessarily contains input of a phenomenological nature
concerning the precise functional form of the anisotropy, as well as
the so-called \lq\lq resonance function\rq\rq.  We find that the data yield
constraints on these inputs, as well as on quantities such as the
Alfv\'en speed and the outer length scale of the turbulent spectrum.
Our approach is similar in spirit to that of
\citet{Malkov:2010yq}, who addressed the problem of anisotropies on
small angular scales. The main methodological differences are that we
use the full general solution of the equation for CR pitch-angle
diffusion, compute the angular diffusion coefficients numerically, and
use more recent data on the direction of the local magnetic field.


In Section~\ref{Model} the general expression for the CR anisotropy as
a function of the CR pitch-angle is derived.  Section~\ref{Results}
presents calculations of this anisotropy, both for Alfv\'enic
turbulence with an anisotropic Goldreich-Sridhar power-spectrum, and
for compressible turbulence with isotropic fast magnetosonic modes. We
then discuss the implications of our findings and the validity of our
assumptions in Section~\ref{Discussion}, and summarize our 
main results in Section~\ref{Conclusions}.

\section{CR anisotropy and pitch-angle diffusion}
\label{Model}



The CR anisotropy at Earth is shaped by our local environment, which,
according to studies
of the polarization of optical light from nearby
stars~\citep{Frisch:2012zj,Frisch:2015hfa}, contains a magnetic field that is
coherent in direction over a few tens of parsecs --- see
Section~\ref{Discussion} for a discussion on the nature and direction
of this field. Since the Larmor radius of TeV--PeV CRs, $r_{\rm L}
\sim (10^{-4} - 1)$\,pc, is relatively small, this field is viewed by
them as a quasi-homogeneous field, that dominates their propagation 
in the vicinity of the Solar System. As noted above, this
is confirmed by the fact that the observed CR anisotropy direction is
compatible with the direction of the local field. Therefore, in the
following, we adopt a model in which CRs propagate in a 
magnetic flux tube of length between $\sim 10$ and a few times $10$
parsecs. \citet{Schwadron2014} find that the energy density
in turbulence on scales of the Larmor radii of TeV--PeV CRs is quite
small compared to that in the coherent field, and, moreover, note that
cross-field diffusion is not expected to influence the global
anisotropy.  Therefore, we assume that the transport of CRs
is described by pitch-angle diffusion along a local, uniform flux
tube, and look for a stationary solution of the transport
equation governing the pitch-angle distribution $f(x,\mu)$ of 
CRs~\citep[e.g.,][]{Vedenov62,Malkov:2010yq,Malkov:2015dda}:
\begin{equation}
  \mu v \frac{\partial f}{\partial x} = \frac{\partial}{\partial \mu} \left( D_{\mu\mu} \frac{\partial f}{\partial \mu} \right)\;,
\label{TransportEqn}
\end{equation}
where $\mu$ is the cosine of the CR pitch-angle $\theta$, $v$ 
the CR velocity (assumed here to equal the speed of light), $x$ the spatial
coordinate along the flux tube, and $D_{\mu\mu}$ is the pitch-angle
diffusion coefficient, which vanishes at $\mu = \pm 1$ and is positive
definite in the range $-1 < \mu < 1$.

\subsection{CR anisotropy}
\label{Expression_Anisotropy}

A standard approach to equations of the form of~(\ref{TransportEqn}) is to
separate the variables and expand in eigenfunctions \citep[e.g.,][]{Bethe38}:
\begin{equation}
  f(x,\mu) = \sum_{\rm i} a_{\rm i} e^{\Lambda_{\rm i} x /v} Q_{\rm i}(\mu) \;,
\label{EqnExpansion}
\end{equation}
where the eigenvalues $\Lambda_{\rm i}$ and eigenfunctions $Q_{\rm i}(\mu)$ obey
\begin{equation}
  \Lambda_{\rm i} \mu Q_{\rm i} = \frac{\partial}{\partial \mu} D_{\mu\mu} \frac{\partial}{\partial \mu} Q_{\rm i} \;,
\label{EqnEigenvalues}
\end{equation}
together with boundary conditions on $Q_{\rm i}$ that ensure
regularity and single-valuedness, i.e., $\pm \Lambda_{\rm i} Q_{\rm
  i} = -D'_{\mu\mu} Q'_{\rm i}$ at $\mu = \pm 1$ (where $'$ means
differentiation with respect to $\mu$).  However, this method is not
guaranteed to work, because the weighting function, $\mu$, on the left-hand side
of Eq.~(\ref{EqnEigenvalues}) has a zero within the relevant range.  The problem then arises
that that the $Q_{\rm i}$ do not form a complete set of functions on the range 
$-1 \leq \mu \leq 1$. Fortunately, this can be remedied simply by
adding a special solution, the {\it diffusion
  solution}, to the expansion in Eq.~(\ref{EqnExpansion})~\citep{FischKruskal1980}:
\begin{equation}
  f_{\rm diff} = a_{\rm diff} \left[x  + g(\mu) \right] \;,
\end{equation}
where $a_{\rm diff}$ is a constant and $g(\mu)$ is a solution of
\begin{equation}
  \frac{\partial}{\partial \mu} D_{\mu\mu} \frac{\partial}{\partial \mu} g = v\mu \;,
\label{EqnGmu}
\end{equation}
that satisfies the regularity conditions at $\mu = \pm 1$.
The general solution can then be written as:
\begin{equation}
  f(x,\mu) = \sum_{\rm i} a_{\rm i} e^{\Lambda_{\rm i} x/v} Q_{\rm i}(\mu) + a_{\rm diff} 
\left[ x + g(\mu) \right] \;,
\label{EqnGenSol}
\end{equation}
and, because $g$ is defined to within an arbitrary, additive constant, we can impose the additional condition
\begin{equation}
  \int_{-1}^{1} {\rm d}\mu \, g(\mu) = 0 \;.
\label{IntGeqZero}
\end{equation}
Note that the $Q_{\rm i}$ are orthogonal with the weighting function $\mu$:
\begin{equation}
  \int_{-1}^{1} {\rm d}\mu \, Q_{\rm i} \mu Q_{\rm j} = 0 \;,
\label{EqnOrthog}
\end{equation}
for $\Lambda_{\rm i} \neq \Lambda_{\rm j}$, and $g$ is orthogonal to
all the $Q_{\rm i}$ except for the isotropic eigenfunction $Q_{0}=1$ 
with eigenvalue $\Lambda_{0} = 0$:
\begin{equation}
  \int_{-1}^{1} {\rm d}\mu \, Q_{\rm i} \mu g = 0 \;,
\end{equation}
for $\Lambda_{\rm i} \neq 0$. The eigenvalues themselves 
can be ordered such that 
$\Lambda_{\rm i+1}>\Lambda_{\rm i}$, and 
occur in pairs with opposite signs:
$\Lambda_{\rm i}=-\Lambda_{\rm -i}$. 

The energy $E$ of the CR enters this problem as a parameter:
if the diffusion coefficient is a function of $E$, then so are
$\Lambda_{\rm i}$, $Q_{\rm i}$, and $g$.

Let us assume that the Earth is located at $x=0$. We wish to impose
boundary conditions \lq\lq far\rq\rq\ from the Earth and examine the anisotropy
expected at Earth. First, note that the CR flux $\mathcal{S}$ is a
conserved quantity: $d\mathcal{S}/dx = 0$, since it is assumed that
there are no sources of CR within the flux tube. Then, using
(\ref{EqnGmu}), (\ref{EqnGenSol}) and (\ref{EqnOrthog}):
\begin{equation}
  \mathcal{S} = \frac{v}{2}
\int_{-1}^{1} {\rm d}\mu \, \mu f = - \frac{a_{\rm diff}}{2} 
\int_{-1}^{1} {\rm d}\mu \, D_{\mu\mu} \left( g' \right)^{2} \;.
\end{equation}
Now consider boundaries at 
$x = \pm d$ ($d > 0$), 
and impose a positive flux, so that $a_{\rm  diff} < 0$.
This procedure is valid as long as there is a 
non-zero component of the global CR density 
gradient when projected onto the $x$-axis, whatever its origin --- 
see Section~\ref{Discussion} for a
more detailed discussion of this point. 
In principle, the solution at the boundaries can contain
the diffusive solution plus an arbitrary isotropic component, plus
components that will either decay or grow exponentially as one moves
from the boundary towards Earth. If $d$ is large, in the sense that 
\begin{equation}
\textrm{exp}\left(-\Lambda_{\rm 1}d/v\right)\ll 1 \,,
\label{Eqnmfpcondition}
\end{equation}
then it is clear from Eq.~(\ref{EqnGenSol}) that the terms with
$\Lambda_{\rm i} > 0$ are important only in a {\em boundary layer}
close to $x = -d$, and those with $\Lambda_{\rm i} < 0$ are important
only in the corresponding layer close to $x = d$.  In this case, within
most of the flux tube, including the position of the Earth, the
solution is accurately represented by
\begin{equation}
  f(x,\mu) = a_{0} + a_{\rm diff} \left[ x + g(\mu) \right] \;.
  \label{Eqn_f}
\end{equation}
The {\em CR anisotropy} in this region is proportional 
to $g$, and equals $a_{\rm diff} \, g(\mu)/a_{0}$ at Earth.
A spatial diffusion coefficient, $\kappa_{\parallel}$
and its associated mean free path $\lambda_{\parallel}$ can be defined
via Fick's Law:
\begin{align}
\kappa_{\parallel}&=-\frac{\mathcal{S}}{\partial f/\partial x}
\nonumber\\
&=\frac{1}{2}\int_{-1}^{1}{\rm d}\mu D_{\mu\mu}\left(g'\right)^2
\\
\lambda_{\parallel}&=\frac{3}{v}\kappa_{\parallel} \,.
\label{MFP_Eqn}
\end{align}

Equation~(\ref{EqnGmu}) with the condition (\ref{IntGeqZero}) can be solved to give
\begin{align}
  g(\mu) &= - \frac{v}{2}\int_0^{\mu} {\rm d}\mu' \, \frac{1-\mu'^{2}}{D_{\mu'\mu'}} \,,
\label{FormulaGmu}
\\
\noalign{\hbox{so that}}
\kappa_{\parallel}&=\frac{v^2}{4}\int_0^1{\rm d}\mu\frac{\left(1-\mu^2\right)^2}{D_{\mu\mu}}
\label{FormulaGmu_Dmumu_Sym}
\end{align}
\citep[cf.][]{hasselmannwibberenz70}. 

Note that Eq.~(\ref{Eqnmfpcondition}) is a sufficient condition for the accuracy of 
(\ref{Eqn_f}), which is usually called the \lq\lq diffusion approximation\rq\rq. There is, in particular, 
no formal restriction on the size of the mean free path $\lambda_{\parallel}$, which can substantially exceed
the length $2d$ of the flux tube without jeopardising the applicability 
of the diffusion approximation. 
The relationship between 
$\lambda_{\parallel}$ and $\Lambda_1$ depends on the functional form of the
pitch-angle diffusion coefficient $D_{\mu\mu}$. In the following, we evaluate these quantities
using the method outlined in \citet{kirketal00}. 
Since we consider only $D_{\mu\mu}$ which are even functions of $\mu$, 
the anisotropy is an odd function. Furthermore, the properties of the 
local turbulence constrain only the shape of the anisotropy, whereas its amplitude depends on the 
externally imposed flux. Therefore, we work with the normalized
quantity
\begin{align}
\hat{g}(\mu)&=g(\mu)/g(1)
\label{NormGmu}
\end{align}
on the range $0\le\mu\le 1$, and, in order to simplify the notation, will drop the circumflex from
here onwards. 
In the special case of isotropic pitch-angle diffusion, $D_{\mu\mu} \propto 1-\mu^{2}$,
one finds $\Lambda_{1}/v=7.26/\lambda_{\parallel}$. This is also the only case in which the anisotropy 
has exactly the form of a dipole: $g(\mu)=\mu$.

\subsection{Pitch-angle diffusion}

In the so-called \lq\lq quasi-linear\rq\rq\ approach, 
the pitch-angle diffusion coefficient is computed from the expression 
\citep[e.g.][]{KulsrudPearce69,Voelk73,Voelk75}:
\begin{align}
  D_{\mu\mu} &= \Omega^{2}  \left( 1 - \mu^{2} \right) \int {\rm d}^{3}k \int_{0}^{\infty} {\rm d}\tau 
\nonumber\\
&\sum_{n=-\infty}^{\infty} \bigg( \frac{n^{2}J_{\rm n}^{2}(z)}{z^{2}} M_{\rm A}({\bf k},\tau) 
+ \frac{k_{\parallel}^{2} J_{\rm n}^{\prime 2}(z)}{k^{2}} M_{\rm P}({\bf k},\tau) \bigg) \;,
\label{ExpressionDmumu_1}
\end{align}
where $\Omega$ is the Larmor frequency, $k_{\parallel}$ and
$k_{\perp}$ are the components of {\bf k} parallel and 
perpendicular to the local field lines, respectively, $z=k_\perp l \eps \sqrt{1-\mu^2}$, 
where $l$ is the outer scale of the turbulence and $\eps = v/(l\Omega) = r_{\rm L}/l$, 
and $J_{\rm n}(z)$ is a Bessel function of the 
first kind ($'$ denotes a derivative with respect to the argument $z$).

The quantities 
$M_{\rm A}$ and $M_{\rm P}$ represent normalized wave power
spectra. Following \citet{Chandran:2000hp}, they 
can be derived from the spatial Fourier transform of the turbulent magnetic field
${\bf B}_{1,{\rm w}}({\bf k},t)$, where ${\bf k}$ is the wave-vector and $t$ the time:
\begin{equation}
  M_{\rm w}({\bf k},\tau ) = \langle {\bf B}_{\rm 1,w}({\bf k},t) \cdot {\bf B}_{\rm 1,w}^{\ast}({\bf k},t+\tau) \rangle / B_{0}^{2} \;.
\end{equation}
Here, $B_0$ is the regular part of the local (within a few tens of parsecs) magnetic field, 
${\rm w} \in \{ {\rm A,P} \}$ denotes the wave mode, and $\langle\dots\rangle$
represents an ensemble-average. These quantities can also be defined in terms of  
time-averages over a single realisation of stationary turbulence, but the distinction
is unimportant for our investigation, since they enter our calculations only in the 
normalized energy spectra and resonance 
functions defined in section~\ref{resonancefunctions}, for which we adopt 
phenomenological descriptions. 
Again following \citet{Chandran:2000hp}, we assume equipartition between the magnetic and kinetic 
energy densities, and equal energy densities in each wave mode, and chose the normalization of the  
fluctuations such that the total magnetic energy in 
turbulence of wavelength shorter than that of the outer scale, $l$, 
equals that in the ambient field $B_0$. 
Shear Alfv\'en waves, which are described by $M_{\rm A}$, 
do not possess a fluctuating component of the magnetic field
in the direction parallel to the ambient field and, therefore, do not contribute to the $n=0$ term 
in the summation. On the other hand, the slow and fast modes in compressible MHD, 
as well as the pseudo-Alfv\'en mode in the model of incompressible MHD do have such a component,
and a corresponding contribution to this term. 
Their power spectra are described by $M_{\rm P}$. Below, we consider a compressible turbulence 
model in which
the fast mode dominates over the slow mode, in which case we replace the suffix \lq\lq P\rq\rq\ by 
\lq\lq F\rq\rq. We also consider 
an incompressible model in which only the pseudo-Alfv\'en mode contributes to this term. In this case, 
because
this mode can be considered as the analog of a slow mode, we replace \lq\lq P\rq\rq\ by \lq\lq S\rq\rq.

\subsubsection{Resonance functions}
\label{resonancefunctions}
In the following, $v_{\parallel} = v \mu$ and $v_{\perp} = v \sqrt{1-\mu^{2}}$ 
denote the components of the CR velocity parallel
and perpendicular to the ambient magnetic field, and $\omega$
is the angular frequency of the waves.
Expression~(\ref{ExpressionDmumu_1}) for $D_{\mu\mu}$ can be written
in terms of a resonance function $R_{\rm n}(k_{\parallel}v_{\parallel}-\omega + n \Omega)$ 
that incorporates the integral over $\tau$:
\begin{equation}
  \begin{aligned}
  &D_{\mu\mu} = \Omega^{2} \left( 1 - \mu^{2} \right) \int {\rm d}^{3}k \sum_{n=-\infty}^{\infty} \bigg( \frac{n^{2}J_{\rm n}^{2}(z)}{z^{2}} \mathcal{I}_{\rm A}({\bf k}) \\
  &+ \frac{k_{\parallel}^{2} J_{\rm n}^{\prime 2}(z)}{k^{2}} \mathcal{I}_{\rm S,F}({\bf k}) \bigg) \times R_{\rm n}(k_{\parallel}v_{\parallel}-\omega + n \Omega) \,,
  \end{aligned}
\label{EqnDmumu}
\end{equation}
where $\mathcal{I}_{\rm A,S,F}$ 
are the 
normalized energy spectra corresponding to the power spectra $M_{\rm A,S,F}$. 

In quasi-linear theory \citep[e.g.,][]{jokipii66}, the resonance
function $R_{\rm n}$ is a delta function. The shortcomings of this theory,
such as the $90^{\circ}$-scattering problem \citep{jonesetal78} are
well known. We note that this theory cannot account for the CR
anisotropy data: as we show in Section \ref{Results}, a broad peak in
$D_{\mu\mu}/(1-\mu^{2})$ around $\mu=0$ is essential to reproduce the
flattening observed in the CR anisotropy data by IceTop and IceCube
for CRs with pitch-angles around $\theta = 90^{\circ}$.  Therefore, in
this study, we consider two generic, phenomenological resonance
functions.

For the first type of resonance function, we choose a Breit-Wigner distribution:
\begin{align}
\MoveEqLeft  R_{\rm n, 1}(k_{\parallel}v_{\parallel}-\omega + n \Omega) 
\nonumber
\\
  & = \mathcal{R}e \left( \int_{0}^{\infty} {\rm d}\tau \, 
\textrm{e}^{-i (k_\parallel v_\parallel - \omega + n \Omega) \tau -\tau/\tau_{\rm w} }\right) 
\nonumber\\
  & = \frac{\tau_{\rm w}^{-1}}{(k_\parallel v_\parallel - \omega + n \Omega)^{2} + \tau_{\rm w}^{-2}} \;
\label{Eqn_Rn1}
\end{align}
where the broadening
  of the resonance, described by $\tau_{\rm w}$ with $w \in {A,F}$, is 
  assumed to be dominated by the Lagrangian
  correlation time for the turbulence. For Alfv\'en and slow (or pseudo-Alfv\'en) modes, 
   $\tau_{\rm
    A} = l^{1/3}/(v_{\rm A} k_{\perp}^{2/3})$ 
\citep{Goldreich:1994zz}, where $v_{\rm A}$ is the Alfv\'en velocity. This 
function corresponds to that used by~\cite{Chandran:2000hp}. 
In the case of fast modes,
\cite{Cho:2002qi} found cascading on a
  time scale $l^{1/2}v_{\rm A}/(k^{1/2} v_{\rm l}^{2})$, where $v_{\rm
    l}$ is the turbulent velocity (at the injection scale). We take
  $v_{\rm l} \approx v_{\rm A}$, because the turbulence in the
  interstellar medium is thought to 
have an Alfv\'en Mach number $\mathcal{M}_{\rm A}\sim 1$. 
Therefore, $\tau_{\rm F} = l/(v_{\rm A} \tilde{k}^{1/2})$, 
where $\tilde{k} = kl$.

For the second type of resonance function, we choose the
  function proposed by~\cite{Yan:2007uc}. The magnetic field strength
  experienced by a cosmic ray depends on its location in space, both
  due to the fluctuations of the small-scale magnetic field {\bf
    B$_{1}$}, and to the fluctuation of the local large-scale field
  {\bf B$_{0}$} within our local flux tube. The latter effect
  therefore includes focussing (or defocussing) of local magnetic
  field lines, which is not taken into account by $R_{\rm n, 1}$. 
  Since the adiabatic invariant $v_{\perp}^{2}/B$ is
  conserved under large-scale fluctuations, the CR pitch-angle varies, 
  which leads to a broadening of
  the resonance. It can be shown that
  variations induced in $v_{\parallel}$ are dominated by the
  variations $\delta B_{\parallel}$ in the parallel magnetic field
  \citep{Voelk75,Yan:2007uc}, in terms of which, it can be written as:
\begin{align}
  \MoveEqLeft
  R_{\rm n, 2}(k_{\parallel}v_{\parallel}-\omega + n \Omega) 
\nonumber\\
&=\mathcal{R}e \Bigg( 
\int_{0}^{\infty} {\rm d}\tau \, 
\textrm{e}^{-i (k_\parallel v_\parallel - \omega + n \Omega) \tau-k_{\parallel}^{2} v_{\perp}^{2}\delta\mathcal{M}_{\rm A} \tau^{2}/4} \Bigg) 
\nonumber
\\
  & = \frac{\sqrt{\pi}}{\left|k_{\parallel}\right| v_{\perp} \delta\mathcal{M}_{\rm A}^{1/2}} 
\exp \left( - \frac{(k_\parallel v_\parallel - \omega + n \Omega)^{2}}{k_{\parallel}^{2} v_{\perp}^{2} \delta\mathcal{M}_{\rm A}}\right)
\label{Eqn_Rn2}
\end{align}
where $\delta\mathcal{M}_{\rm A} = 2\sqrt{\langle
  \delta B_{\parallel}^{2} \rangle / B_{0}^{2}}$ .
\cite{Yan:2007uc} identified $\delta\mathcal{M}_{\rm A}$ with the Alfv\'en Mach
number $\mathcal{M}_{\rm A}$ of the turbulence.
However, in our case, the fluctuations of relevance for 
the shape of the CR anisotropy are those experienced by CRs 
within our local magnetic flux tube, which, for consistency, are 
assumed to be smaller than the Alfv\'en
Mach number of the turbulence. In order to exhibit the sensitivity of
our results to the unknown local value of $\delta\mathcal{M}_{\rm A}$, 
we perform calculations
for five different values: $\delta\mathcal{M}_{\rm A}=0.01$, $0.033$, $0.1$,
$0.33$, and 1.

\subsubsection{Turbulence models}


\begin{deluxetable*}{ccccccc}
\label{tableoverview}
\tablecolumns{7}
\tablecaption{Properties of the turbulence models \label{tableoverview}}
\tablehead{\colhead{Model} & \colhead{Type} & \multicolumn{2}{c}{Spectrum} & \colhead{Resonance function} & \colhead{Section} & \colhead{Figures} \vspace{0.03cm} \\
 \cline{3-4} \vspace{-0.25cm}}
\startdata 
A & GS (incompressible)       & Anisotropic & Heaviside ($\mathcal{I}_{\rm A,S,1}$)   & Narrow -- $R_{\rm n,1}(\delta=v_{\rm A}/v)$      & \ref{I1_Rn1}   & \ref{CHACHA_2} \\
B & GS (incompressible)       & Anisotropic & Exponential ($\mathcal{I}_{\rm A,S,2}$) & Narrow -- $R_{\rm n,1}(\delta=v_{\rm A}/v)$      & \ref{I2_Rn1}   & \ref{CHALAZ}\\
C & GS (incompressible)       & Anisotropic & Heaviside ($\mathcal{I}_{\rm A,S,1}$)  & Broad -- $R_{\rm n,2}(\delta\mathcal{M}_{\rm A})$ & \ref{I1_Rn2}   & \ref{LAZCHA}\\
D & GS (incompressible)       & Anisotropic & Exponential ($\mathcal{I}_{\rm A,S,2}$) & Broad -- $R_{\rm n,2}(\delta\mathcal{M}_{\rm A})$ & \ref{I2_Rn2}   &\ref{LAZLAZ_1}, \ref{LAZLAZ_2}\\
E & Fast modes (compressible) & Isotropic   & $\mathcal{I}_{\rm F} \propto k^{-3/2}$ & Narrow -- $R_{\rm n,1}(\delta=v_{\rm A}/v)$        & \ref{Fast_Rn1} & \ref{FastCHA}\\
F & Fast modes (compressible) & Isotropic   & $\mathcal{I}_{\rm F} \propto k^{-3/2}$ & Broad -- $R_{\rm n,2}(\delta\mathcal{M}_{\rm A})$ & \ref{Fast_Rn2} &  \ref{FastLAZ}
\enddata 
\end{deluxetable*}


One of the leading theories of incompressible MHD
turbulence is that of \citet{Goldreich:1994zz}, which predicts an
anisotropic power-spectrum for both Alfv\'en and pseudo-Alfv\'en modes.
Fluctuations with $|k_\parallel| \gtrsim |k_\perp|^{2/3} l^{-1/3}$ 
are expected to be 
strongly suppressed, in agreement with MHD simulations of \citet{Cho:2001hf}. 
Below, we investigate the effects on the CR anisotropy, using two
phenomenological prescriptions for the suppression taken from the literature: 
$\mathcal{I}_{\rm A,S} = \mathcal{I}_{\rm A,S,1} \propto
k_\perp^{-10/3} h(k_{\parallel}l^{1/3}/k_\perp^{2/3})$ where $h(y)=1$
if $|y| < 1$, and $h=0$ otherwise (see \cite{Chandran:2000hp}), and
$\mathcal{I}_{\rm A,S} = \mathcal{I}_{\rm A,S,2} \propto
k_\perp^{-10/3} \exp (-k_{\parallel}l^{1/3}/k_\perp^{2/3})$ (see
\cite{Cho:2001hf}, \cite{Yan:2002qm}). The latter presents a less
abrupt cutoff in $k_\parallel$ than the former,
which influences the resulting shape of the CR anisotropy,
and the CR mean free path --- see Section~\ref{Results}.

In the case of compressible turbulence, we keep the same expressions
for Alfv\'en and slow modes. We assume that fast magnetosonic modes
have an isotropic energy spectrum, following $\mathcal{I}_{\rm F}({\bf
  k}) \propto k^{-3/2}$, as was found in~\cite{Cho:2002qi}'s
simulations of compressible MHD turbulence. 

To summarize, we consider the anisotropies generated by six distinct turbulence models, whose 
properties are listed in Table~\ref{tableoverview}.

\section{Results}
\label{Results}

For each of the models listed in Table~\ref{tableoverview}, 
we first calculate the (dimensionless)
pitch-angle scattering frequency $\nu(\mu)=2D_{\mu\mu}/(1-\mu^{2})
\times (l/v)$ using Eq.~(\ref{EqnDmumu}). Then, we derive 
the smallest positive eigenvalue $\Lambda_1$, by applying 
the method described in \citet{kirketal00} to 
Eq.~(\ref{EqnEigenvalues}),
and the normalized anisotropy $g(\mu)$, using Eqs.~(\ref{FormulaGmu}) and (\ref{NormGmu}).
(Note that isotropic CR scattering
corresponds to $\nu(\mu)={\rm constant}$ and $g(\mu)=\mu$.)

In terms of an outer turbulence scale $l=l_{100\,{\rm pc}}\times 100\,$pc and a 
magnetic field $B_0=B_{6\,\mu{\rm G}}\times 6\,\mu$G, one has 
$\eps = 1.8 \times 10^{-3} \, E_{\rm CR,\,PeV} \, l_{100\,{\rm
    pc}}^{-1} \, B_{6\,\mu{\rm G}}^{-1}$, where $E_{\rm CR,\,PeV}$
is the CR energy in units of 1\,PeV. 
Therefore, assuming an outer scale-length between 
$1$ and $100\,$pc (see
Sect.~\ref{Discussion}), the relevant range 
of the parameter
$\eps$ is $\sim 10^{-1}-10^{-3}$, for PeV CRs, and an order of magnitude smaller 
for $100\,$TeV CRs. For $\delta=v_{\rm A}/v$, an interstellar Alfv\'en speed
between $10$ and $300\,\textrm{km\,s}^{-1}$ implies values between
$3\times10^{-5}$ and $10^{-3}$, and for the rather uncertain parameter
$\delta\mathcal{M}_{\rm A}$, we adopt the range $10^{-2}$--$1$.

We use analytical expressions for $D_{\mu\mu}$ only in model~A; in the other five models, 
we calculate $D_{\mu\mu}$ by integrating 
Eq.~(\ref{EqnDmumu}) numerically; details are given in Appendix~\ref{appendixA}. 
For Goldreich-Sridhar (GS) turbulence, we compute both the $n=\pm
1$ term of the contribution from Alfv\'en modes, and the $n=0$ term
for pseudo-Alfv\'en modes ---see formulae in Appendix~A. As a safety
check, we have also calculated the $n=\pm 1$ term for pseudo-Alfv\'en modes
for models B, C, and D --- see Eqs.~(\ref{Dmumu_Slow_n1_CHALAZ}),
(\ref{Dmumu_Slow_n1_LAZCHA}), and~(\ref{Dmumu_Slow_n1_LAZLAZ}) --- and 
find that the impact of these terms on $\nu$ and $g$ is negligible.

Our predictions rely on the assumption that the boundary layers at the
ends of the flux tube containing the Earth are narrow. Specifically, 
we interpret Eq.~(\ref{Eqnmfpcondition}) as the requirement 
$\exp (-\Lambda_{\rm 1}d/v ) < 0.10$, which implies
\begin{align}
v/(\Lambda_{1}l) &< 0.43 \, d/l
\nonumber\\
&=0.43\,(\eps/(3.6\times 10^{-3}))\,(d/50\,{\rm pc})\,E_{\rm
  CR,\,PeV}^{-1}\,B_{6\,\mu{\rm G}}
\label{lambda1limit}
\end{align}
Note that if $\eps$ is sufficiently
large, one can fit several times the length of the outer scale within
$d$, which allows for $v/(\Lambda_{1}l) > 0.43$. In this case, our computation of 
$D_{\mu\mu}$ implicitly assumes that the energy density in fluctuations of wavelength 
between $l$ and $d$ is negligible. On the other hand, if $l$ exceeds the chosen value for $d$, 
condition (\ref{lambda1limit}) is too strict, since 
most of the
turbulence power is in modes with $1/|{\bf k}| \sim l$, and the field lines
are expected to remain coherent on scales of a fraction of $l$,
which implies an effective flux tube length $\lesssim l/2$. 
Choosing
 $B_{6\,\mu{\rm G}}=1$, and $d=50$\,pc, we therefore adopt 
\begin{align}
  \frac{v}{\Lambda_{1}l} \leq 0.43 \, \max \left( \frac{1}{2}, \, \frac{\eps}{3.6\times 10^{-3}} \right) \;,
\label{EqnShadedArea}
\end{align}
as the criterion for acceptability of a turbulence model.

\begin{figure*}
  \centerline{\includegraphics[width=0.49\textwidth]{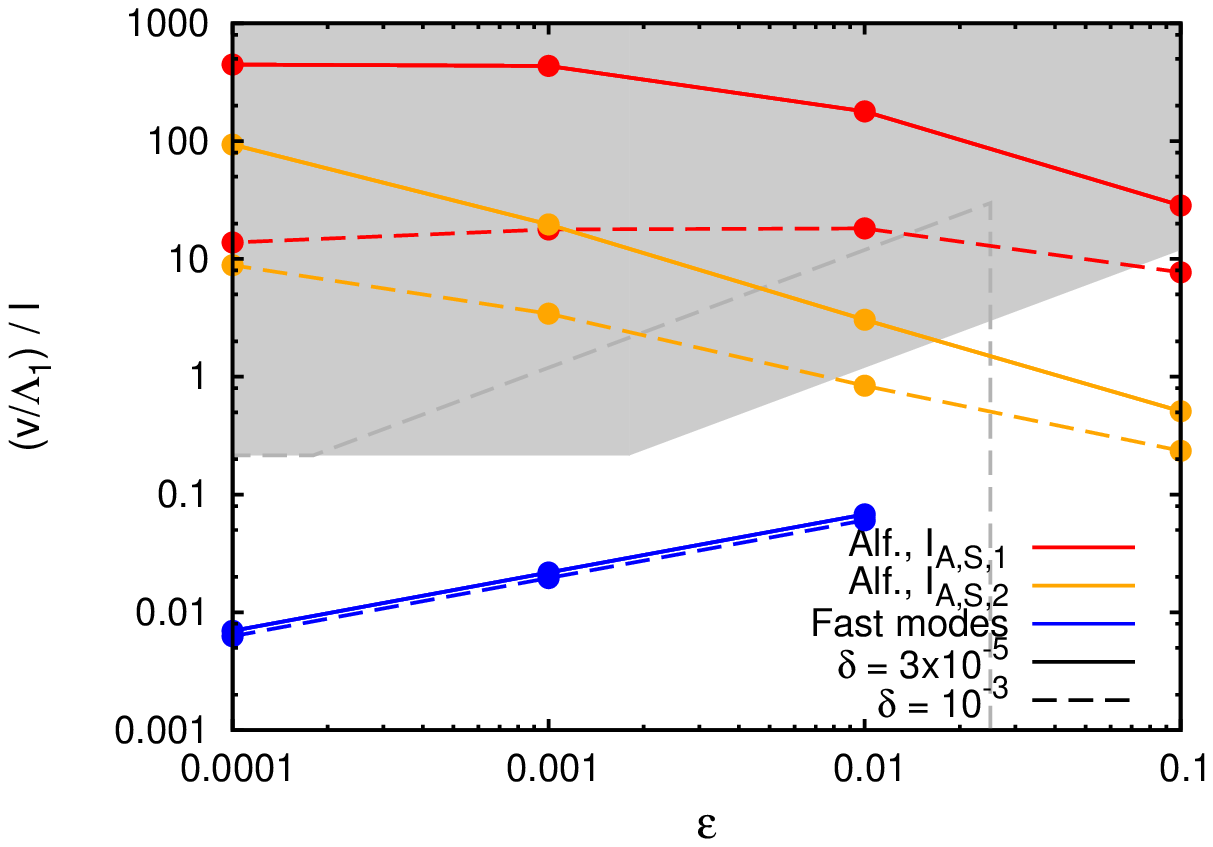}
              \hfil
              \includegraphics[width=0.49\textwidth]{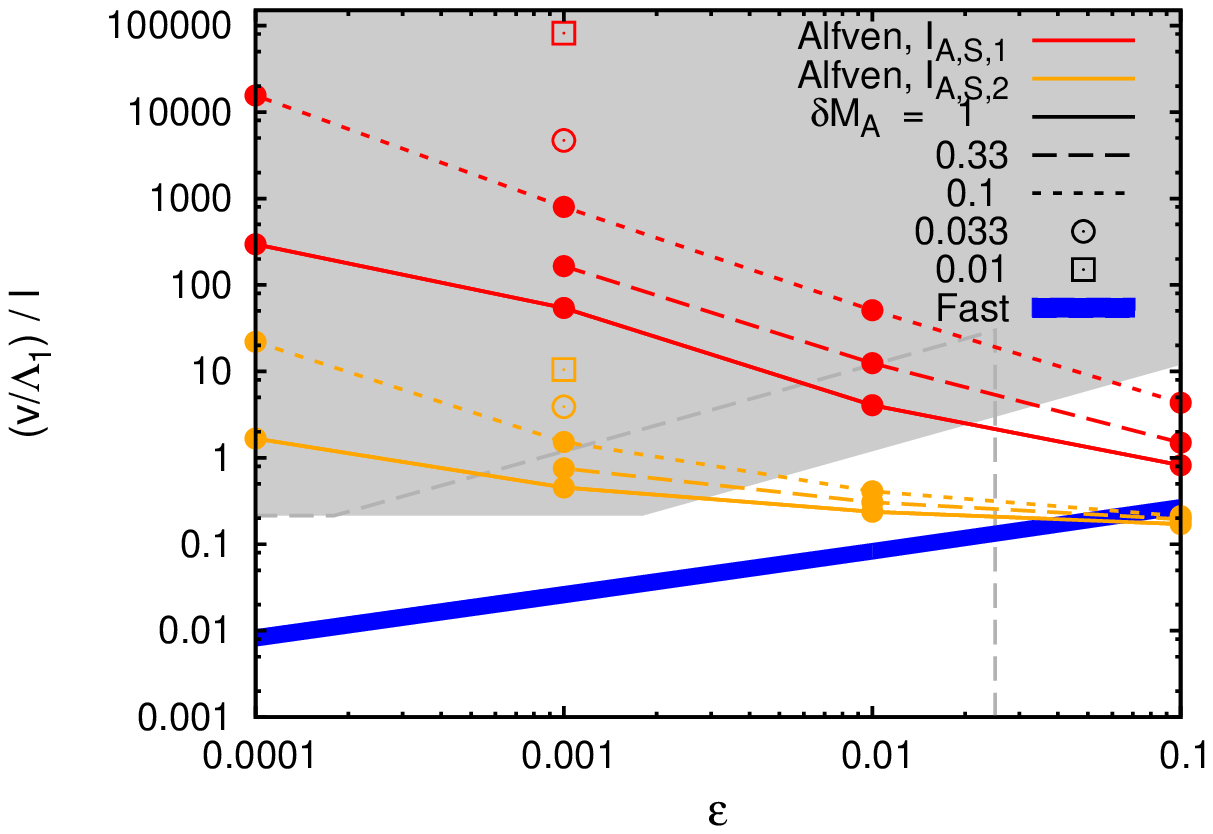}
              }
              \caption{$(v/\Lambda_{1})/l$ versus $\eps$, where
                $\Lambda_{1}$ is the first eigenvalue (see
                Eq.~(\ref{EqnEigenvalues})), for GS turbulence with
                $\mathcal{I}_{\rm A,S} = \mathcal{I}_{\rm A,S,1}$ (red
                lines) or $\mathcal{I}_{\rm A,S,2}$ (orange lines),
                and for isotropic fast modes (blue lines or
                area). {\it Left panel:} $R_{\rm n} = R_{\rm n,1}$ (models A, B \& E);
                {\it Right panel:} $R_{\rm n} = R_{\rm n,2}$ (models C, D \& F). See keys
                for values of $\delta$ and $\delta \mathcal{M}_{\rm
                  A}$. Blue area on the right panel for $\delta
                \mathcal{M}_{\rm A} \in [0.01,1]$. The area which is
                not shaded in grey satisfies
                Eq.~(\ref{EqnShadedArea}), see text for
                parameters. It corresponds to the region where the CR
                anisotropy at Earth for PeV CRs is described by
                Eq.~(\ref{FormulaGmu}). The area shaded in grey is
                the excluded region where the observer is within the
                boundary layers as defined in
                Sect.~\ref{Expression_Anisotropy} (see
                Eq.~(\ref{Eqnmfpcondition})). The region above the
                dashed grey line corresponds to the excluded region
                for 100\,TeV CRs.}
\label{FirstEigenvalue}
\end{figure*}
For a range of the parameters
$\delta$ and $\delta\mathcal{M}_{\rm A}$,
Fig.~\ref{FirstEigenvalue} plots $v/\left(\Lambda_1 l\right)$ as a
function of $\eps$ for all six models listed in
Table~\ref{tableoverview}: A, B and E in the left panel, and
C, D and F in the right panel.  Models that fail to meet
criterion (\ref{EqnShadedArea}) fall in the grey shaded region (for
CRs of energy $1\,$PeV), or above the dashed grey line (for CRs of
$100\,$TeV). In the latter case, the additional constraint
$\eps<0.025$ is included, which is an approximate implementation of the
requirement that $r_{\rm L}<l$ also for CRs of several PeV.
Model~A is the most strongly affected by this criterion: out of the eight cases chosen, 
only $\{\eps,\delta\} = \{10^{-1},10^{-3}\}$ is outside the shaded
area. The fact that the observer is in the boundary layers (see
Sect.~\ref{Expression_Anisotropy}) for such a large fraction of
the parameter space is due to the fact that $\nu \ll 1$ over a
large range of $\mu$. The CR mean free path $\lambda_{\parallel}$ is
also quite large ($\lambda_{\parallel}/l$ versus $\eps$ is plotted in
Fig.~\ref{MFP_All_Figure} ---left panel, in
Appendix~B), ranging from $\simeq 70\,l$ to
$\simeq 2.0\times 10^{3}\,l$, substantially larger than the
outer scale of the turbulence. In most cases,
$\Lambda_{1}\lambda_{\parallel}/v \sim 10$, which implies that the distance between Earth
and the boundaries of the flux tube must be $\approx {\rm
  a~few} \times \lambda_{\parallel}/10$ for criterion
Eq.~(\ref{Eqnmfpcondition}) to be satisfied.

Our results on the anisotropy are presented in several ways. First,
we quantify the ability of each model to give rise to 
a \lq\lq hole\rq\rq\ in the CR flux --- which in our treatment
can only occur axisymmetrically, i.e., 
around a direction parallel or antiparallel to the magnetic field --- 
by plotting angular half-width, $\theta_{1/2}$, defined such that:
\begin{equation}
  g \left( \cos \theta_{1/2} \right) = 1/2 \;.
\label{Definition_Half_Width}
\end{equation}
%

\begin{figure*}
  \centerline{\includegraphics[width=0.49\textwidth]{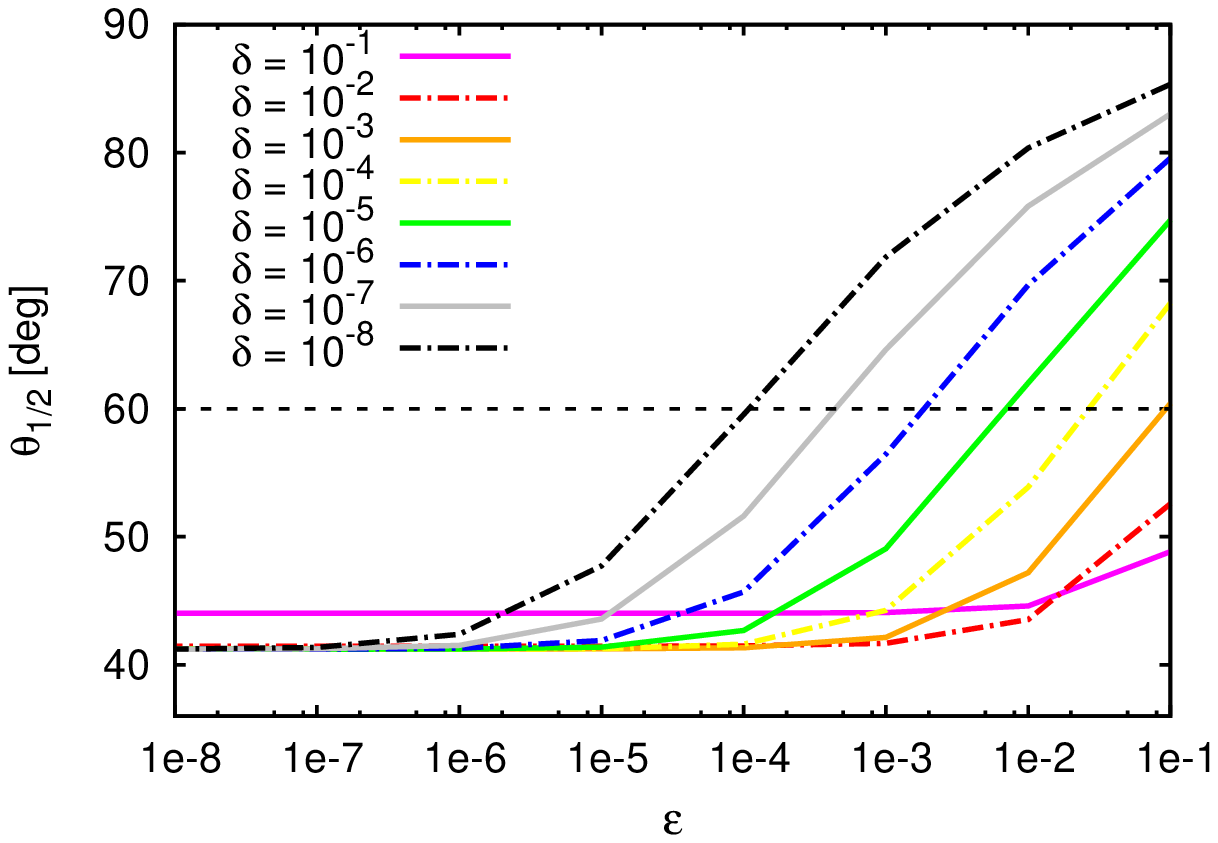}
              \hfil
              \includegraphics[width=0.49\textwidth]{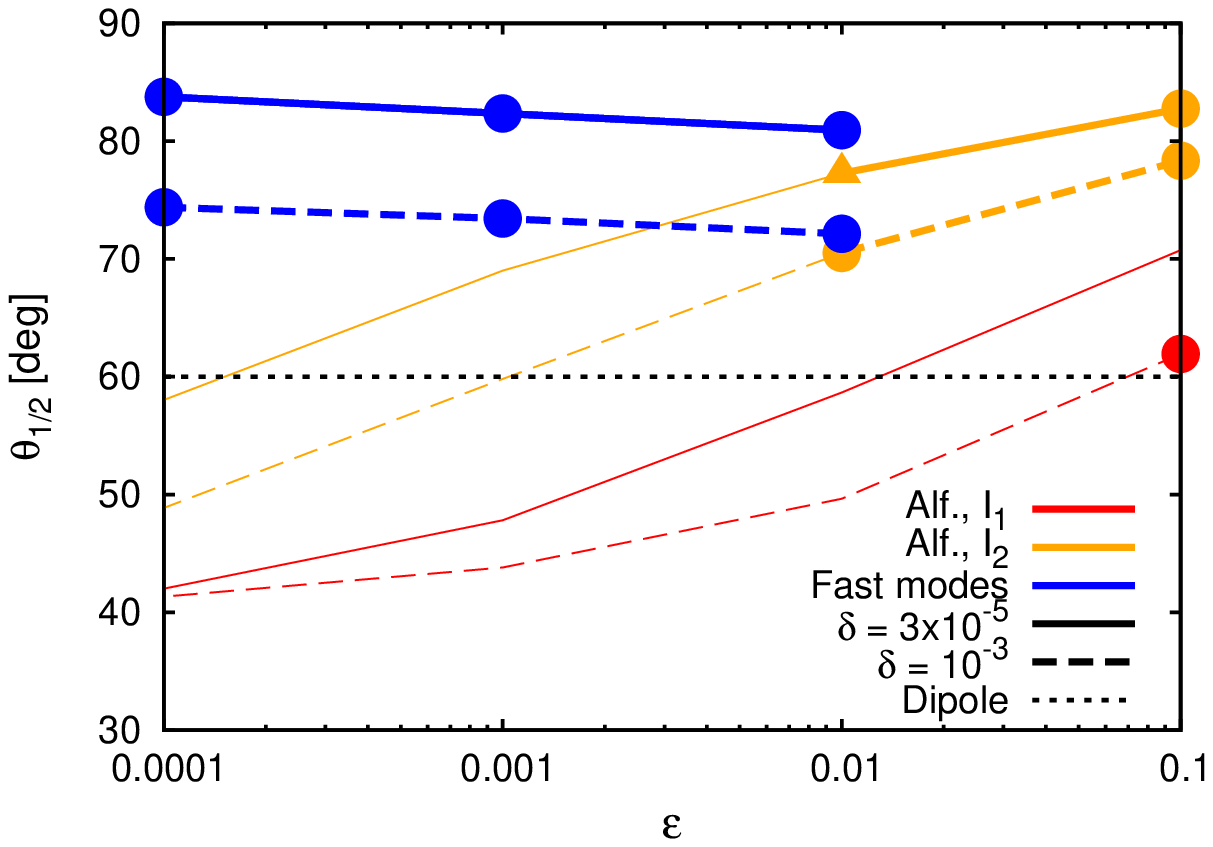}
              }
  \centerline{\includegraphics[width=0.49\textwidth]{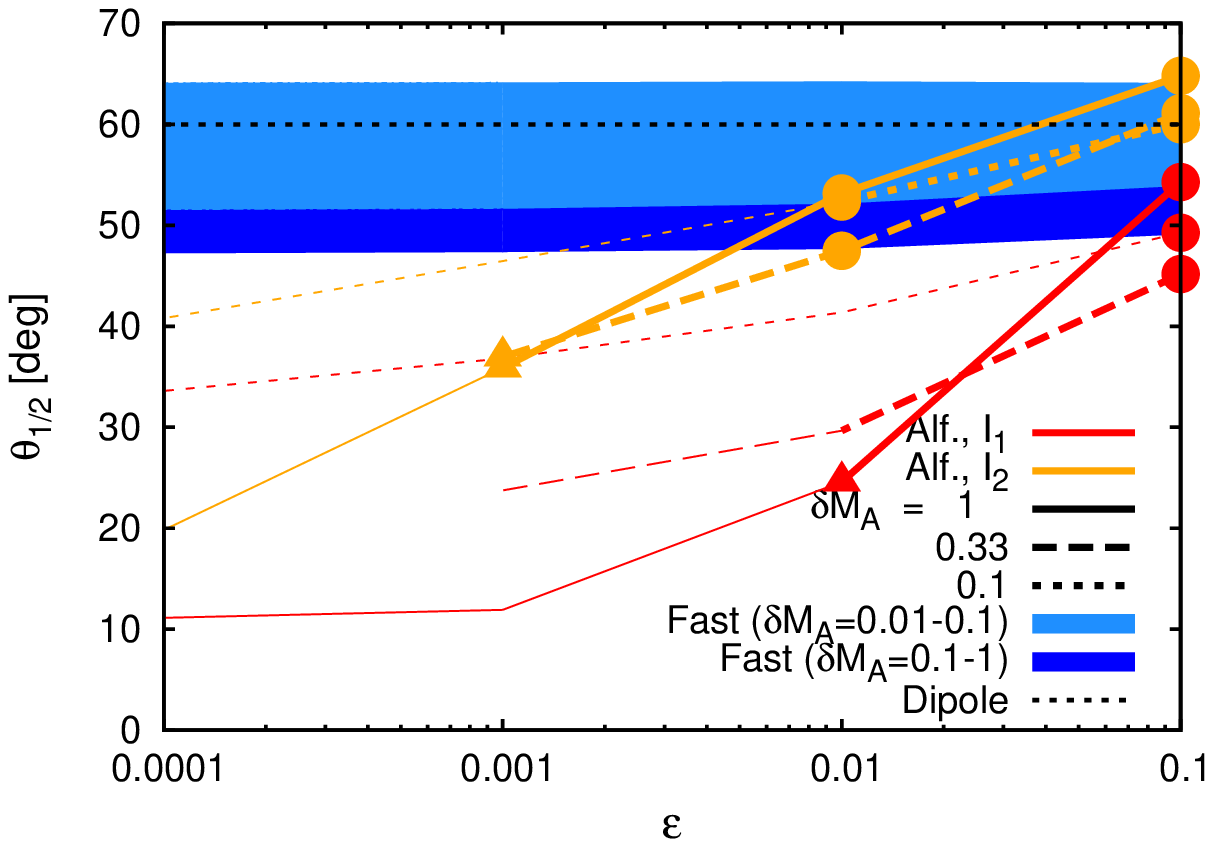}
              \hfil
              \includegraphics[width=0.49\textwidth]{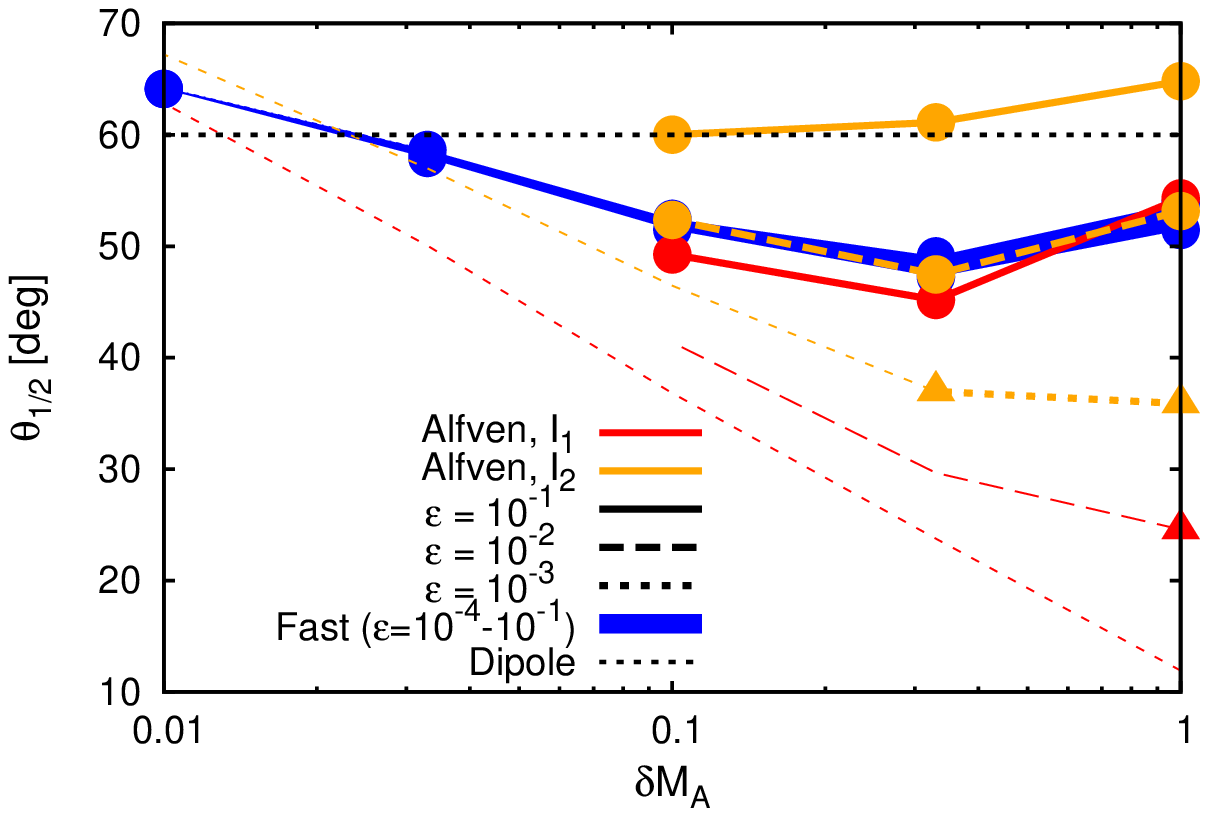}
              }
              \caption{Half-width of the anisotropy, $\theta_{1/2}$,
                for GS turbulence and isotropic fast modes, using
                $R_{\rm n}=R_{\rm n,1}$ ({\it upper row}) or $R_{\rm
                  n,2}$ ({\it lower row}). For a dipole anisotropy,
                $\theta_{1/2} = 60^{\circ}$ (thin black dotted
                lines). {\it Upper left panel:} GS turbulence with
                $\mathcal{I}_{\rm A,S}=\mathcal{I}_{\rm A,S,1}$ (model A) and
                $\delta \in [10^{-8},10^{-1}]$ (see key), using the
                analytical formulae of
                \cite{Chandran:2000hp} for
                $D_{\mu\mu}$. {\it Upper right panel and lower
                  panels:} $D_{\mu\mu}$ is calculated numerically. Red
                lines for GS turbulence with
                $\mathcal{I}_{\rm A,S,1}$ (models A and C), orange lines for $\mathcal{I}_{\rm
                  A,S,2}$ (models B and D), and blue lines or areas for isotropic fast
                modes (models E and F). Filled circles correspond to
                calculations with parameter values outside the area
                shaded in grey, triangles correspond to those below the grey dashed line in
                Fig.~\ref{FirstEigenvalue}. Otherwise, no symbols are plotted. 
Thick lines correspond to lines that are partially
                outside the shaded area in
                Fig.~\ref{FirstEigenvalue}. {\it
                  Upper right panel:} solid (dashed) lines for
                $\delta = 3\times 10^{-5}$ ($\delta =
                10^{-3}$). {\it Lower row:} $\theta_{1/2}$ as a
                function of $\eps$ ({\it left}), and as a function of
                $\delta \mathcal{M}_{\rm A}$ ({\it right}). See key in
                the lower left (resp. right) panel for the dependence
                of $\theta_{1/2}$ on $\delta \mathcal{M}_{\rm A}$
                (resp. $\eps$).}
\label{Half_Width}
\end{figure*}
This quantity is plotted in Fig.~\ref{Half_Width}. In the upper left
panel, the analytic formulae given by \citet{Chandran:2000hp} for
model~A, are exploited to present the half-width as a function of
$\eps$ for a very wide range of $\delta$.  The upper right panel (for
models A, B and E) and lower left panel (for models C, D and F) also
show $\theta_{1/2}$ as a function of $\eps$, but for a more limited
range of $\delta$ and $\delta\mathcal{M}_{\rm A}$. Finally, the lower
right panel illustrates the functional dependence of $\theta_{1/2}$ on
$\delta\mathcal{M}_{\rm A}$ for models C, D and F, for various values
of $\eps$. In each plot, we also show the result for a dipole
anisotropy ($\theta_{1/2}=60^{\circ}$), and highlight those models that 
lie in the allowed region in Fig.~\ref{FirstEigenvalue}.

Each of the following seven figures presents panels containing details of the 
predicted anisotropy for the models in Table~\ref{tableoverview}. Along with 
plots of the dimensionless scattering frequency $\nu(\mu)$ 
and the normalized anisotropy $g(\mu)$ against
$\mu$, we also present 2D sky maps of the anisotropy
and comparisons with the IceTop measurements.
To construct these, we assume the interstellar
  magnetic field (which is measured with a precision of $\approx
  20^{\circ} - 30^{\circ}$) lies in the direction $(l_{\rm
    Gal}=47^{\circ},\, b_{\rm Gal}=25^{\circ})$, as given in Table~1
  of~\citet{Frisch:2012zj}. This choice leads to a 
minimum of the anisotropy at the same right ascension as that observed. 
A more sophisticated fit to the data would take into account the 
uncertainty in the magnetic field direction explicitly, see Sect.~\ref{Discussion}.
At 400\,TeV and 2\,PeV median energy, IceTop provides the variation
 $\Delta N/\langle N\rangle$ of
the CR flux
  as a function of right ascension, averaged over declinations between
  $-75^\circ$ and $-35^\circ$ \citep{Aartsen:2012ma}. To compare these data 
with our predictions, we 
integrate our (axisymmetric) anisotropy over the
range of phase and $\mu$ corresponding to IceTop's range of declination,
add a constant to ensure that the average over right ascension vanishes, and 
normalize the minimum value to coincide with the IceTop data.

\subsection{Goldreich-Sridhar turbulence}
\label{AlfvenModes}

Let us first consider incompressible Alfv\'enic turbulence with a GS power-spectrum.

\subsubsection{Model A ($\mathcal{I}_{\rm A,S} = \mathcal{I}_{\rm A,S,1}$ and $R_{\rm n}=R_{\rm n,1}$)}
\label{I1_Rn1}

\begin{figure*}
  \centerline{
              \includegraphics[width=0.55\textwidth]{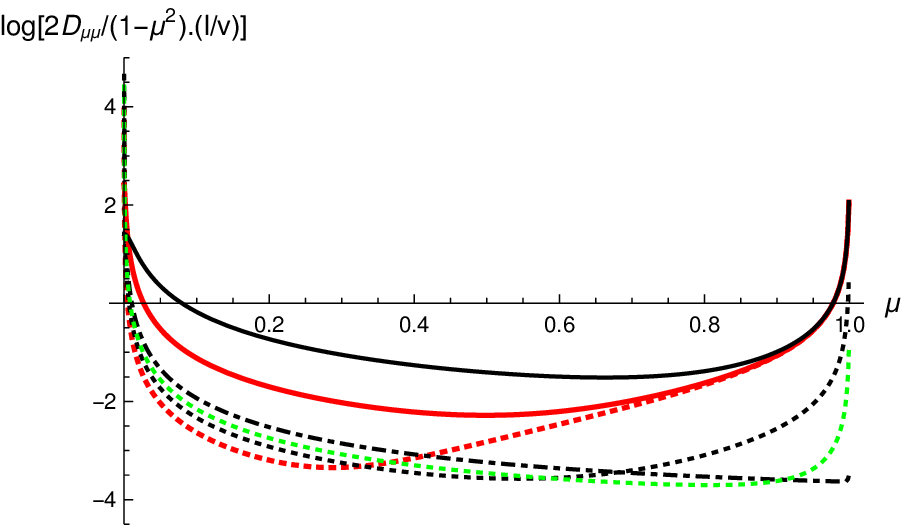}
              \hfil
              \includegraphics[width=0.48\textwidth]{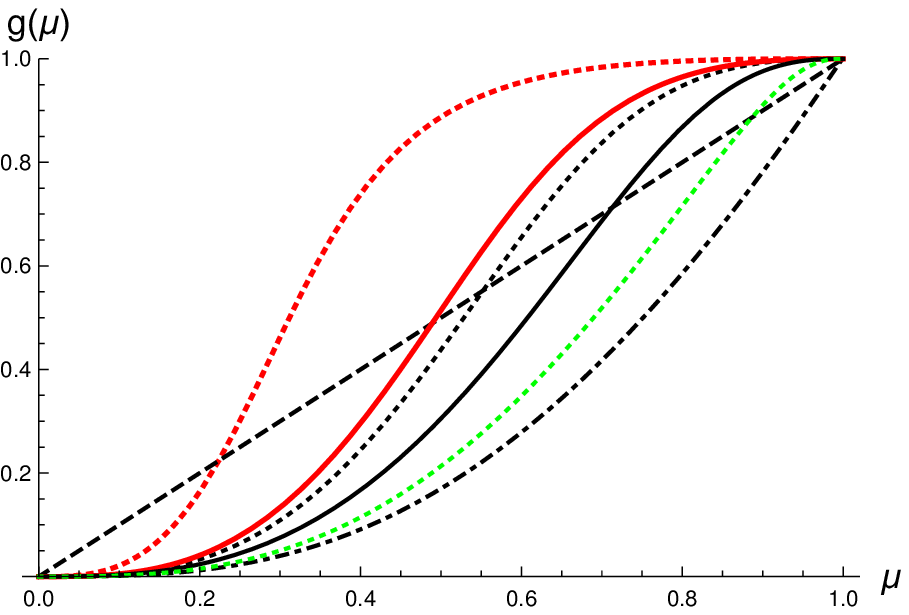}
              }
  \centerline{
              \includegraphics[width=0.33\textwidth]{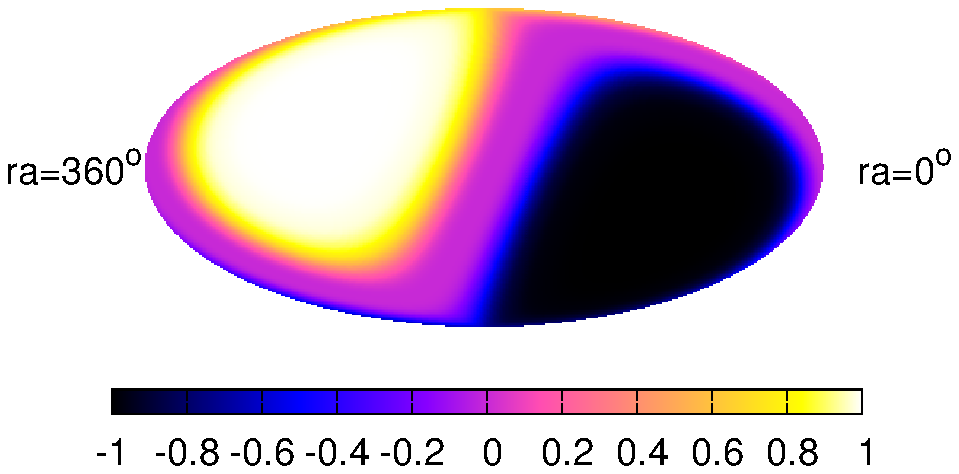}
              \hfil
              \includegraphics[width=0.33\textwidth]{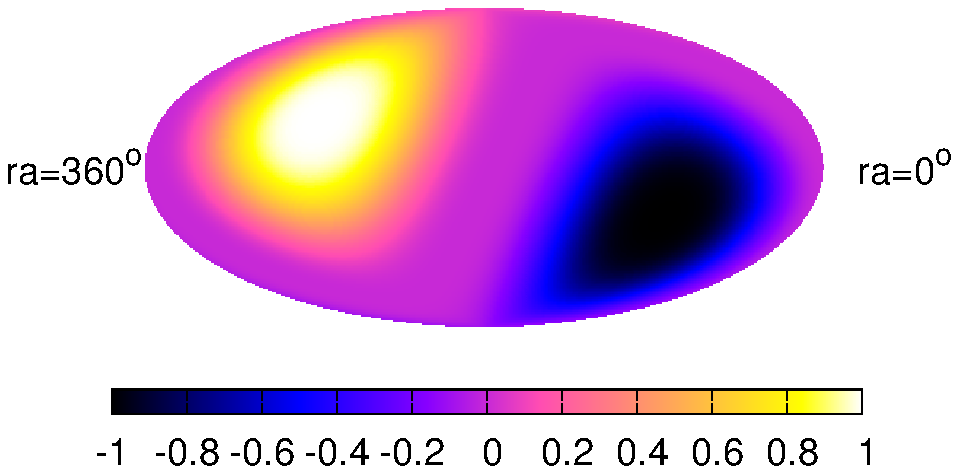}
              \hfil
              \includegraphics[width=0.30\textwidth]{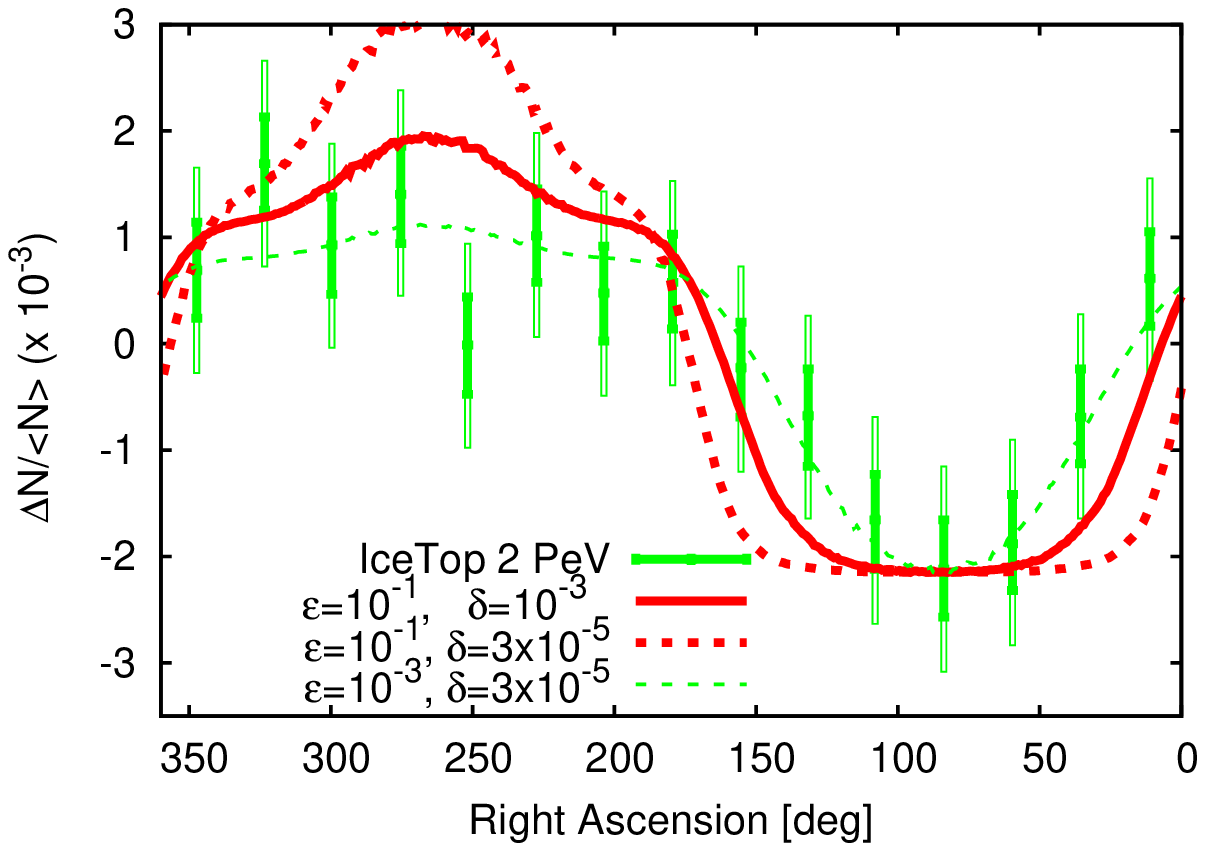}
              }
              \caption{Model~A, GS turbulence with $\mathcal{I}_{\rm
                  A,S} = \mathcal{I}_{\rm A,S,1}$, and using $R_{\rm
                  n,1}$. {\it Upper left panel:} $\nu$ as
                a function of $\mu$, for
                $\{\eps,\delta\}=\{10^{-1},3\times 10^{-5}\}$ (thick
                red dotted line), $\{10^{-2},3\times 10^{-5}\}$ (black
                dotted), $\{10^{-3},3\times 10^{-5}\}$ (green dotted),
                $\{10^{-5},3\times 10^{-5}\}$ (black dashed-dotted),
                $\{10^{-1},10^{-2}\}$ (black solid), and
                $\{10^{-1},10^{-3}\}$ (thick red solid). {\it Upper
                  right panel:} $g(\mu)$, using the same line
                types. {\it Lower left and middle panels:} $g(\mu)$ in
                equatorial coordinates for $\delta=3\times 10^{-5}$,
                and $\eps=10^{-1}$ ({\it left}) or $\eps=10^{-3}$
                ({\it middle}). {\it Lower right panel:} relative CR
                intensity $\Delta N/\langle N \rangle$ at $-75^{\circ}
                \leq {\rm dec} \leq -35^{\circ}$, as a function of
                right ascension, compared with IceTop 2\,PeV data
                set~\citep{Aartsen:2012ma}.}
\label{CHACHA_2}
\end{figure*}

In Fig.~\ref{CHACHA_2}, we plot $\nu$ and $g$ versus $\mu$ in the upper
panels for $\delta=3\times 10^{-5}$, using red, black, 
and green dotted lines, for 
$\eps=10^{-1}$, $\eps=10^{-2}$, and
$\eps=10^{-3}$, respectively.
The transition from the minimum of $\nu$ being
located at small $\mu$ to it being located at $\mu \rightarrow 1$
corresponds to a transition in the shape of $g(\mu)$: it changes from
a function with $\theta_{1/2}$ larger than for a dipole to a function
with small cold spots $\theta_{1/2}\approx 40^{\circ}$. The value of
$\mu$ where $\nu$ reaches its minimum is important, because it defines
the region where $g^{\prime}(\mu) \propto 1/\nu$ reaches its maximum
value. At any $\mu$ where $\nu$ is much greater than its minimum
value, $g(\mu)$ looks flat. As can be seen in the upper right panel of
Fig.~\ref{CHACHA_2}, for the red dotted line (i.e. at large
$\eps=10^{-1}$) $g$ is close to 1 in a broad region $\mu \gtrsim
0.6$. If this case happens to be in a region of parameter space where
the CR anisotropy is given by $g$ (i.e. Eq.~(\ref{Eqnmfpcondition})
is satisfied), the anisotropy would consist of a large hot spot and a
large cold spot, along the direction of magnetic field lines. When the
CR energy decreases, the size of the hot and cold spots decreases,
due to the aforementioned shift in of the minimum of $\nu$ to $\mu
\rightarrow 1$ (see the black dotted and green dotted lines in the upper
right panel). The black dashed-dotted line corresponds to an even lower
CR energy, $\eps = 10^{-5}$, for the same $\delta$. $\theta_{1/2}$ is
smaller than for $\eps = 10^{-3}$. We find that decreasing $\eps$ to
values smaller than $10^{-5}$ does not appreciably change the shape
of $g$. Correspondingly, $\theta_{1/2}$ remains at its minimum in this region. 
In summary, the
hot and cold spots become very wide and flat 
when $\eps$ is large. The half width of the hot and cold spots reaches a
minimum value of about $40^{\circ}$ when $\eps$ is small. This
behaviour of $\theta_{1/2}$ with $\eps$ can be clearly seen in
Fig.~\ref{Half_Width} (upper left panel), where we plot $\theta_{1/2}$
as a function of $\eps$, for eight values of $\delta \in
[10^{-1},10^{-8}]$. The larger $\delta$ is, the larger is the value of
$\eps$ above which $\theta_{1/2}$ significantly departs from its
minimum value. At $\eps$ fixed, increasing $\delta$ has qualitatively
the same effect as decreasing $\eps$ at $\delta$ fixed. The thick red
solid (resp. black solid) line in the upper panels of
Fig.~\ref{CHACHA_2} corresponds to $\delta=10^{-3}$
(resp. $\delta=10^{-2}$) and $\eps=10^{-1}$. By comparing the red
dashed, red solid, and black solid lines, one can see that
$\theta_{1/2}$ decreases and tends towards the limiting behaviour
found above for small $\eps$.

We now consider two rather \lq\lq extreme\rq\rq\ cases,
$\{\eps,\delta\}=\{10^{-1},3\times 10^{-5}\}$ and $\{10^{-3},3\times
10^{-5}\}$ and plot the corresponding anisotropy in equatorial
coordinates in the lower left and centre panels of
Fig.~\ref{CHACHA_2}, respectively.  In the first case,
$\theta_{1/2}>60^{\circ}$, whereas in the second
$\theta_{1/2}<60^{\circ}$, as can be seen in the upper right panel of
Fig.~\ref{Half_Width}.  Both cases lie in the forbidden region of
Fig.~\ref{FirstEigenvalue}, i.e., the actual anisotropy at Earth is 
determined by the (unknown) boundary conditions at the ends of our
flux tube, and not by $g$.  Nevertheless, it is useful to study them,
because they exhibit particularly clear features that are also
present in the more realistic cases. In each panel, a large magenta
region where $g$ is nearly constant and $\simeq 0$ is visible. It
corresponds to directions on the sky that are almost
perpendicular to the local field lines ($\mu \approx 0$). For a
dipole, the magenta region would cover a much smaller area, and be
barely visible with this color key. In the lower centre panel, the
cold and hot spots are quite small. If the CR anisotropy were given by
such a function, an experiment such as IceTop/IceCube, which observes
part of the Southern hemisphere, would see only a single tight cold
spot ($\simeq 40^{\circ}$ in size), surrounded by a region of
approximately constant flux covering the rest of the visible sky. In
the lower right panel of Fig.~\ref{CHACHA_2} we show the relative CR
intensity $\Delta N/\langle N \rangle$ as a function of right
ascension, at declinations $-75^{\circ} \leq {\rm dec} \leq
-35^{\circ}$, as would be observed if the anisotropy were given by
$g$, for three cases: $\{\eps,\delta\}=\{10^{-1},3\times 10^{-5}\}$,
$\{10^{-1},10^{-3}\}$, and $\{10^{-3},3\times 10^{-5}\}$. We keep the
same line types and colors as in the two upper panels. Superimposed on
this, we plot the 2\,PeV data from IceTop~\citep{Aartsen:2012ma} using
green errorbars. For the case shown in the lower centre panel, the
hot spot is almost completely out of sight in the Northern hemisphere:
$\Delta N/\langle N \rangle$ displays a small cold spot and is rather
flat elsewhere (see the thin green dotted line in the lower right
panel).  
Although the model is ruled out by the criteria of Fig.~\ref{FirstEigenvalue},
this line nevertheless provides a good fit to the 2\,PeV data,
whereas for both cases with $\eps=10^{-1}$, the size of the cold spot is too large
to fit the data. For $\eps=10^{-1}$, $\delta=3\times 10^{-5}$, 
the half-width of the hot spot is so large 
 that it would spill over into the Southern hemisphere and 
be visible to IceTop as a big bump in $\Delta
N/\langle N \rangle$ at ${\rm RA}\approx 240^{\circ}-280^{\circ}$.
This case is also ruled out by Fig.~\ref{FirstEigenvalue}.

In summary, there is not enough freedom 
to permit $v/\Lambda_{1}$ to be sufficiently small,
whilst, at the same time, keeping $\theta_{1/2}$ small. Indeed, $\eps$ cannot be increased
to values much larger than $10^{-1}$ because the CR gyroradius should
remain smaller than the outer scale of the turbulence. 
Also, $\delta$ cannot be increased further: $\delta =
10^{-3}$ already corresponds to a rather large value of $v_{\rm A}
\simeq 300$\,km\,s$^{-1}$, compared to that expected in
our local interstellar medium. We show with red lines in
Fig.~\ref{Half_Width} (upper right panel) the values of $\theta_{1/2}$
for these 8 cases. The red dot corresponds to the only case outside
the shaded area in Fig.~\ref{FirstEigenvalue}. Its $\theta_{1/2}$ is
close to $60^{\circ}$.
However, $v/\Lambda_{1}$ is strongly dependent on $\mathcal{I}_{\rm A,S}$, 
and may be substantially smaller, see below.

\subsubsection{Model B ($\mathcal{I}_{\rm A,S} = \mathcal{I}_{\rm A,S,2}$ and $R_{\rm n}=R_{\rm n,1}$)}
\label{I2_Rn1}

\begin{figure*}
  \centerline{\includegraphics[width=0.32\textwidth]{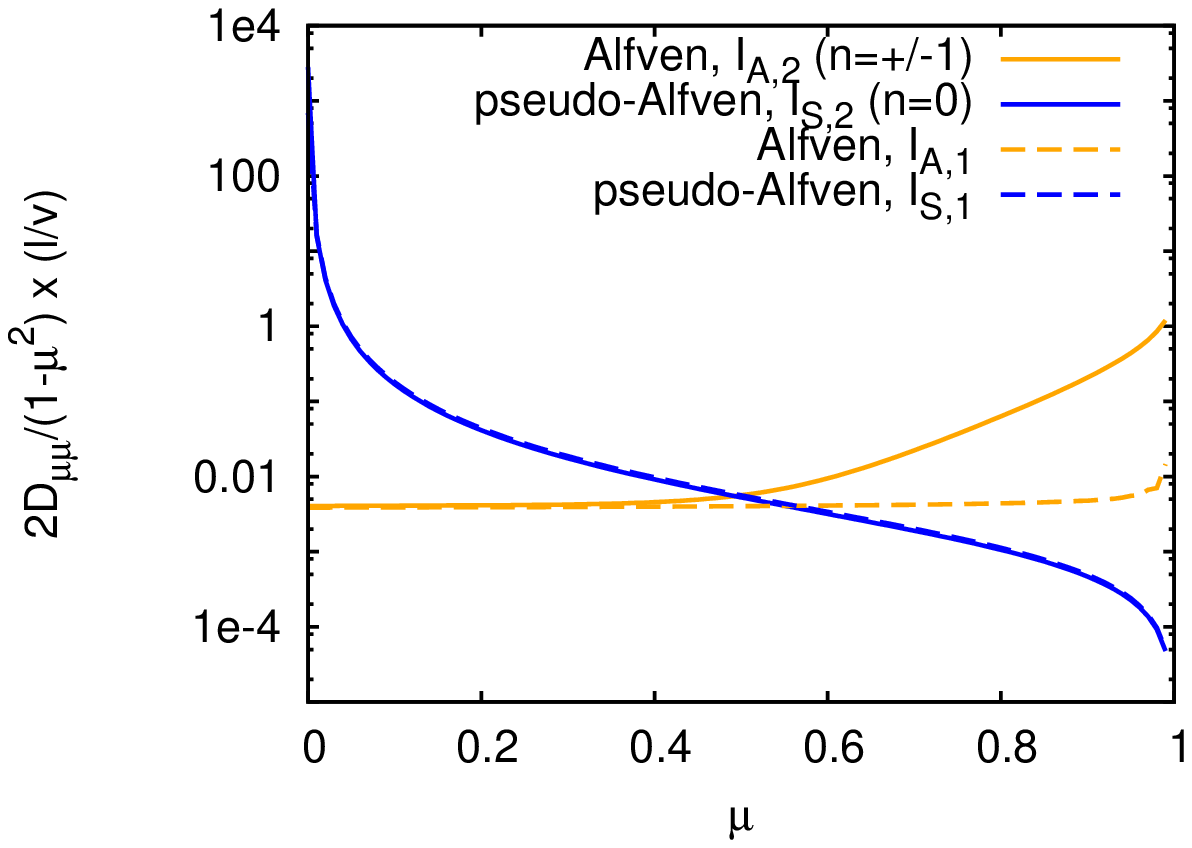}
              \hfil
              \includegraphics[width=0.32\textwidth]{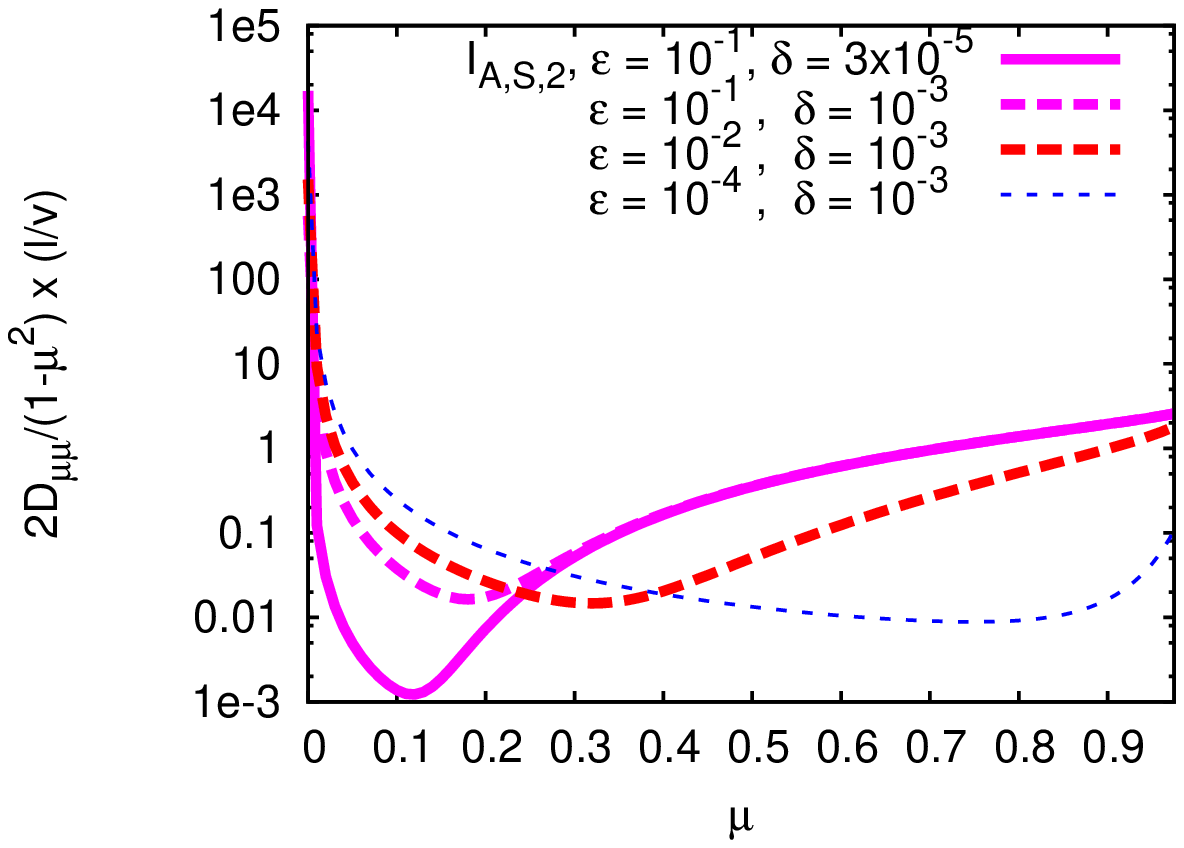}
              \hfil
              \includegraphics[width=0.32\textwidth]{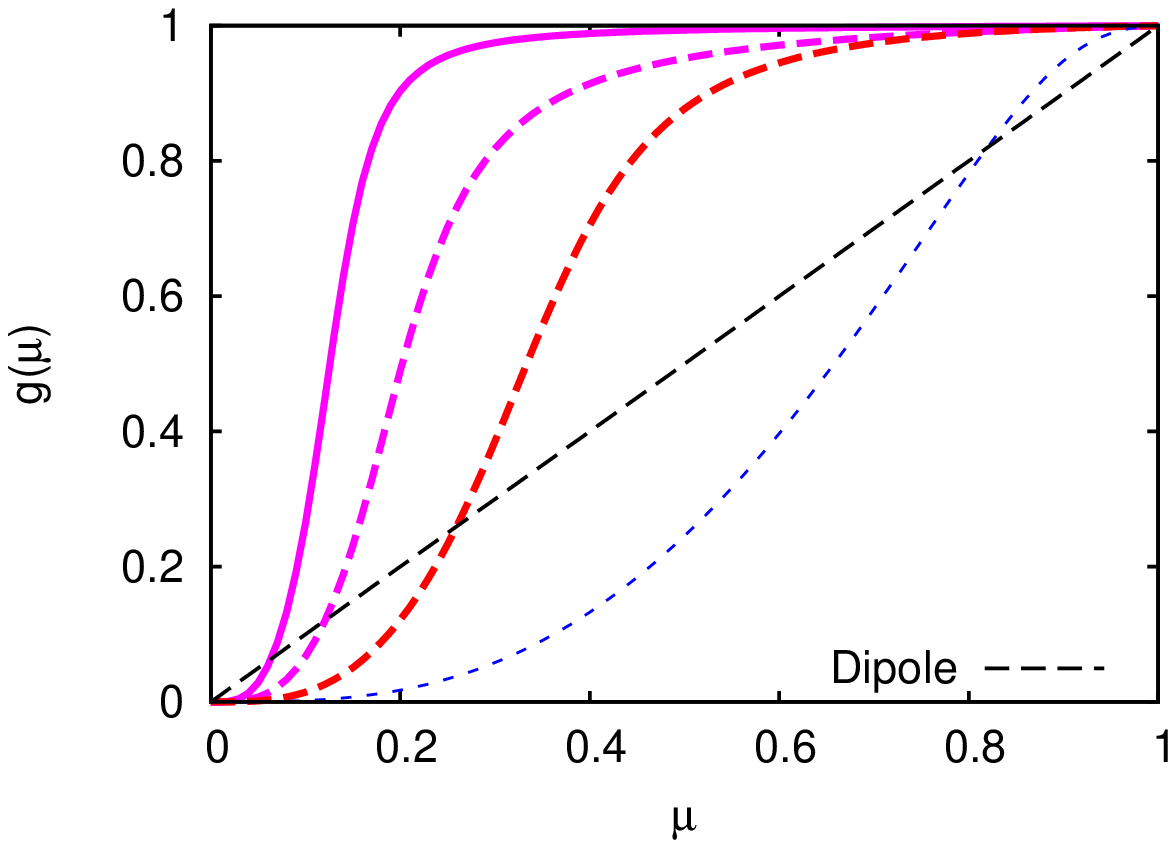}
              }
  \centerline{\includegraphics[width=0.33\textwidth]{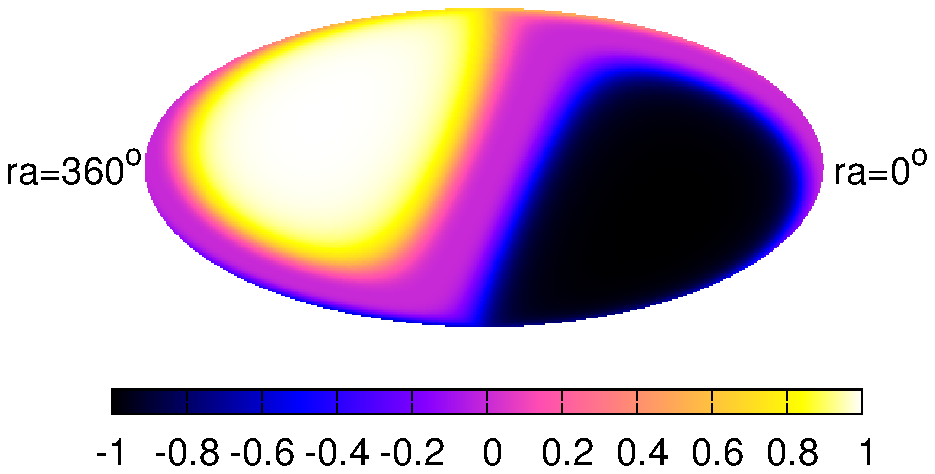}
              \hfil
              \includegraphics[width=0.33\textwidth]{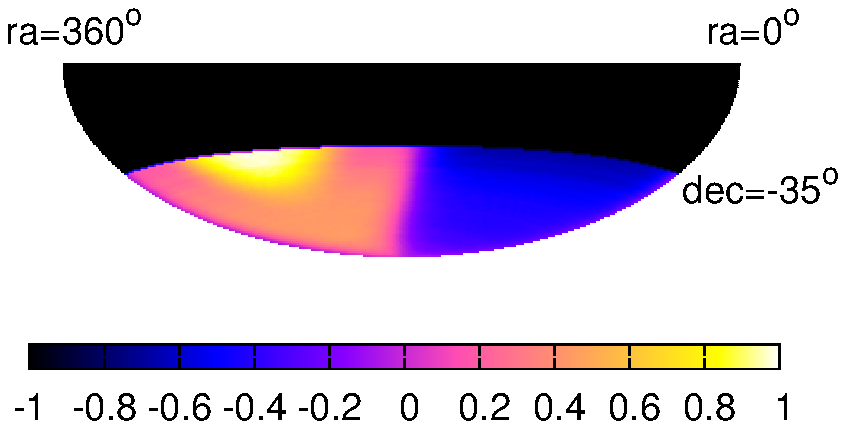}
              \hfil
              \includegraphics[width=0.30\textwidth]{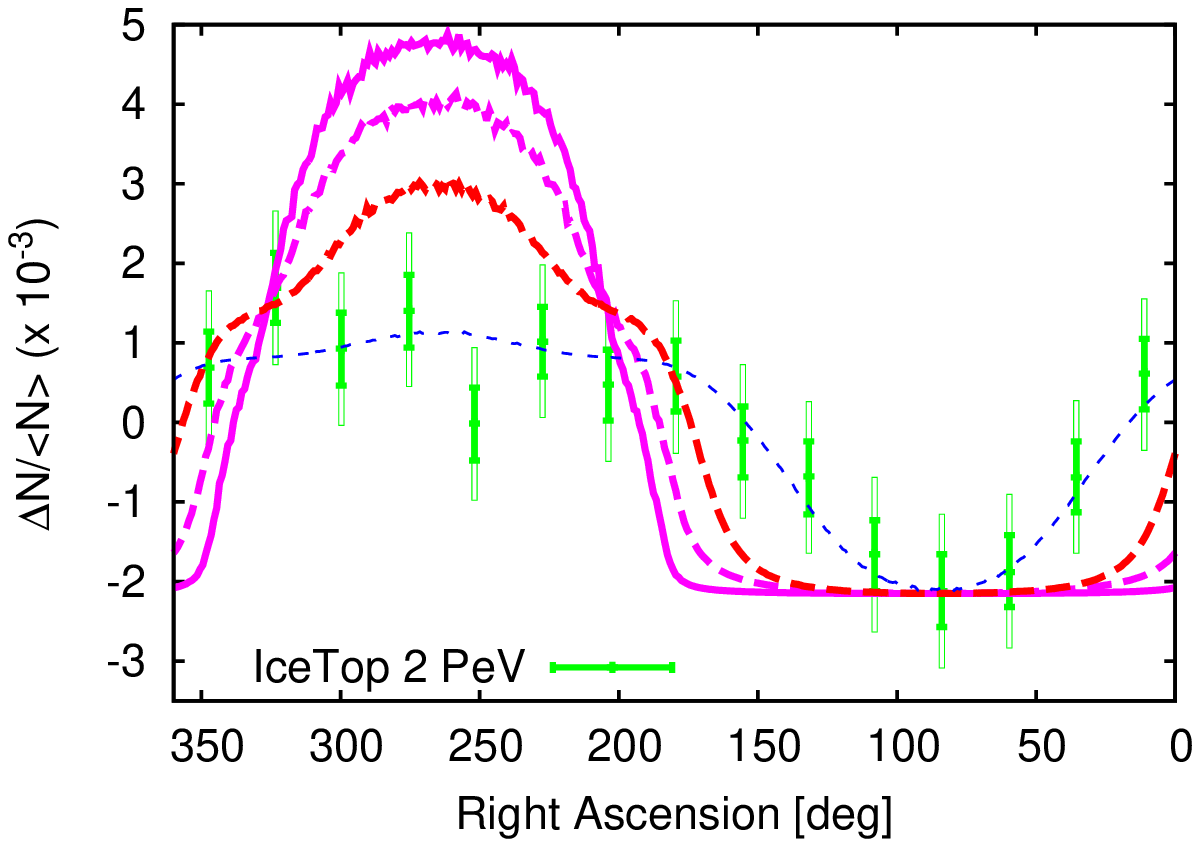}
              }
              \caption{Model~B, GS turbulence with $\mathcal{I}_{\rm
                  A,S}=\mathcal{I}_{\rm A,S,2}$, and using $R_{\rm
                  n,1}$. {\it Upper middle panel:} $\nu $
                as a function of $\mu$, for
                $\{\eps,\delta\}=\{10^{-1},3\times 10^{-5}\}$,
                $\{10^{-1},10^{-3}\}$, $\{10^{-2},10^{-3}\}$, and
                $\{10^{-4},10^{-3}\}$ --- see key. (The same line types are used in
                the upper right and lower right panels). {\it Upper
                  right panel:} $g$ as a function of $\mu$. {\it Lower
                  left panel:} anisotropy $g(\mu)$ in equatorial
                coordinates, for $\{10^{-2},10^{-3}\}$. {\it Lower
                  middle panel:} predicted CR anisotropy, with
                extremum amplitude renormalized to $\pm 1$, in the
                field of view of IceTop, for
                $\{10^{-2},10^{-3}\}$. The anisotropy is calculated
                here with respect to the averaged flux in each
                declination band. {\it Lower right panel:} relative CR
                intensity $\Delta N/\langle N \rangle$ at $-75^{\circ}
                \leq {\rm dec} \leq -35^{\circ}$, as a function of
                right ascension, and compared with IceTop 2\,PeV
                data~\citep{Aartsen:2012ma}. {\it Upper left panel:}
                comparison of the contributions from Alfv\'en modes
                (orange lines) and pseudo-Alfv\'en modes (blue lines)
                to $\nu$, for GS turbulence with
                $\mathcal{I}_{\rm A,S}=\mathcal{I}_{\rm A,S,2}$ (solid
                lines) and with $\mathcal{I}_{\rm A,S,1}$ (dashed
                lines), both for $R_{\rm n}=R_{\rm n,1}$ (models A \& B) and
                $\{\eps,\delta\} = \{10^{-3},10^{-3}\}$.}
\label{CHALAZ}
\end{figure*}

We now consider GS turbulence with a more gradual (exponential) form
of anisotropy $\mathcal{I}_{\rm A,S} = \mathcal{I}_{\rm A,S,2}$
(model~B) and calculate $D_{\mu\mu}$ numerically (see
Eqs.~(\ref{Dmumu_Alf_n1_CHALAZ}) and~(\ref{Dmumu_Slow_n0_CHALAZ})) for
eight cases: $\eps = 10^{-1},10^{-2},10^{-3},10^{-4}$, with
$\delta=3\times 10^{-5}$ and $\delta=10^{-3}$. As can be seen in
Fig.~\ref{FirstEigenvalue} and Fig.~\ref{MFP_All_Figure} (left
panels), this model more easily satisfies the constraint on
$\Lambda_1$, signalling an increased CR scattering rate. This arises
from a larger contribution to $\nu$ by Alfv\'en modes at medium and
large values of $|\mu|$. In Fig.~\ref{CHALAZ} (upper left panel), we
compare the contributions from Alfv\'en (orange lines) and
pseudo-Alfv\'en (blue lines) modes to $\nu$, in models~A (dashed
lines) and~B (solid lines), using $\{\eps,\delta\} =
\{10^{-3},10^{-3}\}$. The blue dashed and blue solid lines are nearly
coincident, i.e., the contribution from pseudo-Alfv\'en modes is the
same in each model. In contrast, the contribution from Alfv\'en modes
in model~B is larger by up to two orders of magnitude, in the range
$0.5 \lesssim \mu \leq 1$. Therefore, taking an exponential cutoff in
$k_{\parallel}$ instead of a step function makes a significant
difference for CR scattering off Alfv\'en waves; the small
but finite power present in modes with \lq\lq large\rq\rq\
$k_{\parallel}$, when an exponential cutoff is used instead of a sharp
Heaviside function has a marked impact.

According to Fig.~\ref{FirstEigenvalue} (left panel), three out of the
eight cases calculated for model~B are outside the shaded area:
$\{\eps,\delta\} = \{10^{-1},3 \times 10^{-5}\}$,
$\{10^{-1},10^{-3}\}$, and $\{10^{-2},10^{-3}\}$. We plot $\nu$ and
$g(\mu)$ in the upper centre and upper right panels of
Fig.~\ref{CHALAZ}, using thick magenta solid lines, thick magenta
dashed lines, and thick red dashed lines, for $\{10^{-1},3 \times
10^{-5}\}$, $\{10^{-1},10^{-3}\}$, and $\{10^{-2},10^{-3}\}$,
respectively.  The scattering frequency $\nu$ has a large peak at
$\mu=0$, followed by a minimum in the range $\mu=0.1-0.3$, after which
it recovers at larger $\mu$. This leads to an anisotropy $g(\mu)$
that is rather flat in a small range of $\mu \lesssim 0.1$, and then
rapidly increases towards unity. As can be seen in Fig.~\ref{Half_Width}
(upper right panel), the size of the hot (and cold) spot is quite
large for these three cases (indicated by large orange dots in
the upper right panel), even larger than for model~A 
(red lines) at the same value of $\eps$ and for the dipole anisotropy. 
We calculate in
Fig.~\ref{CHALAZ} (lower right panel) the resulting $\Delta N/\langle
N \rangle$ as a function of right ascension, at $-75^{\circ} \leq {\rm
  dec} \leq -35^{\circ}$, and compare it with IceTop 2\,PeV data. The
cold spot size and shape are incompatible with the data, as is also 
the presence of a large maximum around RA~$\approx 250^{\circ}$. Sky
maps for the anisotropy in the \lq\lq least bad\rq\rq\ case,
$\{10^{-2},10^{-3}\}$, are shown in the lower left and lower centre
panels of Fig.~\ref{CHALAZ}: full sky (left) and Southern sky in the
field of view of IceTop (centre). Model~B is clearly ruled out by
the data.

We point out that, although the size of the cold spot decreases with
$\eps$ (see the orange lines in the upper right panel of
Fig.~\ref{Half_Width}), all cases with $\theta_{1/2}<60^{\circ}$ are
located well inside the forbidden region shaded in grey in
Fig.~\ref{FirstEigenvalue}. For cases inside this region, we use
thinner lines and do not show results with large dots in
Fig~\ref{Half_Width}. The orange triangle in this figure corresponds
to the case that lies both inside the shaded region and below the
grey dashed line in Fig.~\ref{FirstEigenvalue}. In the upper centre,
upper right, and lower right panels of Fig.~\ref{CHALAZ}, we plot the
case $\{10^{-4},10^{-3}\}$ with thin blue dotted lines. Here again,
the cold spot is small enough to fit the 2\,PeV data, but only where
the anisotropy is not given by $g$, i.e., in where
condition~(\ref{Eqnmfpcondition}) is not satisfied.

\subsubsection{Model C ($\mathcal{I}_{\rm A,S} = \mathcal{I}_{\rm A,S,1}$ and $R_{\rm n}=R_{\rm n,2}$)}
\label{I1_Rn2}

We now turn to the broad resonance function, $R_{\rm n,2}$, and start with
$\mathcal{I}_{\rm A,S,1}$ (model~C). We calculate $D_{\mu\mu}$ from
Eqs.~(\ref{Dmumu_Alf_n1_LAZCHA}) and~(\ref{Dmumu_Slow_n0_LAZCHA}), for
13 combinations of $\eps \in \{10^{-1},10^{-2},10^{-3},10^{-4}\}$ and
$\delta \mathcal{M}_{\rm A} \in \{1,0.33,0.1,0.033,0.01\}$ (see the 
red symbols in the right panel of Fig.~\ref{FirstEigenvalue}).

\begin{figure*}
  \centerline{\includegraphics[width=0.32\textwidth]{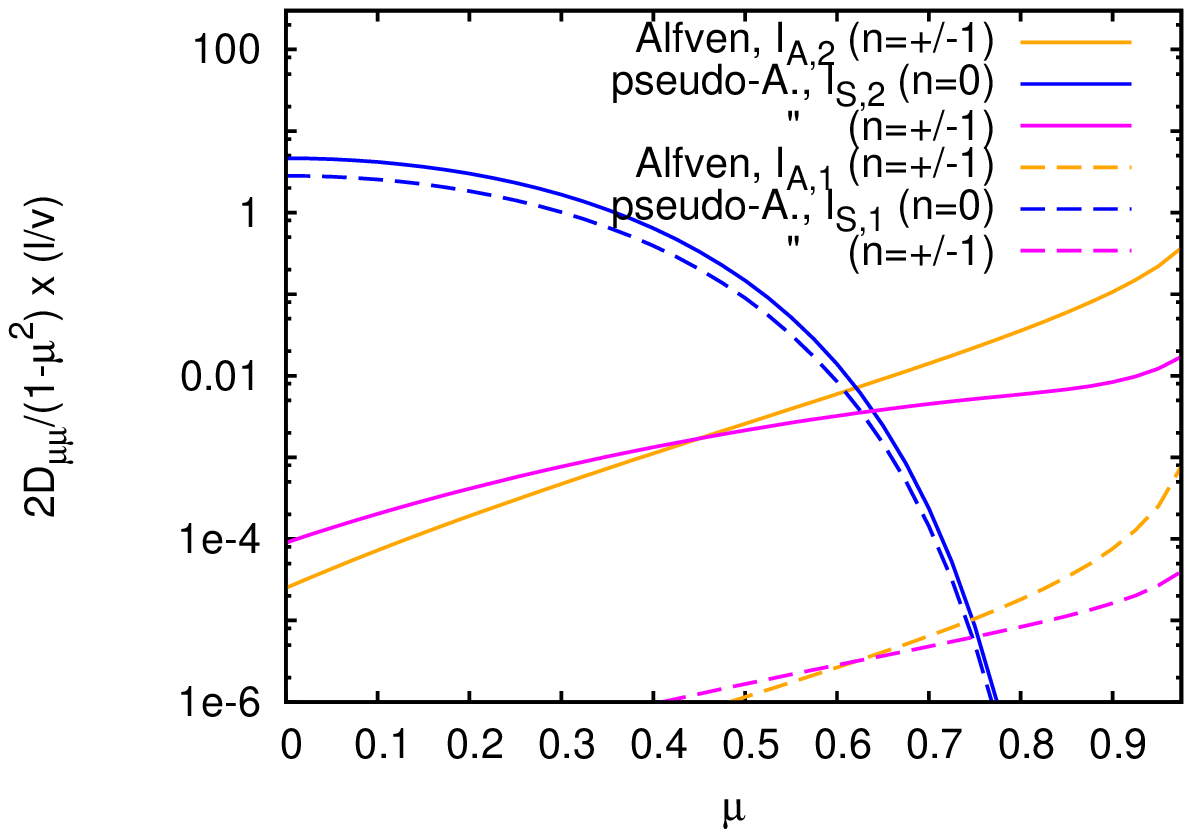}
              \hfil
              \includegraphics[width=0.32\textwidth]{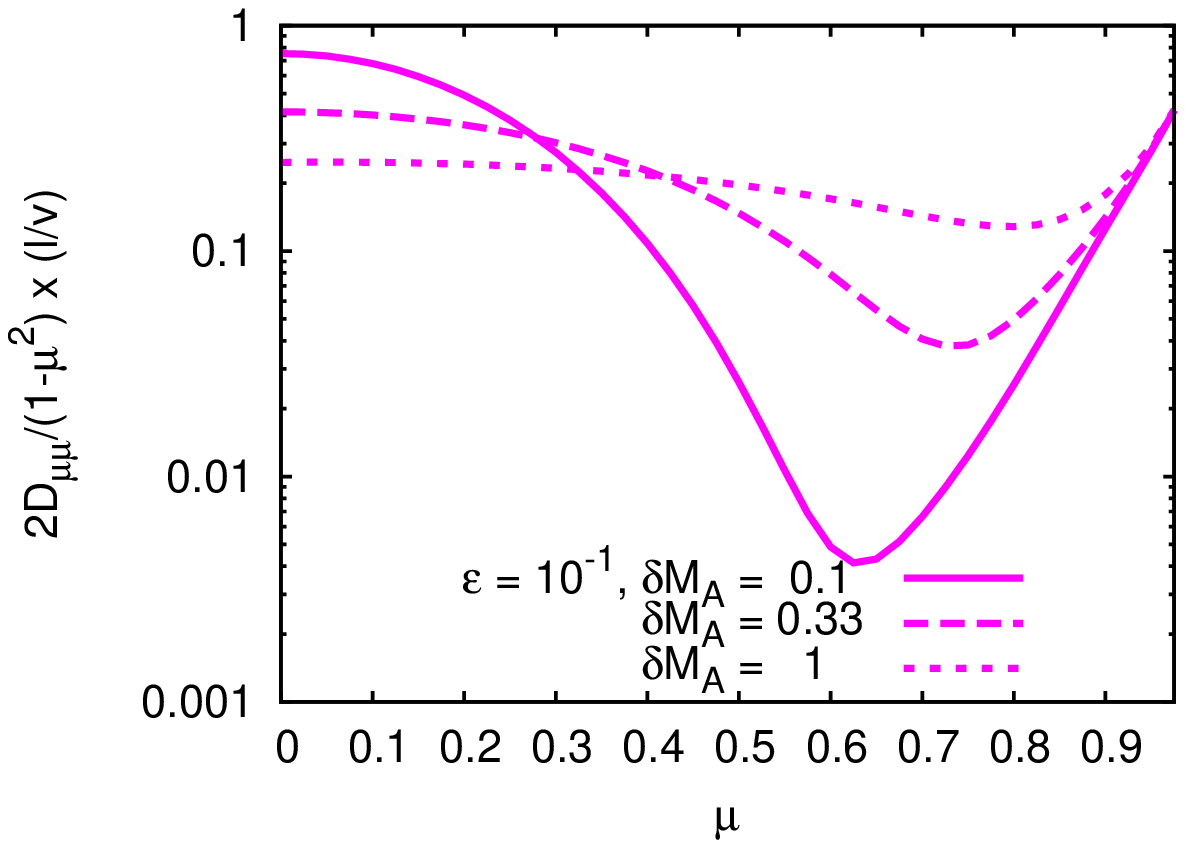}
              \hfil
              \includegraphics[width=0.32\textwidth]{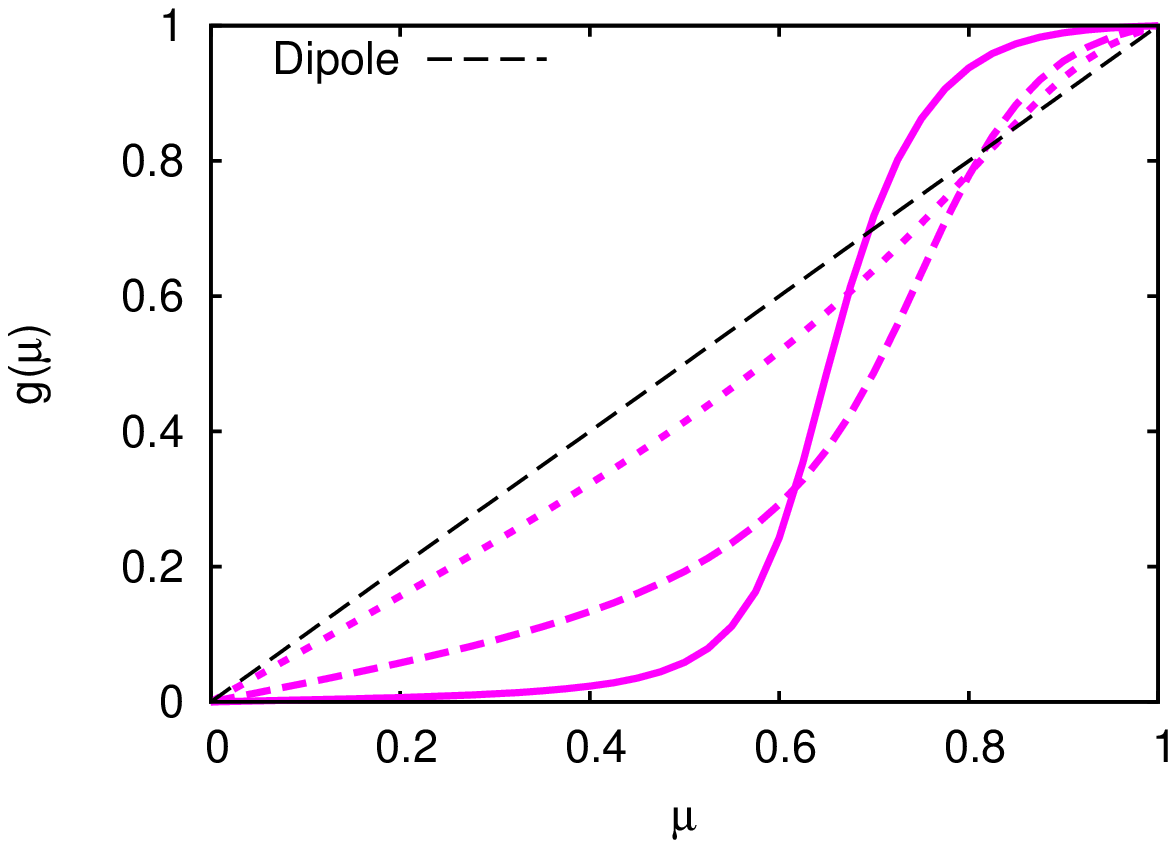}
              }
  \centerline{\includegraphics[width=0.33\textwidth]{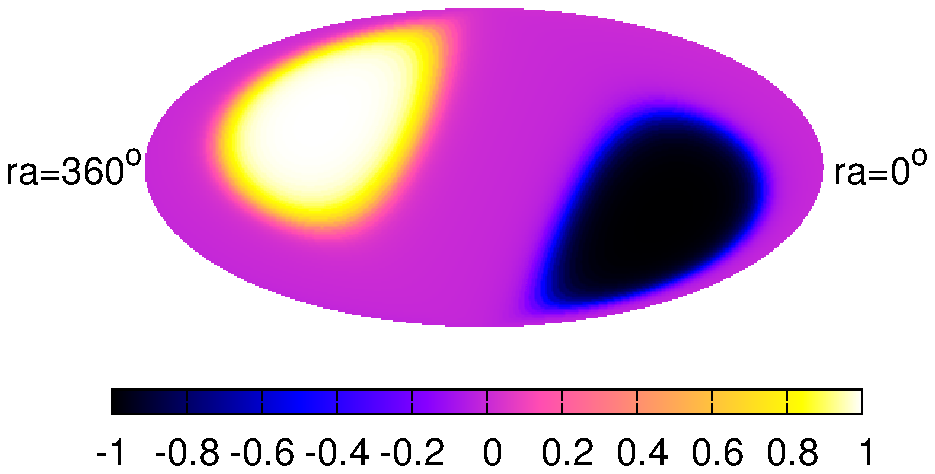}
              \hfil
              \includegraphics[width=0.33\textwidth]{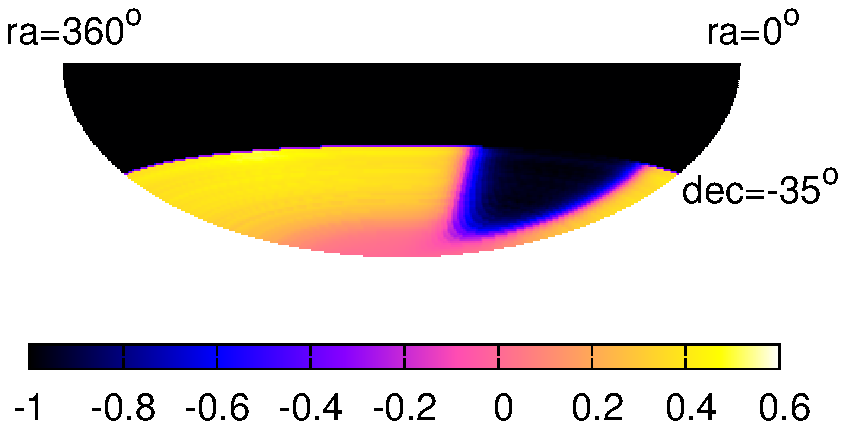}
              \hfil
              \includegraphics[width=0.30\textwidth]{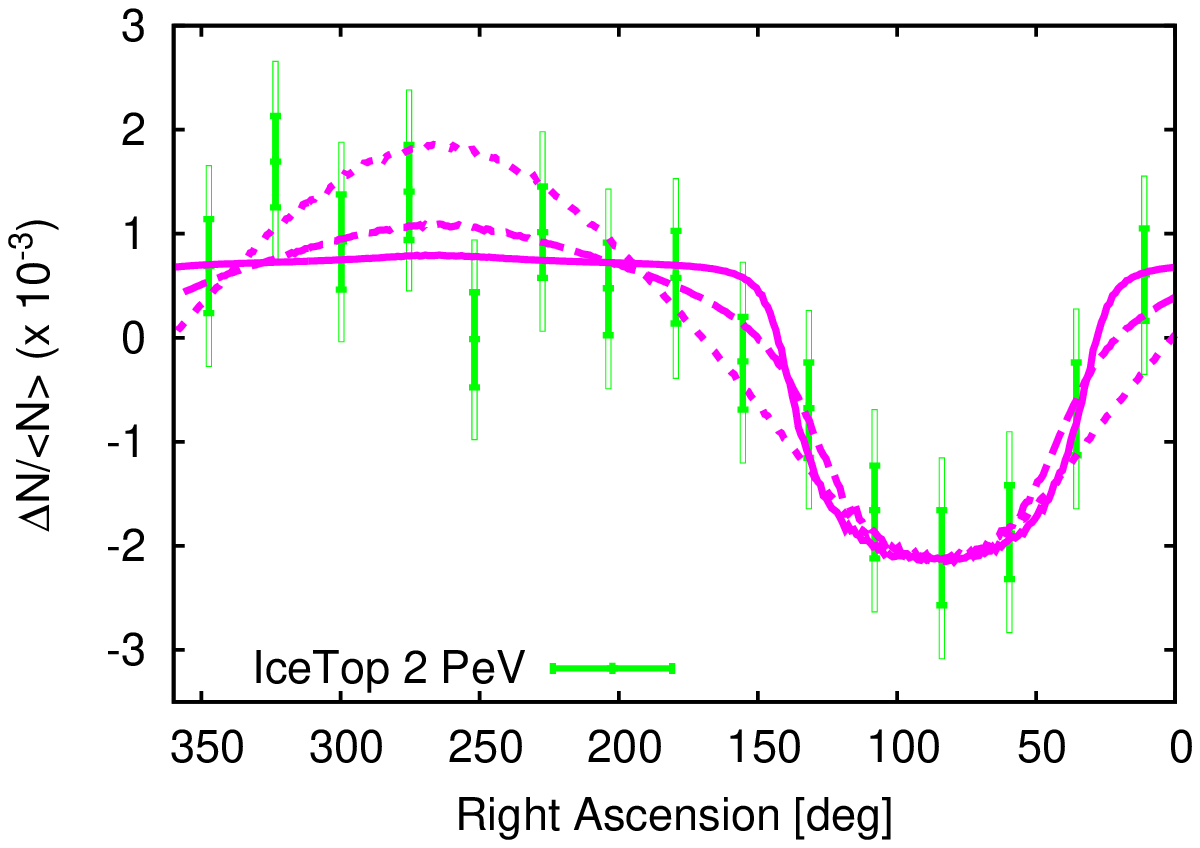}
              }
              \caption{Model C, GS turbulence with $\mathcal{I}_{\rm
                  A,S}=\mathcal{I}_{\rm A,S,1}$, and using $R_{\rm
                  n,2}$. $\eps=10^{-1}$ in all panels, except in the
                upper left one. {\it Upper middle panel:} $\nu$ as a
                function of $\mu$, for $\delta \mathcal{M}_{\rm A} \in
                \{ 0.1,0.33,1\}$, see key. (Same line types used in the
                upper right and lower right panels.) {\it Upper right
                  panel:} $g$ as a function of $\mu$. {\it Lower left
                  panel:} anisotropy $g(\mu)$ in equatorial
                coordinates, for $\delta \mathcal{M}_{\rm A} =
                0.1$. {\it Lower middle panel:} predicted CR
                anisotropy, with extremum amplitude renormalized to
                $\pm 1$, in the field of view of IceTop, for $\delta
                \mathcal{M}_{\rm A} = 0.1$. The anisotropy is
                calculated here with respect to the averaged flux in
                each declination band. {\it Lower right panel:}
                relative CR intensity $\Delta N/\langle N \rangle$ at
                $-75^{\circ} \leq {\rm dec} \leq -35^{\circ}$, as a
                function of right ascension, and compared with IceTop
                2\,PeV data~\citep{Aartsen:2012ma}. {\it Upper left
                  panel:} comparison of the contributions from
                Alfv\'en modes (orange lines for $n=\pm 1$) and
                pseudo-Alfv\'en modes (blue lines for $n=0$, and
                magenta lines for $n=\pm 1$) to $\nu$,
                for GS turbulence with $\mathcal{I}_{\rm
                  A,S}=\mathcal{I}_{\rm A,S,1}$ (dashed lines) and
                with $\mathcal{I}_{\rm A,S,2}$ (solid lines), both for
                $R_{\rm n}=R_{\rm n,2}$ (models C \& D) and
                $\{\eps,\delta\mathcal{M}_{\rm A}\}=\{10^{-3},0.1\}$.}
\label{LAZCHA}
\end{figure*}

In Fig.~\ref{LAZCHA} (upper left panel), we plot the contribution to the
scattering frequency $\nu$ of 
the $n=0$ 
term for pseudo-Alfv\'en waves using a blue, dashed line,
and that of the $n=\pm 1$ terms for Alfv\'en waves
using an orange dashed line (the solid lines will be 
explained in the next subsection).  The
magenta dashed line corresponds to the $n=\pm 1$ contribution from
pseudo-Alfv\'en modes (Eq.~(\ref{Dmumu_Slow_n1_LAZCHA})), which is
found to be subdominant
at all $\mu$ 
when $\delta \mathcal{M}_{\rm A}
\gtrsim 0.1$. As in model~A, scattering around $\mu=0$ is
predominantly provided by pseudo-Alfv\'en waves, and scattering at
large $|\mu|$ is dominated by scattering off Alfv\'en waves. The
height of the peak at $\mu=0$ is smaller, but this is of little
relevance for the shape of the CR anisotropy. A more important
difference between the broad and narrow resonance functions
$R_{\rm n,1}$ (A) and $R_{\rm n,2}$ (C) is that the width
of the peak is larger in the latter,
unless $\delta \mathcal{M}_{\rm A} \ll 0.01$. This has the important
consequences of both reducing the 
size of the hot/cold spots and reducing $v/\Lambda_1$,
as can be seen in Figs.~\ref{FirstEigenvalue} and \ref{Half_Width}.

Out of the 13 cases, three are in the allowed, non-shaded area of
Fig.~\ref{FirstEigenvalue}: $\eps=10^{-1}$ with $\delta
\mathcal{M}_{\rm A} = 1,0.33,0.1$. In the upper centre, upper right, 
and lower right panels of Fig.~\ref{LAZCHA}, we plot, for these three
cases, $\nu(\mu)$, $g(\mu)$, and $\Delta N/\langle N \rangle$ (see the
line types in the key of the upper centre panel). With $\delta
\mathcal{M}_{\rm A} = 1$ (dotted magenta line), the contribution from
pseudo-Alfv\'en modes around $\mu=0$ is so wide that $\nu$ is rather
flat on $0 \leq \mu \leq 1$. The corresponding anisotropy is then not
far from being a dipole --- see the upper right panel, and it does not
fit the small cold spot in IceTop data --- see the lower right panel. On
the other hand, both $\delta \mathcal{M}_{\rm A} = 0.1$ and $\delta
\mathcal{M}_{\rm A} = 0.33$ provide a good fit to the 2\,PeV data:
they have a moderately low minimum in $\nu$ at sufficiently large
values of $\mu \simeq 0.6-0.75$ (see upper centre panel), around the
transition from scattering off pseudo-Alfv\'en modes to scattering off
Alfv\'en modes. This leads to a rather flat $g$ at $\mu \lesssim
0.5$--$0.6$, and to smaller hot and cold spots around $\mu=\pm 1$, in
good agreement with the IceTop 2\,PeV data (see the solid and dashed
magenta lines in the lower right panel). We show in the lower left and
centre panels the full-sky anisotropy (left) and the anisotropy in the
field of view of IceTop (middle) for one of these two good cases,
$\{\eps,\delta \mathcal{M}_{\rm A}\} = \{10^{-1},0.1\}$. The presence
of a flat CR flux in a broad region in $\mu$ outside the cold and hot
spots is visible in both panels. This conceals the hot spot in the
Northern hemisphere, while providing a rather flat CR flux outside a
tight cold spot, see the lower middle panel.

In the lower panels of Fig.~\ref{Half_Width}, $\theta_{1/2}$ is plotted
with red lines, as a function of $\eps$ (left), and as a function of
$\delta \mathcal{M}_{\rm A}$ (right). Although a good fit to the
2\,PeV data, the values of $\theta_{1/2}\approx 45^{\circ}-50^{\circ}$
at $\eps=10^{-1}$ are not small enough to account for the smaller cold
spot in the 400\,TeV data. Decreasing $\eps$ causes $\theta_{1/2}$ to
drop to values much lower than $40^{\circ}$. As a result, $\{10^{-2},1\}$
is the only point with $\eps<10^{-1}$ that satisfies
Eq.~(\ref{Eqnmfpcondition}) for 100\,TeV, i.e., which is below the
grey dashed line in Fig.~\ref{FirstEigenvalue} (right panel), whilst,
at the same time providing an acceptable fit to the 400\,TeV
data. However, $\eps=10^{-1}$, $\delta\mathcal{M}_{\rm A}=1$ does not
provide a good fit to the higher-energy data set.

\subsubsection{Model D ($\mathcal{I}_{\rm A,S} = \mathcal{I}_{\rm A,S,2}$ and $R_{\rm n}=R_{\rm n,2}$)}
\label{I2_Rn2}

\begin{figure*}
  \centerline{\includegraphics[width=0.32\textwidth]{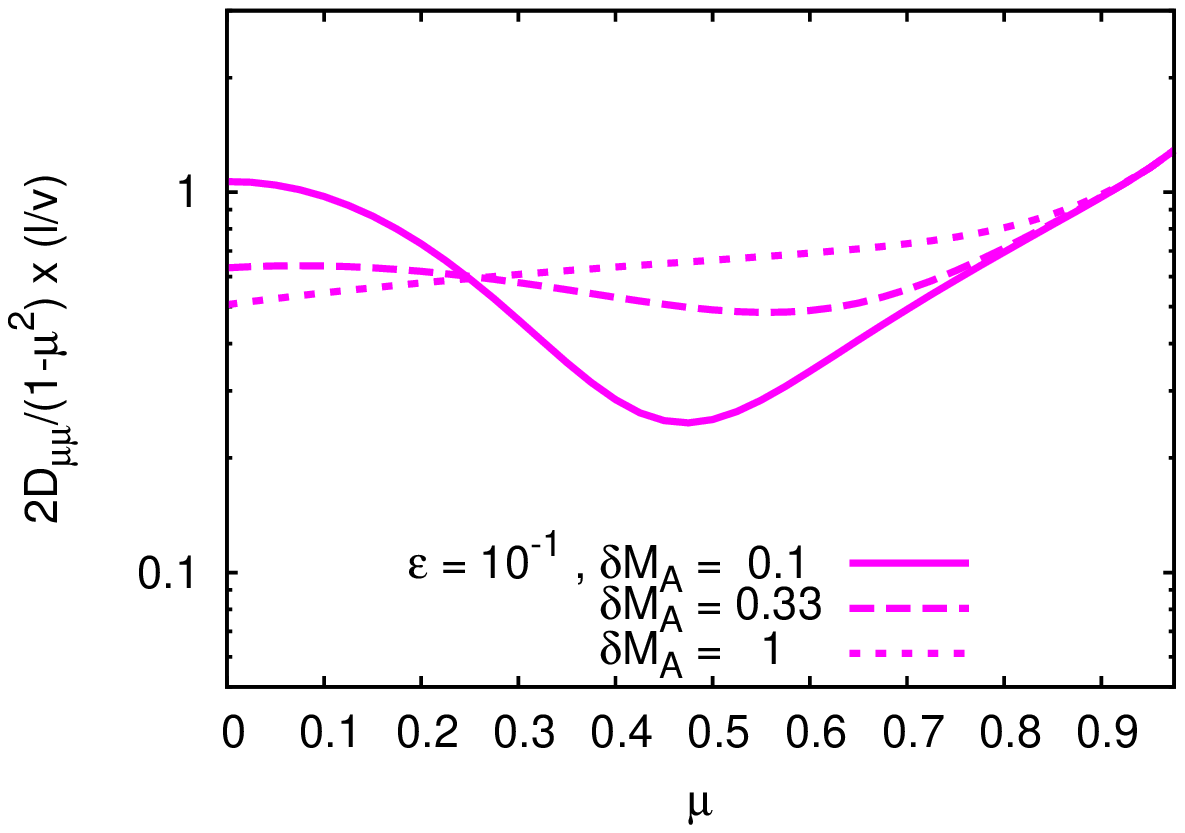}
              \hfil
              \includegraphics[width=0.32\textwidth]{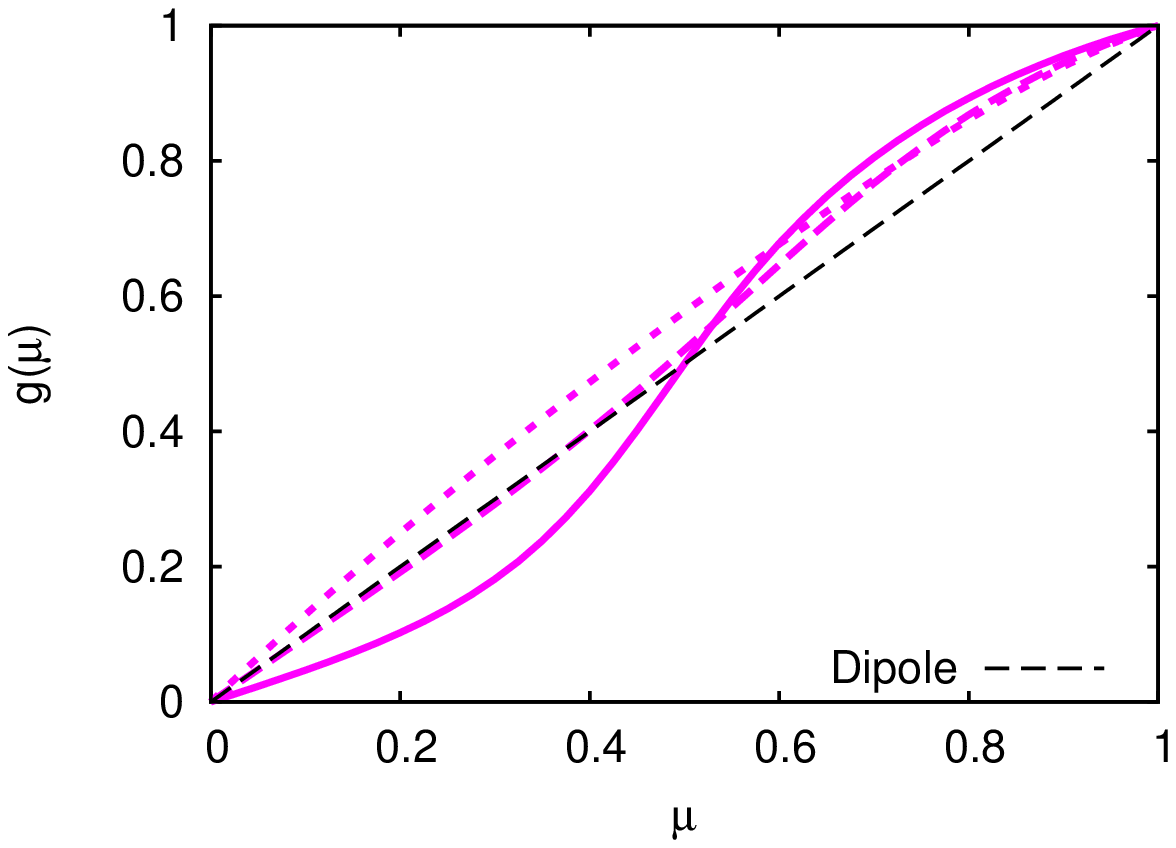}
              \hfil
              \includegraphics[width=0.32\textwidth]{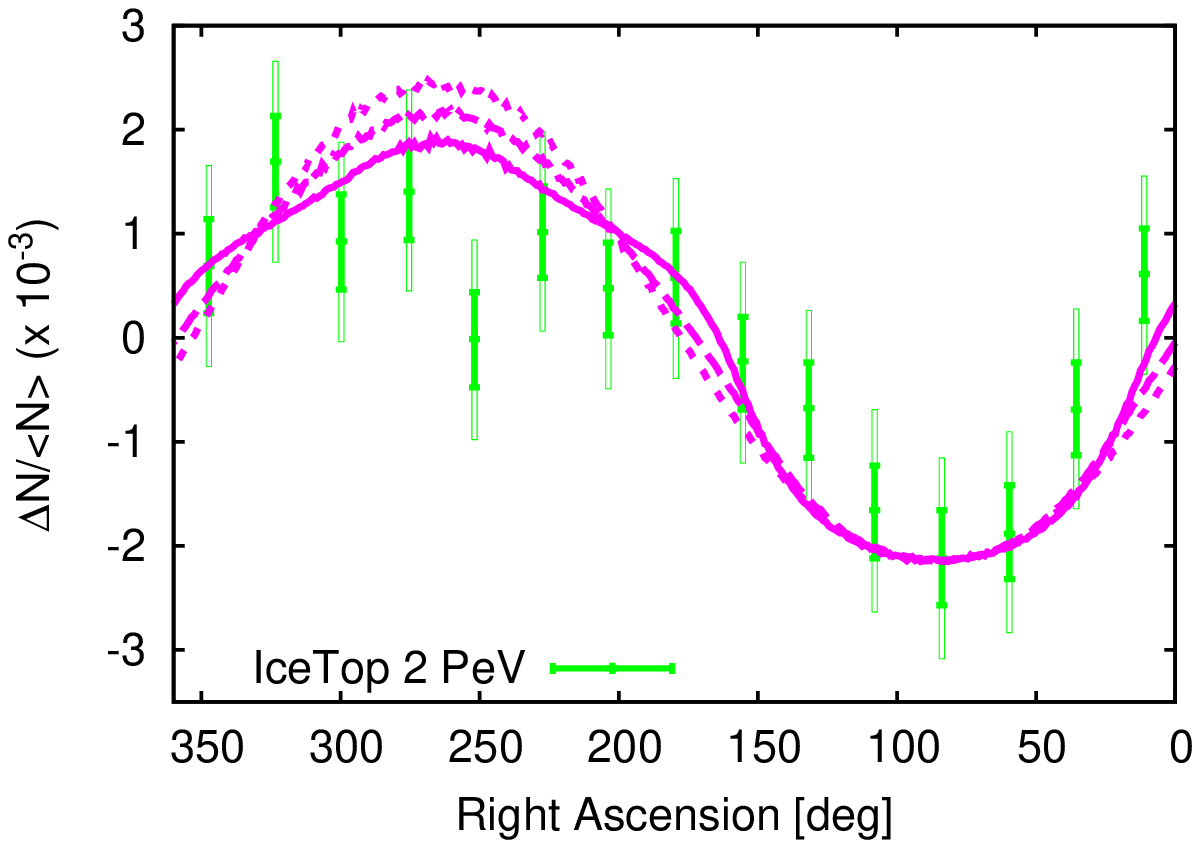}
              }
  \centerline{\includegraphics[width=0.32\textwidth]{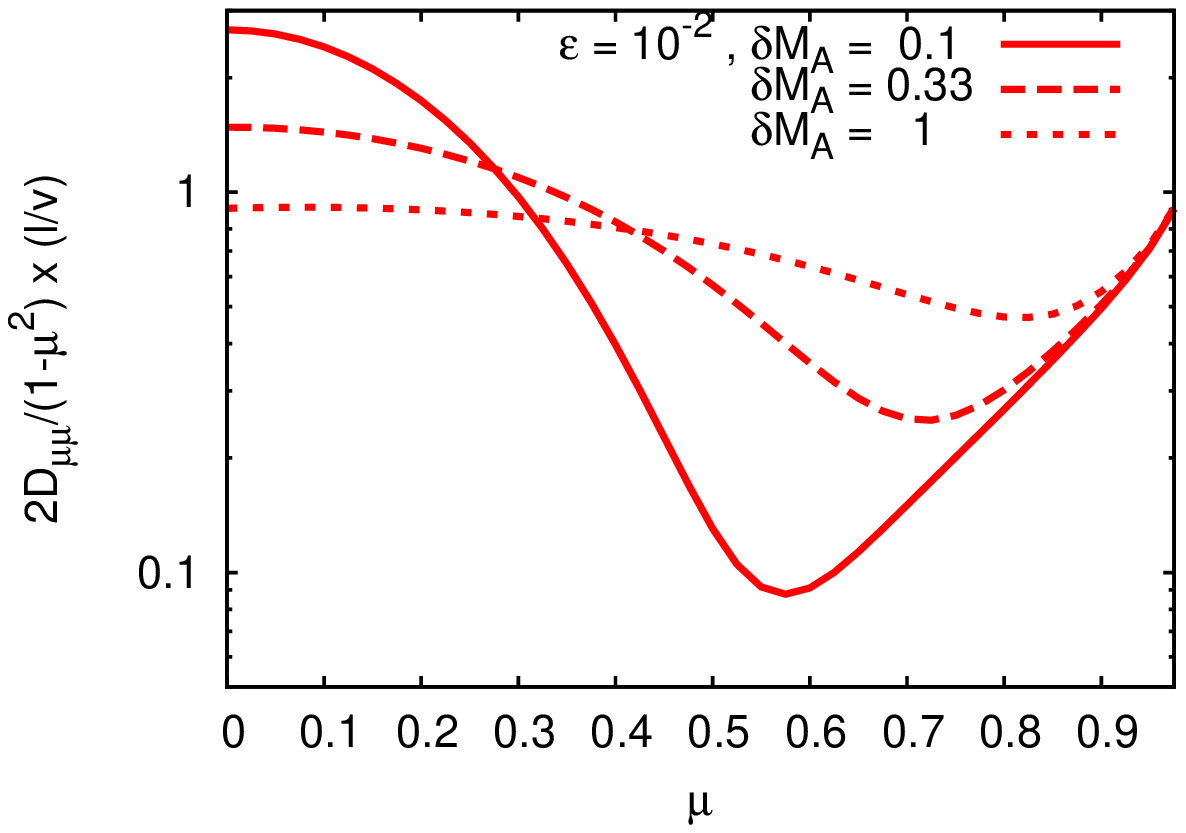}
              \hfil
              \includegraphics[width=0.32\textwidth]{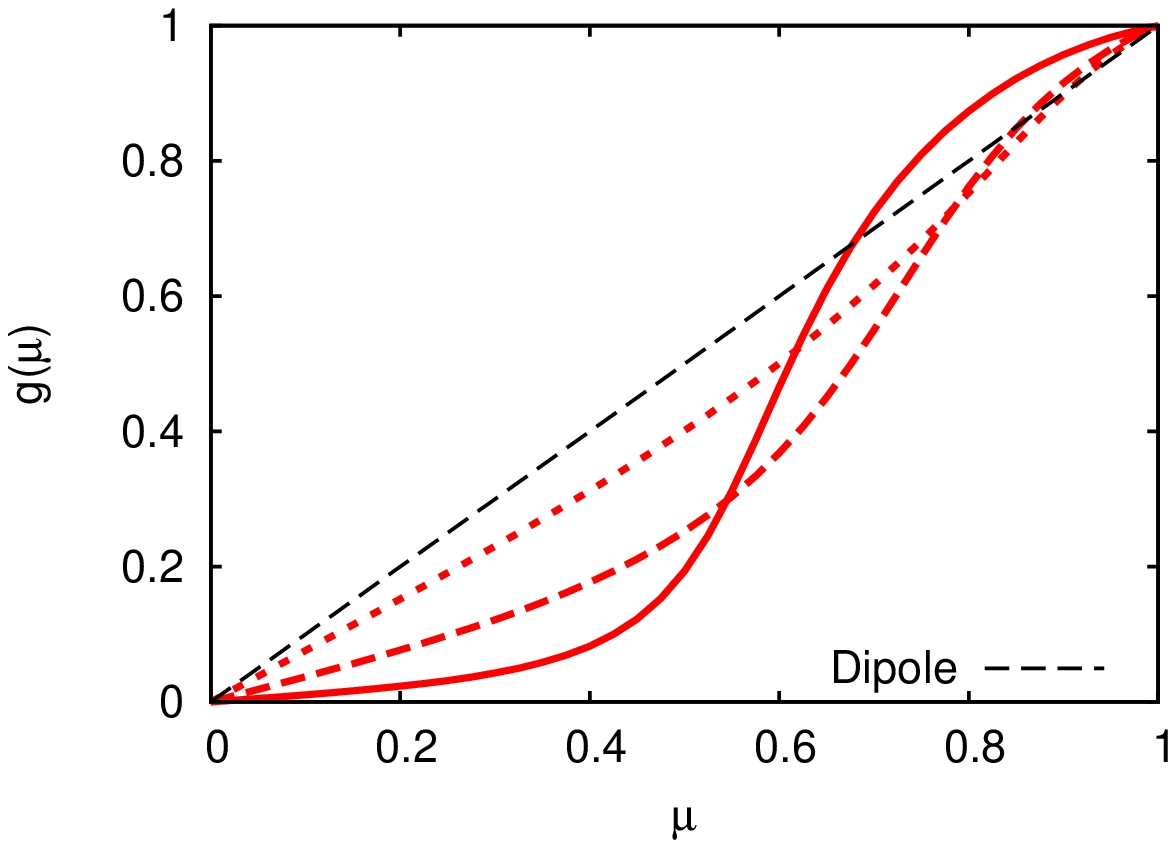}
              \hfil
              \includegraphics[width=0.32\textwidth]{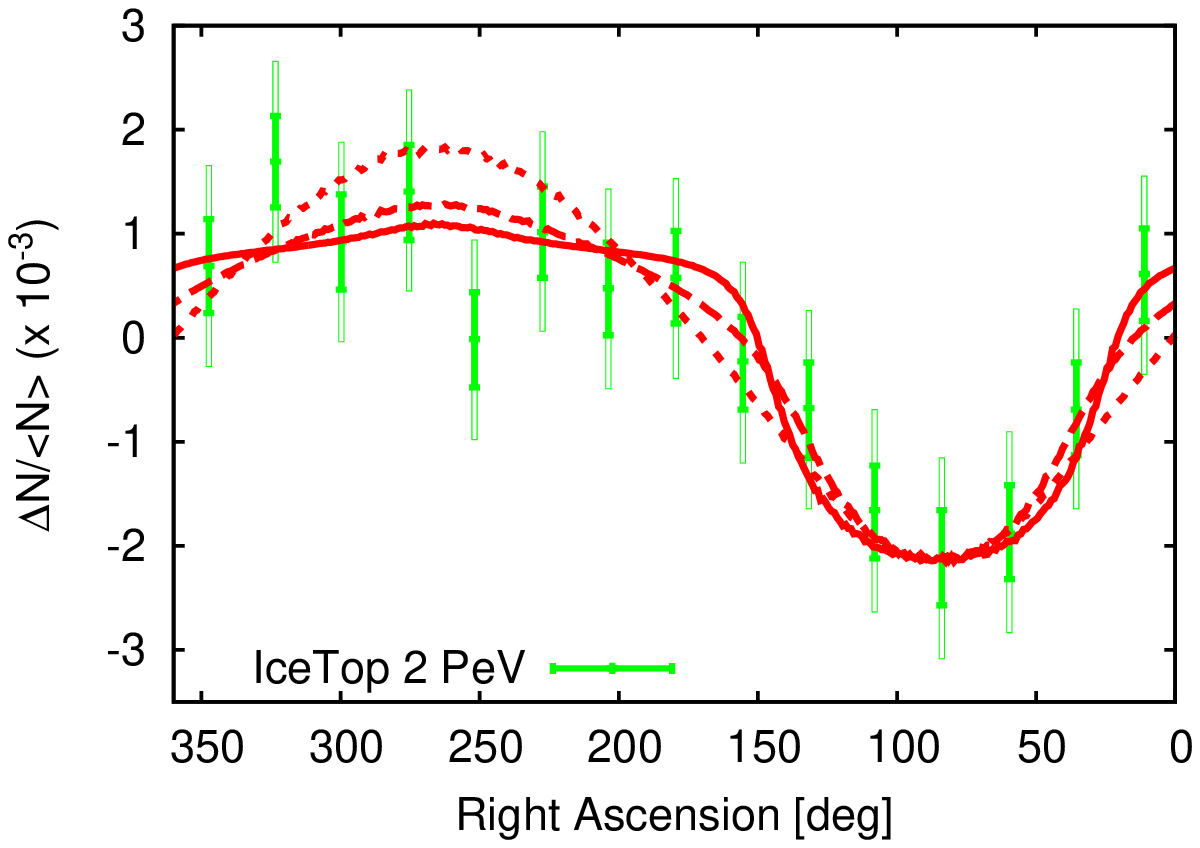}
              }
  \centerline{\includegraphics[width=0.32\textwidth]{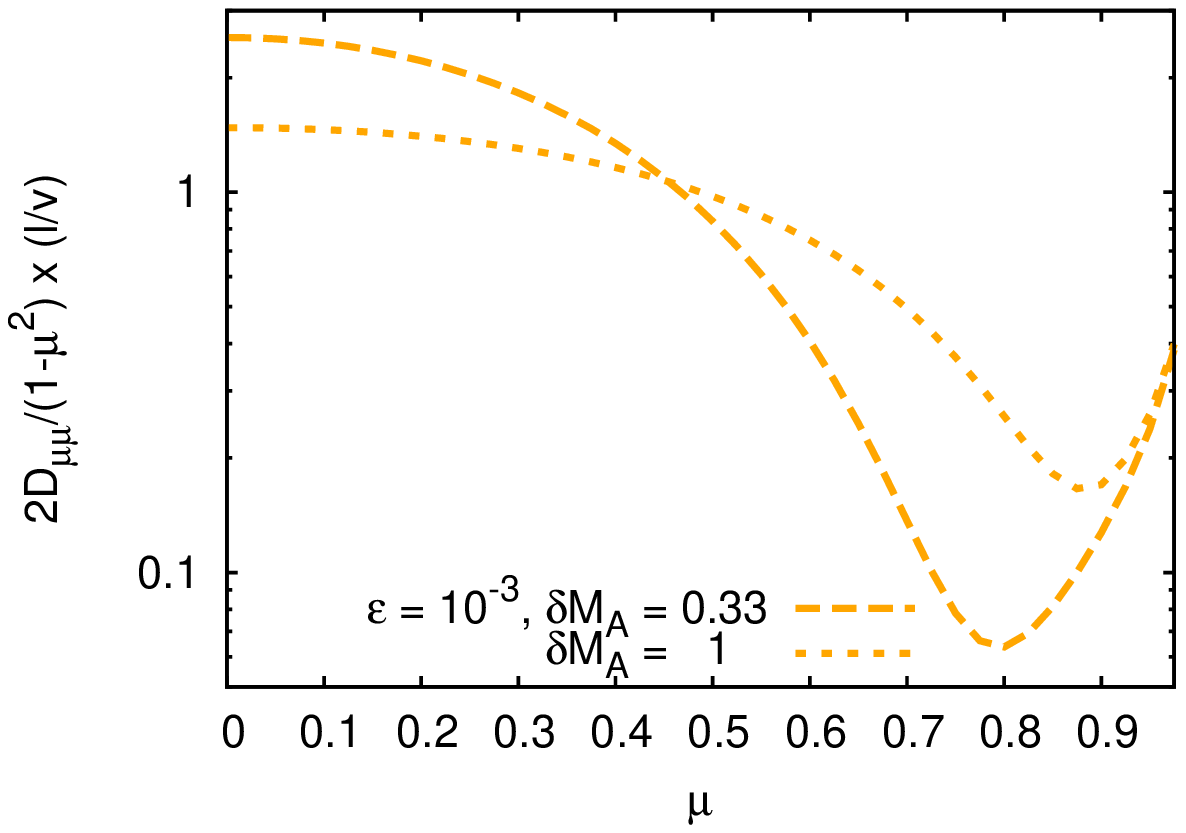}
              \hfil
              \includegraphics[width=0.32\textwidth]{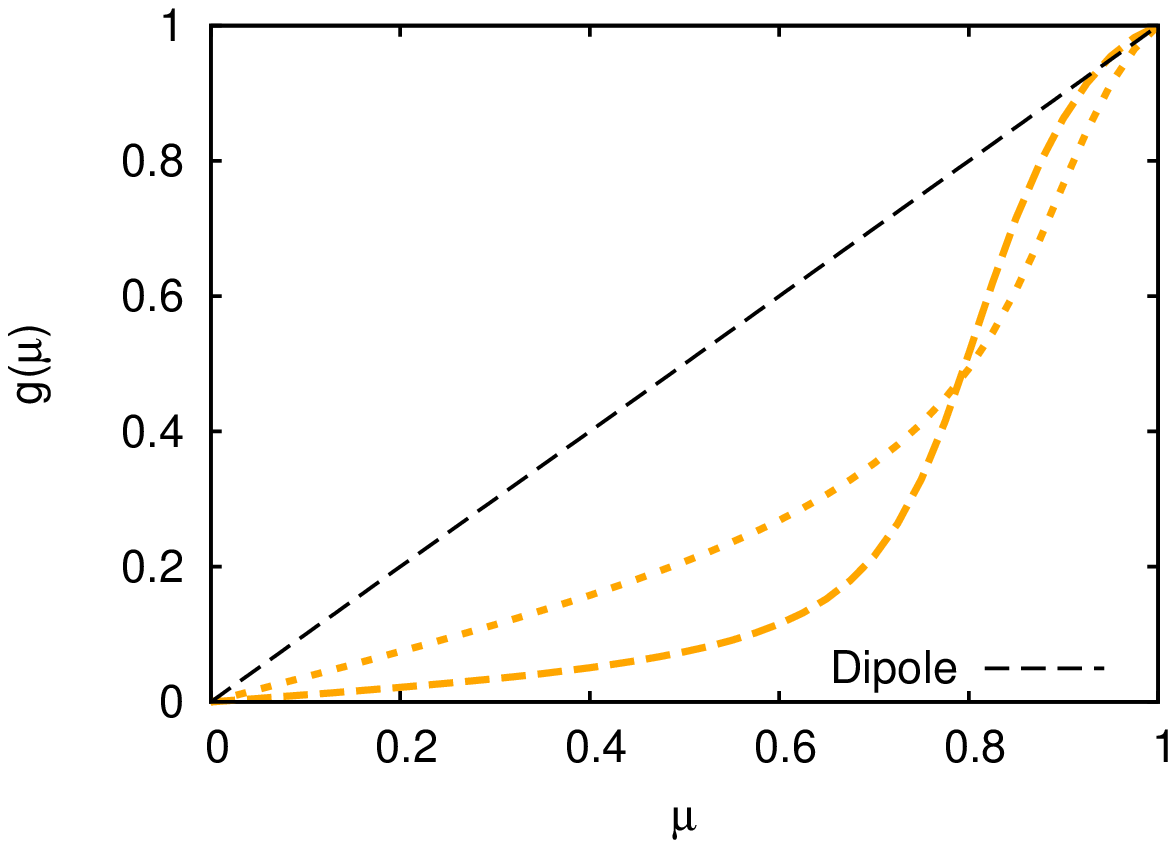}
              \hfil
              \includegraphics[width=0.32\textwidth]{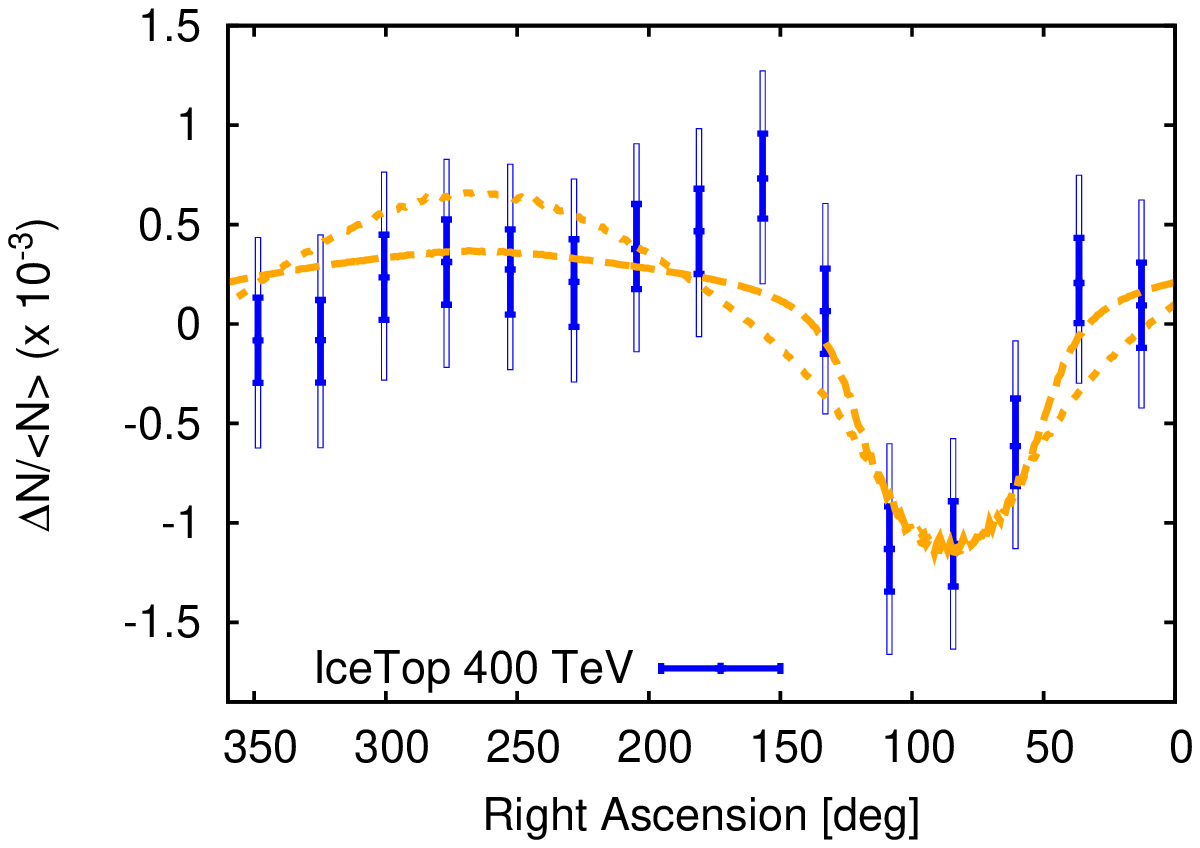}
              }
              \caption{Model D, GS turbulence with $\mathcal{I}_{\rm
                  A,S}=\mathcal{I}_{\rm A,S,2}$, and using $R_{\rm
                  n,2}$: $\nu(\mu)$ ({\it left column}),
                $g(\mu)$ ({\it middle column}), and relative CR
                intensity $\Delta N/\langle N \rangle$ in the
                declination band $-75^{\circ}$ to $-35^{\circ}$, as a
                function of right ascension, and compared with IceTop
                2\,PeV or 400\,TeV data sets
                from~\cite{Aartsen:2012ma} ({\it right column}). {\it
                  Upper row:} $\eps=10^{-1}$; {\it Middle row:}
                $\eps=10^{-2}$; {\it Lower row:} $\eps=10^{-3}$. Each
                line type corresponds to a different value of $\delta
                \mathcal{M}_{\rm A} \in \{ 0.1,0.33,1\}$. Identical
                line types in each row, see keys in the left panels.}
\label{LAZLAZ_1}
\end{figure*}
\begin{figure*}
  \centerline{\includegraphics[width=0.33\textwidth]{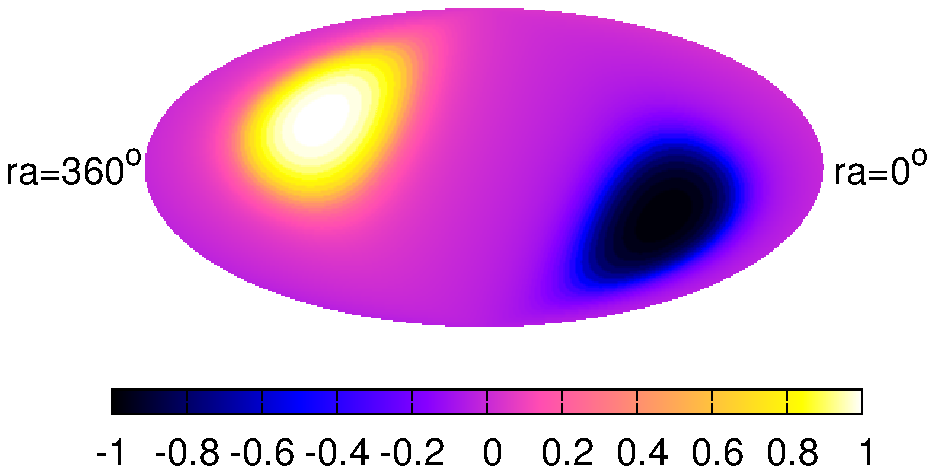}
              \hfil
              \includegraphics[width=0.33\textwidth]{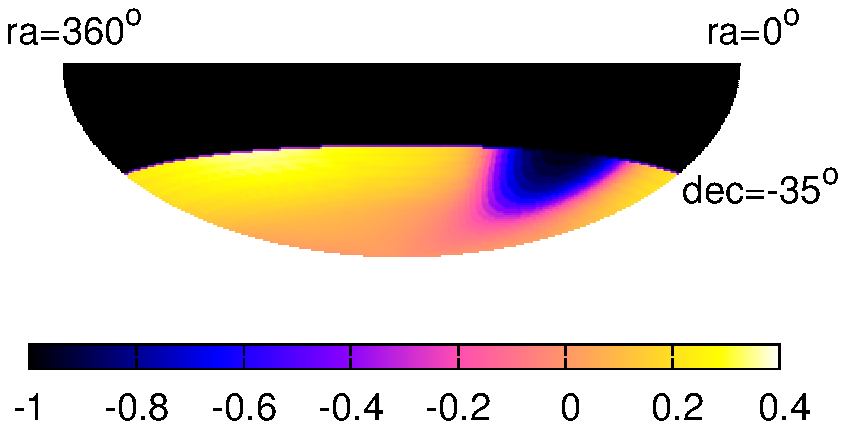}
              \hfil
              \includegraphics[width=0.30\textwidth]{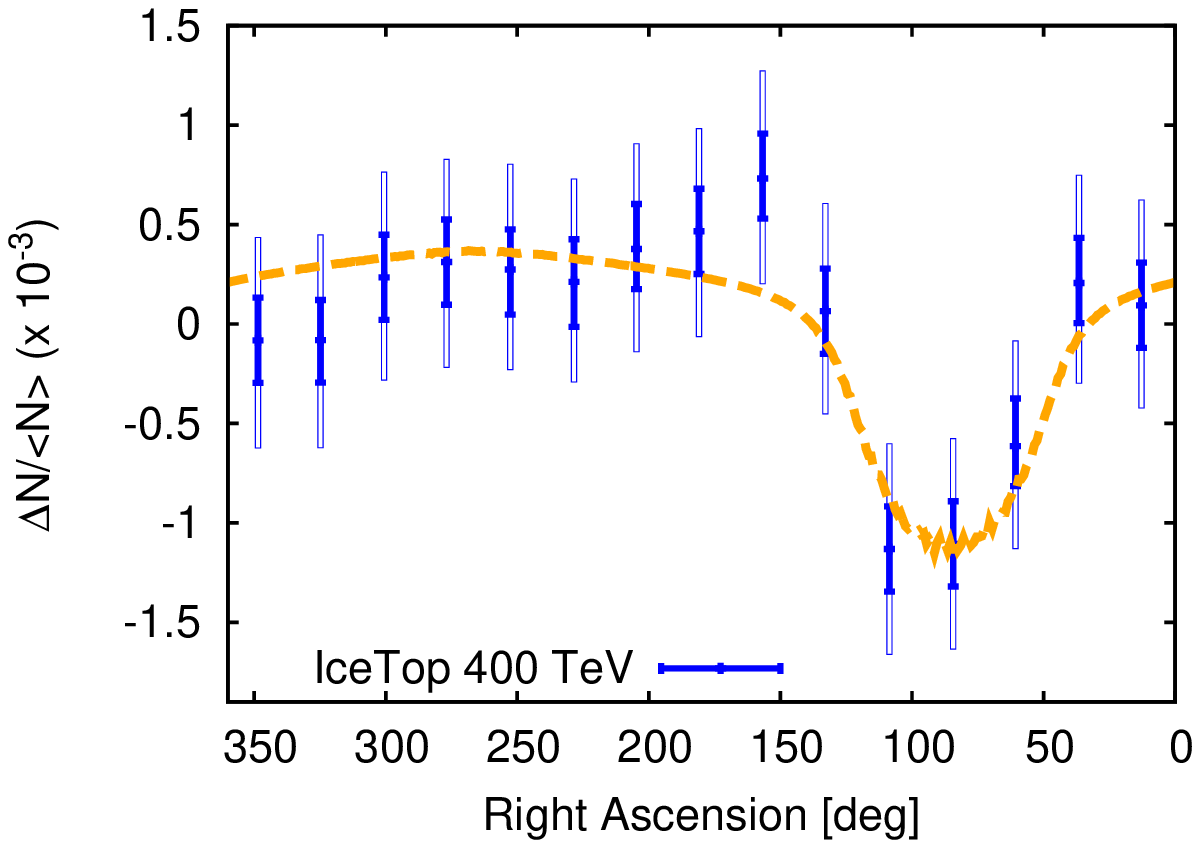}
              }
  \centerline{\includegraphics[width=0.33\textwidth]{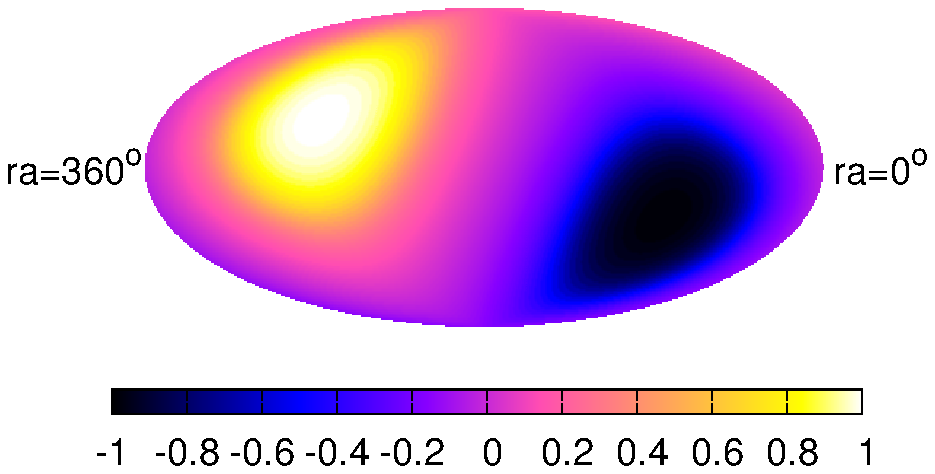}
              \hfil
              \includegraphics[width=0.33\textwidth]{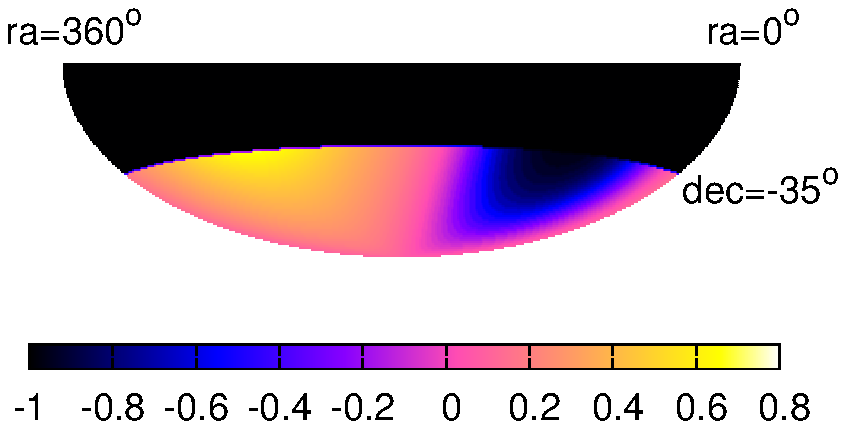}
              \hfil
              \includegraphics[width=0.30\textwidth]{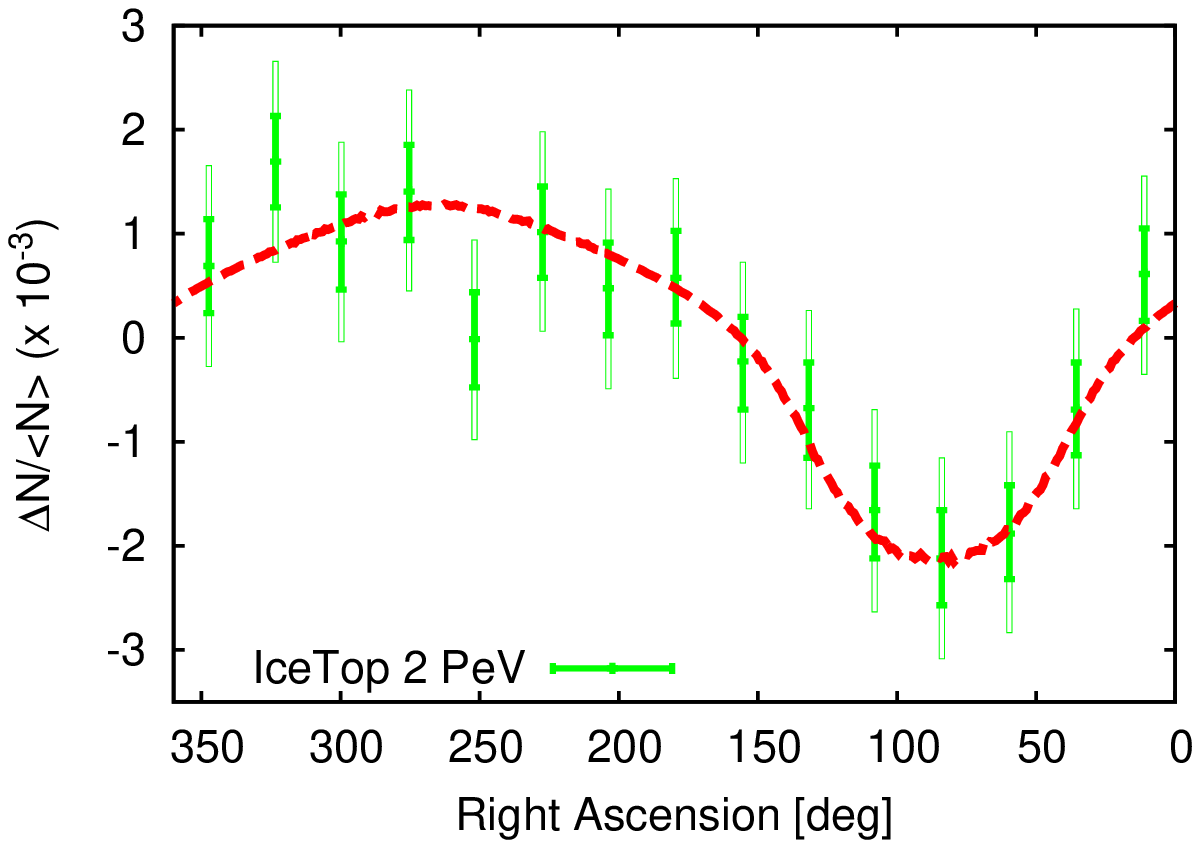}
              }
              \caption{Large-scale CR anisotropy for model~D, GS
                turbulence with $\mathcal{I}_{\rm
                  A,S}=\mathcal{I}_{\rm A,S,2}$, and using $R_{\rm
                  n,2}$ with $\delta \mathcal{M}_{\rm A} = 0.33$. {\it
                  Upper row:} $\eps = 10^{-3}$. {\it Lower row:} $\eps
                = 10^{-2}$, i.e. the CR energy is 10 times larger than
                in the upper row. {\it Left column:} anisotropy
                $g(\mu)$ in equatorial coordinates. {\it Middle
                  column:} predicted CR anisotropy, with extremum
                amplitude renormalized to $\pm 1$, in the field of
                view of IceTop. The anisotropy is calculated here with
                respect to the averaged flux in each declination
                band. {\it Right column:} relative CR intensity
                $\Delta N/\langle N \rangle$ at $-75^{\circ} \leq {\rm
                  dec} \leq -35^{\circ}$, as a function of right
                ascension, compared with IceTop
                data~\citep{Aartsen:2012ma} at 400\,TeV ({\it upper row})
                and 2\,PeV ({\it lower row}).}
\label{LAZLAZ_2}
\end{figure*}

We now calculate $D_{\mu\mu}$, using Eqs.~(\ref{Dmumu_Alf_n1_LAZLAZ})
and~(\ref{Dmumu_Slow_n0_LAZLAZ}), for the same 13 combinations of
$\eps$ and $\delta \mathcal{M}_{\rm A}$ as in Sect.~\ref{I1_Rn2}.

In the upper left panel of Fig.~\ref{LAZCHA}, we plot for
$\{\eps,\delta \mathcal{M}_{\rm A}\}=\{10^{-3},0.1\}$ the $n=\pm 1$
contribution to $\nu$ from Alfv\'en modes and the $n=0$
contribution from pseudo-Alfv\'en modes using orange and blue {\em
  solid} lines, respectively. We also show the $n=\pm 1$ contribution
from pseudo-Alfv\'en modes (Eq.~(\ref{Dmumu_Slow_n1_LAZLAZ})) using a
magenta solid line. At $\delta \mathcal{M}_{\rm A} \geq 0.1$, the
magenta line is below the orange and blue solid lines at all $\mu$,
as was also found for model~C. At
$\delta \mathcal{M}_{\rm A} = 0.01,0.033$, the $n=\pm 1$ contribution
from pseudo-Alfv\'en modes dominates in only a small range of $\mu$,
and its impact on $g$ is weak. By comparing the solid orange
and blue lines to the dashed orange and blue ones, we note that using
$\mathcal{I}_{\rm A,S,2}$ (model~D) instead of $\mathcal{I}_{\rm A,S,1}$ (model~C) 
again
strongly increases the CR scattering off Alfv\'en waves, as was the case
with the narrow resonance function
$R_{\rm n,1}$ (models~A and~B): at all $\mu$, the solid orange line is
two to three orders of magnitude above the dashed orange
one. Scattering off pseudo-Alfv\'en waves is only slightly 
increased by this change (see the blue lines).

This increased scattering is manifested in Fig.~\ref{FirstEigenvalue}
(right panel) by a reduction in $v/\Lambda_1$ of one to four 
orders of magnitude, so that larger portions of the orange lines 
and six out of the computed
13 cases are in the permitted region: $\eps=10^{-1}$
or~$10^{-2}$, with $\delta \mathcal{M}_{\rm A}=1,\,0.33$ or~0.1. 
Two cases,
namely
$\eps=10^{-3}$ with $\delta \mathcal{M}_{\rm A}=0.33$
or~1, are in the allowed region for 100\,TeV CR (see the grey dashed
line in Fig.\ref{FirstEigenvalue}), and thence can be compared with
the 400\,TeV data set. The allowed region for 400\,TeV CRs is 
somewhat
smaller than for 100\,TeV CRs. However, it is still interesting to
consider these two points at $\eps=10^{-3}$, because our coarse
resolution in $\eps$ (one point per decade) does not allow us to test
more favourable points at larger $\eps$ (e.g. $\eps = 2 \times
10^{-3}$), on the same lines with $\delta \mathcal{M}_{\rm A}=0.33$
or~1.

In Fig.~\ref{LAZLAZ_1}, we plot for these eight cases: $\nu$ 
(left column), $g$ (centre column), and $\Delta N/\langle
N \rangle$ compared with IceTop 2\,PeV or 400\,TeV data (right
column). Each $\delta \mathcal{M}_{\rm A}$ has its own line type, see
keys in the left column plots. In the first row, $\eps=10^{-1}$: for
$\delta \mathcal{M}_{\rm A} = 1$~and~0.33, the scattering rate is
about the same at all $\mu$ (left), and the anisotropy is close to a
dipole (centre). The case $\delta \mathcal{M}_{\rm A}=0.1$ is 
also close to a dipole, so that none of these three cases fits IceTop
data (see the three orange dots on the
orange solid line in Fig.~\ref{Half_Width}, lower right panel). 
For
smaller $\eps$, $\theta_{1/2}$ decreases significantly: in
the second row of Fig.~\ref{LAZLAZ_1}, $\eps=10^{-2}$, and
both $\delta \mathcal{M}_{\rm A} = 0.33$~and~0.1 fit the IceTop
2\,PeV data. On the other hand, $\delta \mathcal{M}_{\rm A} = 1$ 
remains close
to being a dipole. In the third row, $\eps=10^{-3}$, and the case
$\delta \mathcal{M}_{\rm A} = 0.33$ provides a sufficiently small cold
spot to fit the 400\,TeV data reasonably well. For $\delta
\mathcal{M}_{\rm A} = 1$, the predicted cold spot is too wide.

In Fig.~\ref{LAZLAZ_2}, we study the dependence of the anisotropy on
CR energy for a fixed set of parameters of the turbulence and
resonance function. We set $\delta \mathcal{M}_{\rm A}$ to $0.33$. 
Since $l$ is fixed, a ten-fold increase in CR energy corresponds to a 
ten-fold increase of $\eps$. Accordingly, we set 
$\eps=10^{-3}$ in the upper row, and
$\eps=10^{-2}$ in the lower row. The left column contains all-sky
plots of $g(\mu)$, the centre column the anisotropy within IceTop field
of view at ${\rm dec} \leq -35^{\circ}$, and the right column, $\Delta
N/\langle N \rangle$. The IceTop data, and their energy-dependence are well
reproduced for these values of $\eps$ and the turbulence parameters: at
$\eps=10^{-3}$, the 400\,TeV data fits well both the small cold
spot and the large flatter region. On increasing the CR energy by
a factor of 10 (close to the nominal factor five difference between 
the \lq\lq 400\,TeV\rq\rq\ 
and \lq\lq 2\,PeV\rq\rq\ data sets) 
the size of the cold spot increases and the higher-energy data is
then also well fitted (see lower row).

Calculations of $\nu$ and $g$ for all cases located in the shaded
areas in Figure~\ref{FirstEigenvalue}, which were not presented in
Sections~\ref{I2_Rn1}--\ref{I2_Rn2}, can be found in
Appendix~\ref{appendixC}, Figure~\ref{Extra_CHALAZ_LAZCHA_LAZLAZ}.

\subsection{Fast magnetosonic modes}
\label{FastModes}

We now treat the case in which the turbulent fluctuations are dominated by 
fast magnetosonic waves, which have an isotropic spectrum with intensity 
$\propto k^{-3/2}$.

\subsubsection{Model E ($R_{\rm n}=R_{\rm n,1}$)}
\label{Fast_Rn1}

\begin{figure*}
  \centerline{\includegraphics[width=0.49\textwidth]{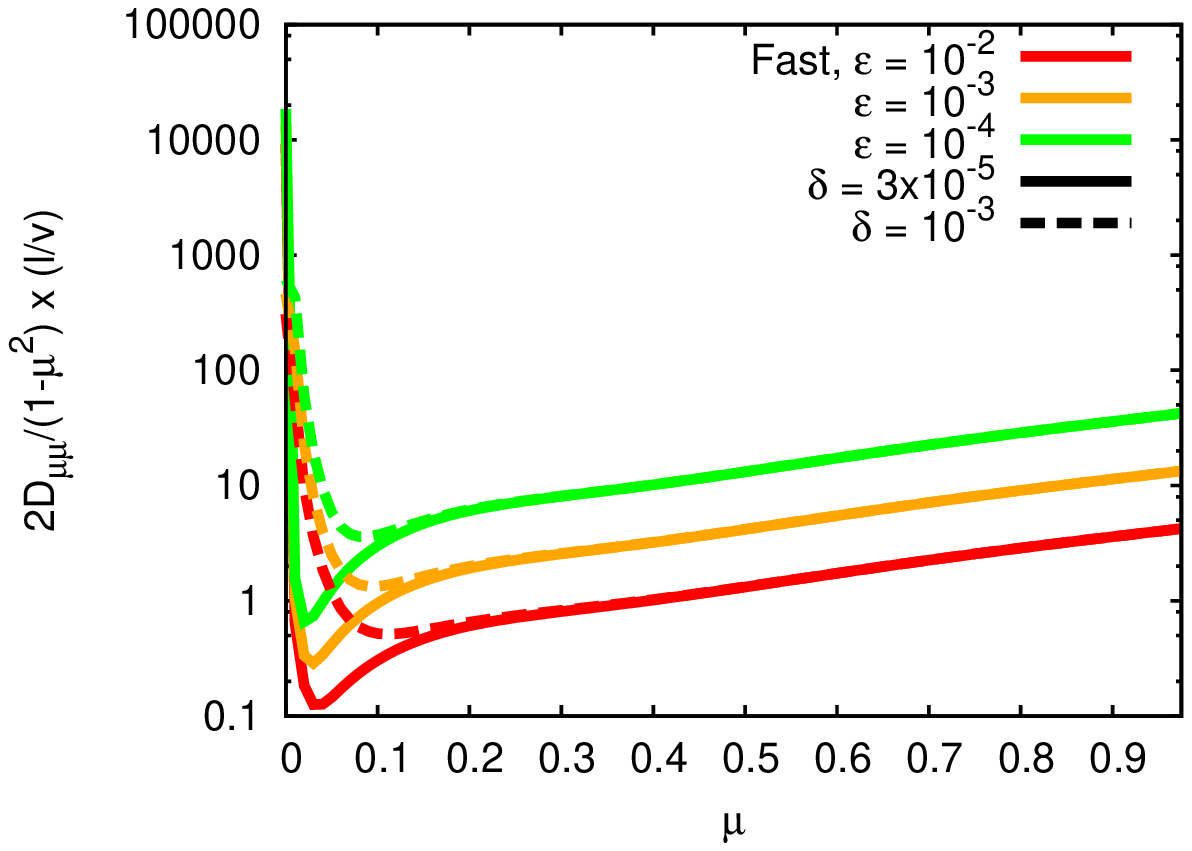}
              \hfil
              \includegraphics[width=0.49\textwidth]{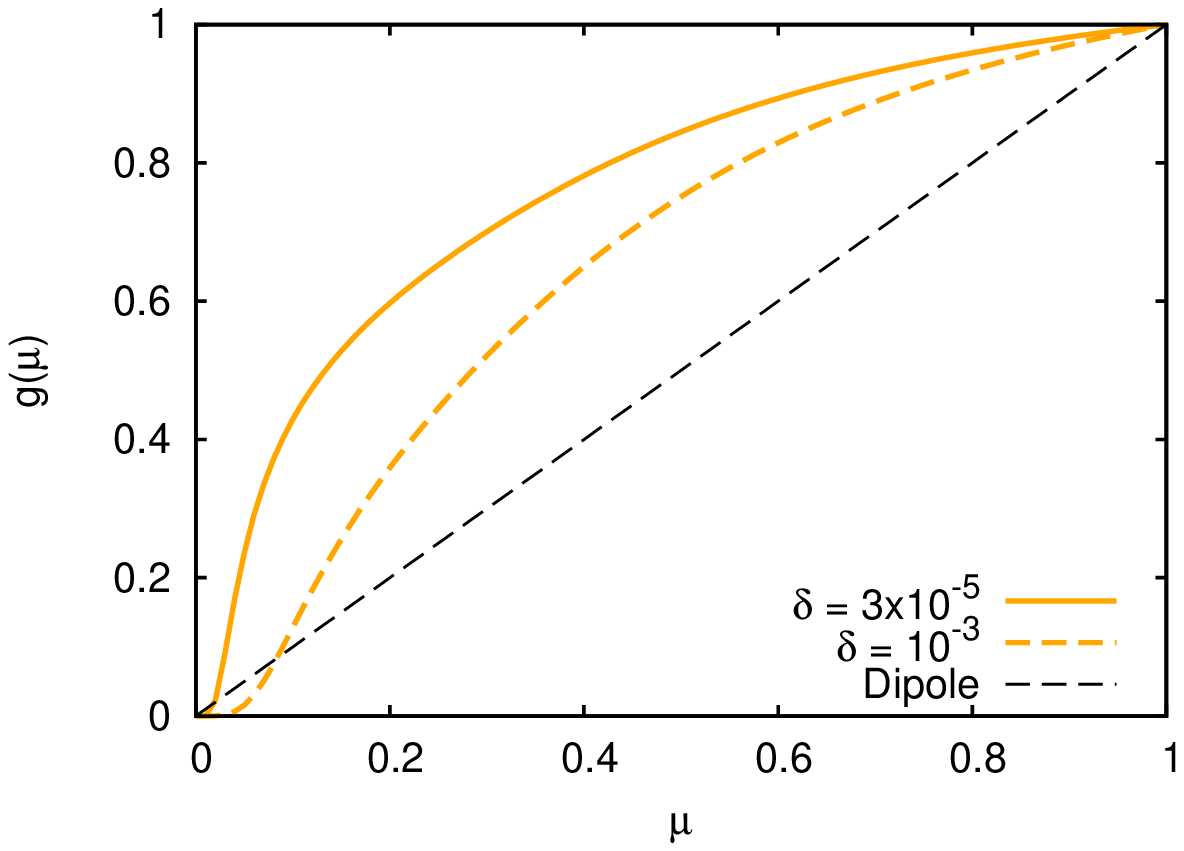}
              }
  \centerline{\includegraphics[width=0.33\textwidth]{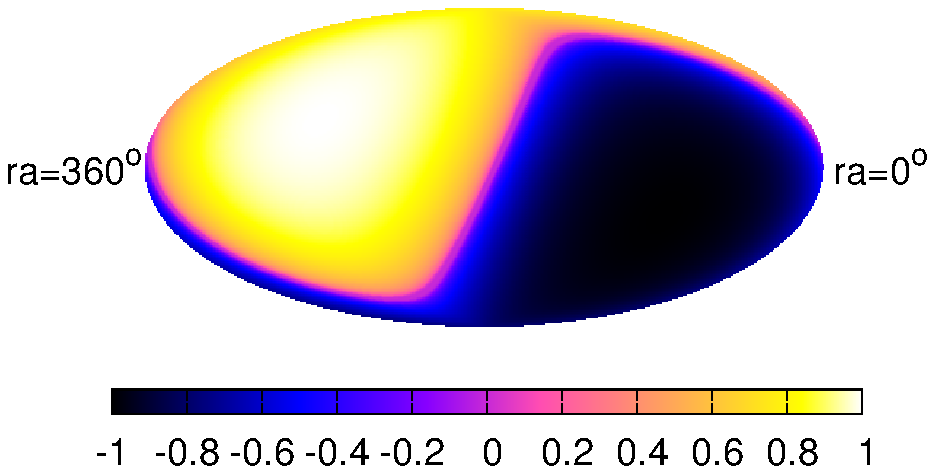}
              \hfil
              \includegraphics[width=0.33\textwidth]{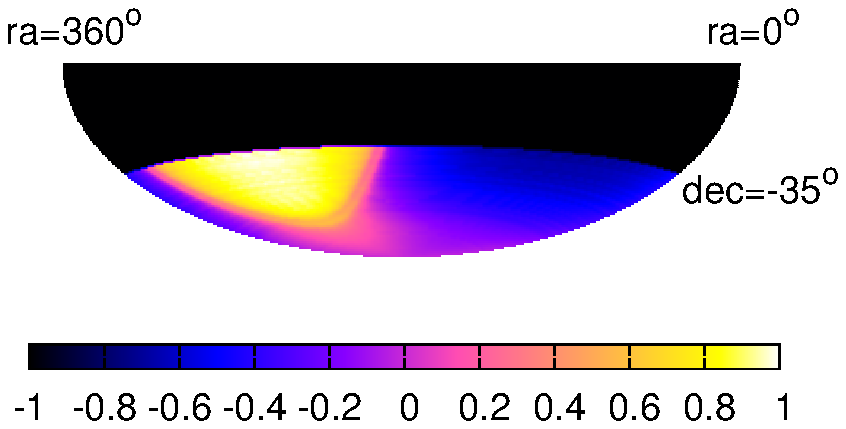}
              \hfil
              \includegraphics[width=0.30\textwidth]{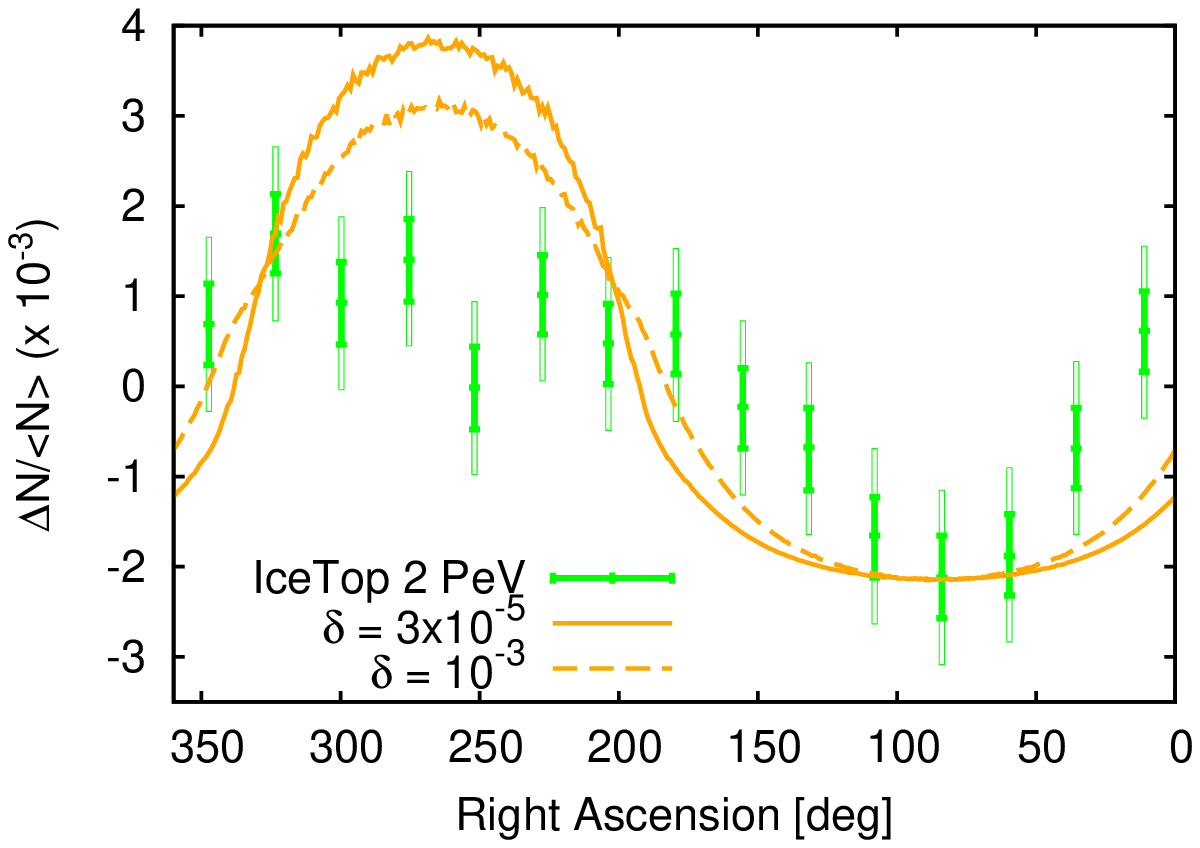}
              }
              \caption{Model E, isotropic fast modes, using $R_{\rm
                  n,1}$. {\it Upper left panel:} $\nu$ as
                a function of $\mu$. Red lines for $\eps=10^{-2}$,
                orange for $\eps=10^{-3}$, and green for
                $\eps=10^{-4}$. Solid (dashed) lines for
                $\delta= 3 \times 10^{-5}$
                ($\delta=10^{-3}$). {\it Upper right panel:} $g$
                versus $\mu$. {\it Lower left panel:} anisotropy
                $g(\mu)$ in equatorial coordinates, for $\delta =
                3\times 10^{-5}$ (and $\eps =10^{-3}$). {\it Lower
                  middle panel:} predicted CR anisotropy, with
                extremum amplitude renormalized to $\pm 1$, in the
                field of view of IceTop. Same values of $\eps$ and
                $\delta$. The anisotropy is calculated here with
                respect to the averaged flux in each declination
                band. {\it Lower right panel:} relative CR intensity
                $\Delta N/\langle N \rangle$ at $-75^{\circ} \leq {\rm
                  dec} \leq -35^{\circ}$, as a function of right
                ascension, and compared with IceTop 2\,PeV
                data~\citep{Aartsen:2012ma}. $\delta \in \{ 3 \times
                10^{-5},10^{-3}\}$, see key.}
\label{FastCHA}
\end{figure*}

Using the narrow resonance function $R_{n,1}$, we
calculate numerically the $n=0,\pm 1$ contributions of fast modes
to $D_{\mu\mu}$ -- see Eqs.~(\ref{Dmumu_Fast_n0_CHALAZ})
and~(\ref{Dmumu_Fast_n1_CHALAZ}), for 6 cases: $\eps \in
\{10^{-2},10^{-3},10^{-4}\}$, with $\delta = 3\times 10^{-5}$ or
$10^{-3}$.

The scattering rate $\nu $ is shown in Fig.~\ref{FastCHA} (upper left
panel). There is a large, narrow peak at $\mu=0$, as in the previous
models of GS turbulence (A and~B) with this resonance function, which
is due here to the $n=0$ contribution from fast modes. As in GS
turbulence, the peak broadens when $\delta$ is increased, as can be
seen by comparing the solid and dashed lines in the upper left panel.
The minimum is located at the value of $|\mu|$ above which the $n=\pm
1$ term starts to dominate over the $n=0$ term. An important
difference compared to the corresponding models of GS turbulence is
that the $n=\pm 1$ contribution of fast modes is typically a few
orders of magnitude larger than that of the Alfv\'en modes of GS
turbulence, leading to values of $v/\Lambda_{1}$ that are about two to
four orders of magnitude smaller than those for GS turbulence, as
shown in the left panel of Fig.~\ref{FirstEigenvalue}, where we plot
$(v/\Lambda_{1})/l$ versus $\eps$ with blue lines. For all the
parameters chosen, this quantity lies in the permitted region, outside
the shaded area.  The {\em shape} of the function $\nu(\mu)$ does not
change significantly with energy (Fig.~\ref{FastCHA}, upper left
panel).  Its absolute value is larger for smaller $\eps$, as
expected. Consequently, $g(\mu)$ also does not change significantly
with energy within the energy range studied.  We plot $g$ in the upper
right panel: both lines, for $\delta = 3\times 10^{-5}$ and $\delta =
10^{-3}$, are well above the black dashed line for a dipole. This
implies that the hot and cold spots are much wider than those for a
dipole, see also the blue lines in upper right panel of
Fig.~\ref{Half_Width}. This is in clear contradiction with existing
observations. We plot in the lower row of Fig.~\ref{FastCHA} a full-sky 
map for $\delta = 3\times 10^{-5}$ (left panel), the anisotropy
within IceTop field of view for the same value of $\delta$ (centre),
and a comparison of $\Delta N/\langle N \rangle$ with IceTop 2\,PeV
data for both values of $\delta$ (right). These maps show no
resemblance to the observed anisotropy, which rules out this model.

\subsubsection{Model F ($R_{\rm n}=R_{\rm n,2}$)}
\label{Fast_Rn2}

\begin{figure*}
  \centerline{\includegraphics[width=0.32\textwidth]{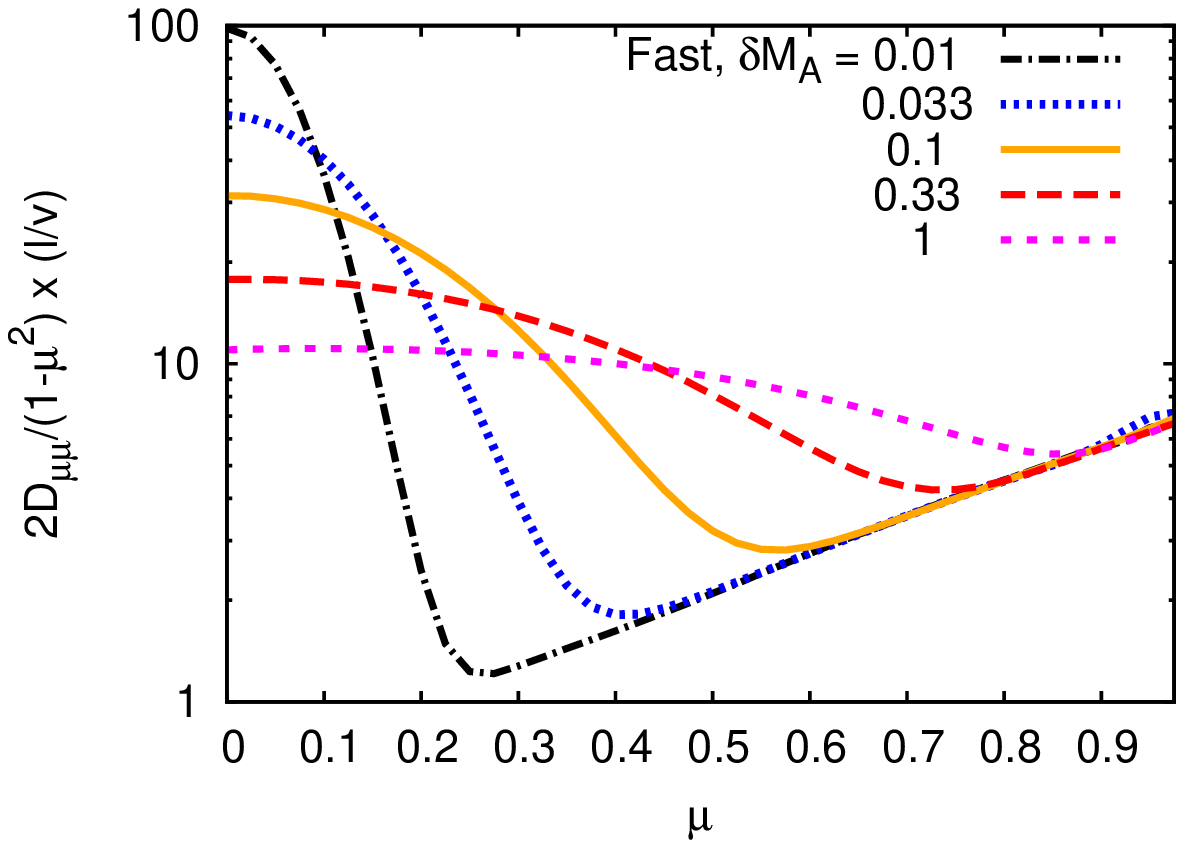}
              \hfil
              \includegraphics[width=0.32\textwidth]{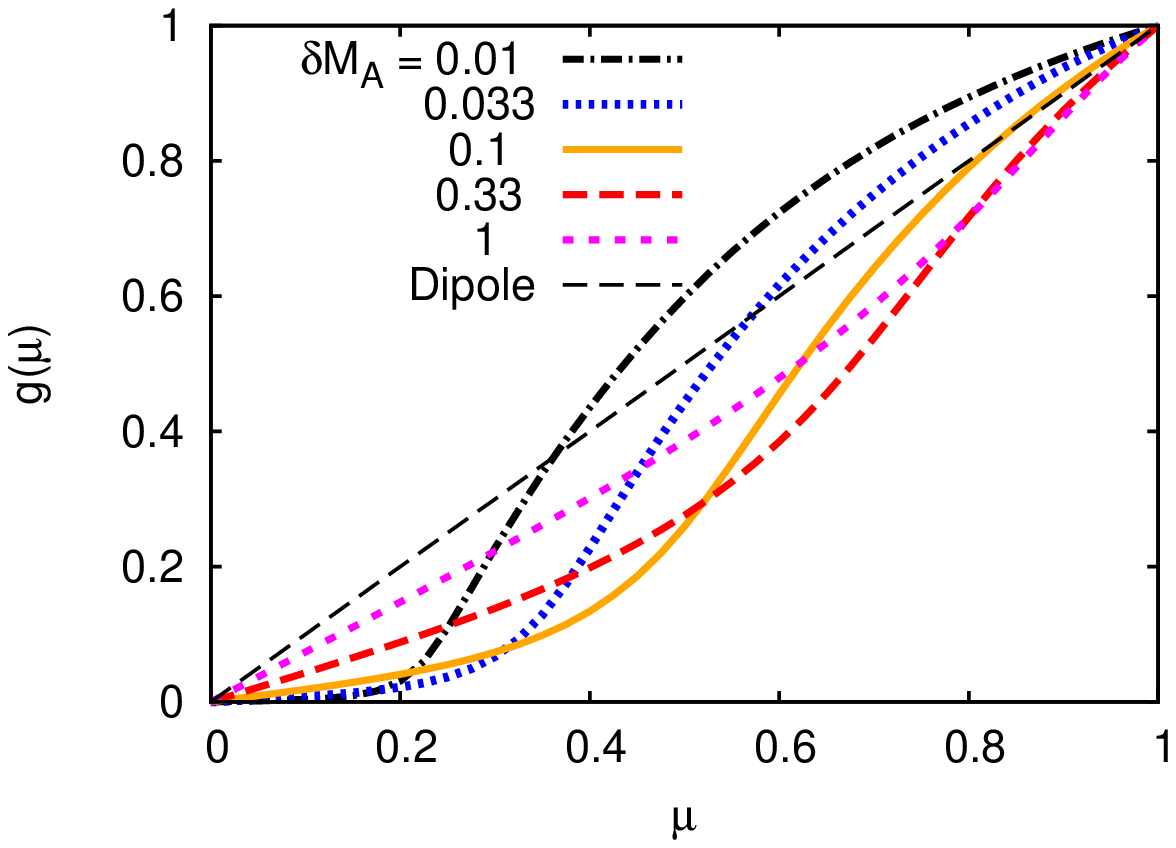}
              \hfil
              \includegraphics[width=0.32\textwidth]{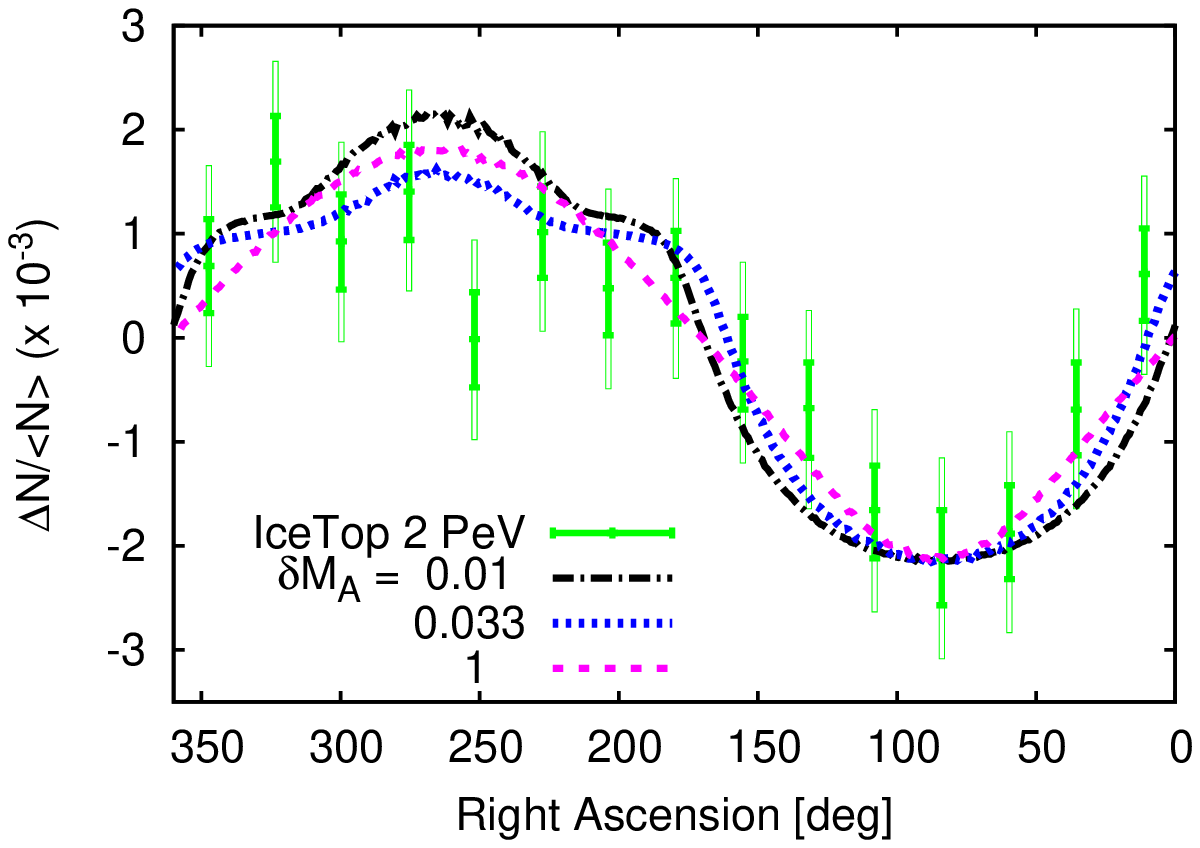}
              }
  \centerline{\includegraphics[width=0.32\textwidth]{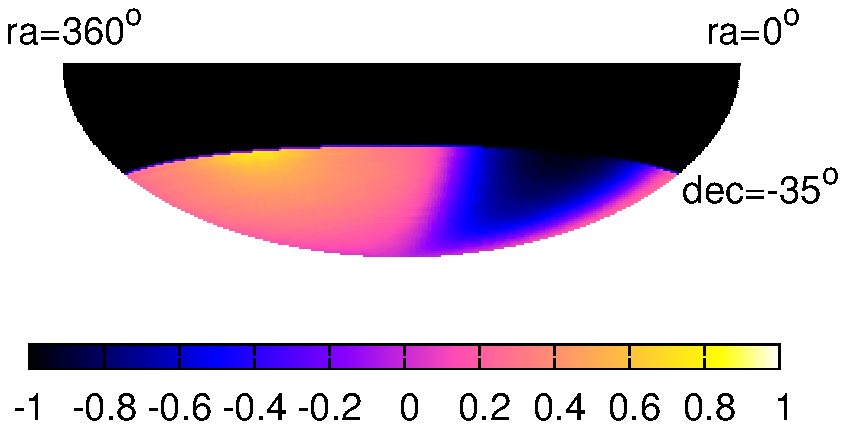}
              \hfil
              \includegraphics[width=0.32\textwidth]{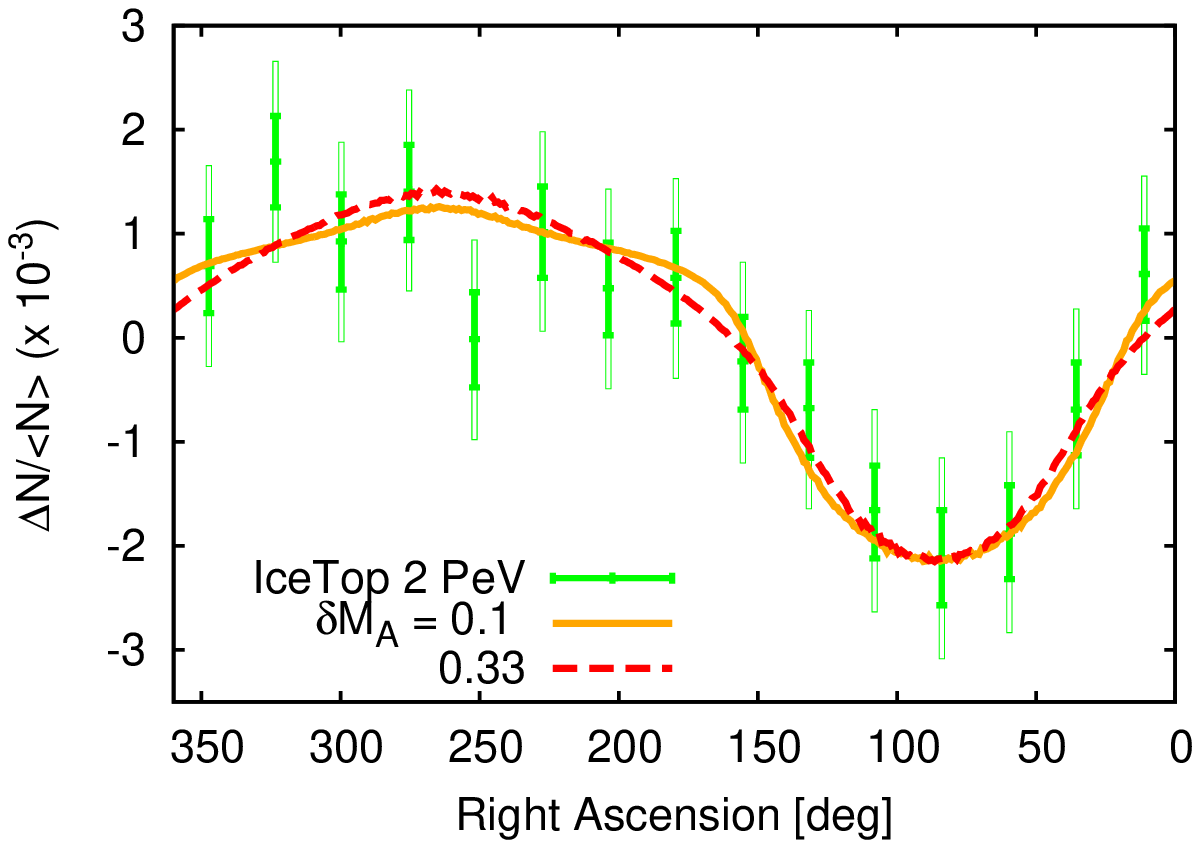}
              \hfil
              \includegraphics[width=0.32\textwidth]{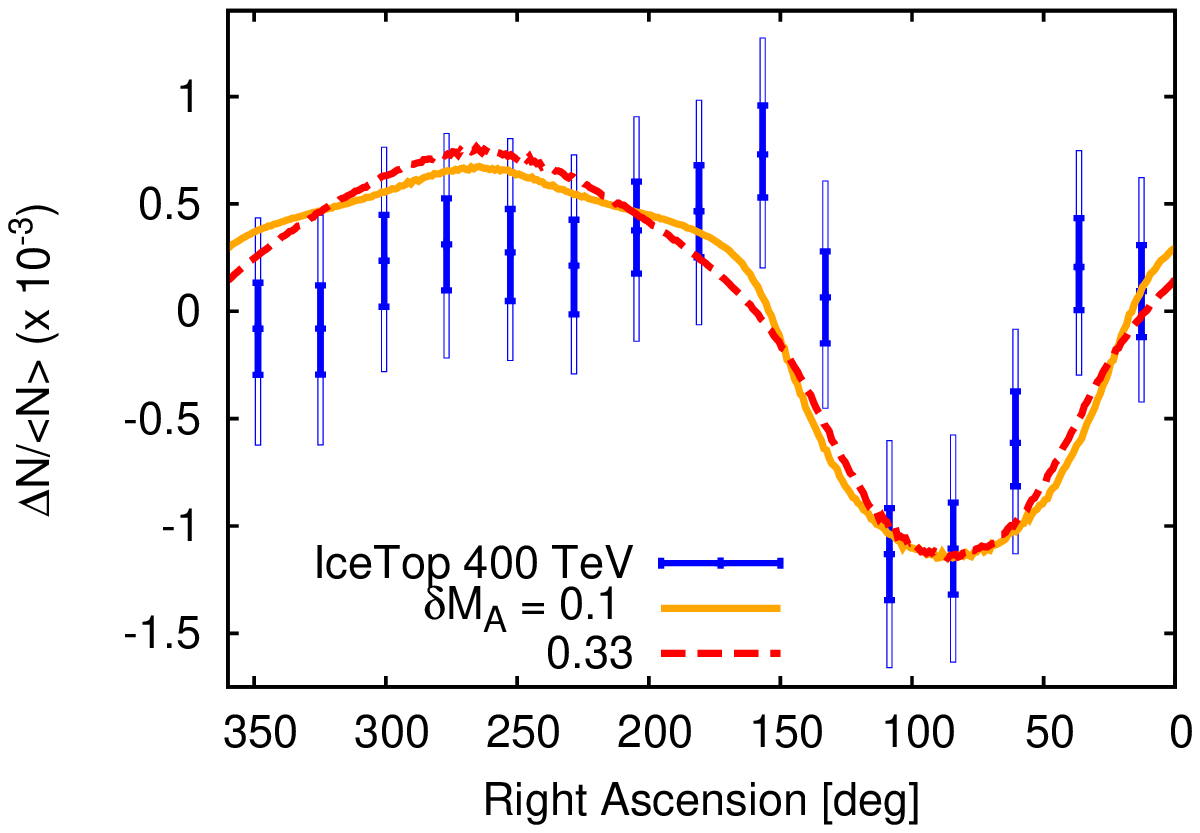}
              }
              \caption{Model F, isotropic fast modes, using $R_{\rm
                  n,2}$. $\eps=10^{-3}$ in all panels, and each line
                type corresponds to a different value of $\delta
                \mathcal{M}_{\rm A} \in \{ 0.01,0.033,0.1,0.33,1 \}$,
                see keys. {\it Upper left panel:} $\nu$ as a function
                of $\mu$. {\it Upper middle panel:} $g$ as a function
                of $\mu$. {\it Upper right panel:} relative CR
                intensity $\Delta N/\langle N \rangle$ at $-75^{\circ}
                \leq {\rm dec} \leq -35^{\circ}$, as a function of
                right ascension, and compared with IceTop 2\,PeV
                data~\citep{Aartsen:2012ma} for $\delta
                \mathcal{M}_{\rm A} \in \{ 0.01,0.033,1 \}$. {\it
                  Lower left panel:} predicted CR anisotropy, with
                extremum amplitude renormalized to $\pm 1$, in the
                field of view of IceTop for $\delta \mathcal{M}_{\rm
                  A} = 0.1$. The anisotropy is calculated here with
                respect to the averaged flux in each declination
                band. {\it Lower middle and lower right panels:}
                $\Delta N/\langle N \rangle$ for $\delta
                \mathcal{M}_{\rm A} \in \{ 0.1,0.33\}$, compared with
                IceTop 2\,PeV ({\it middle}) and 400\,TeV ({\it
                  right}) data.}
\label{FastLAZ}
\end{figure*}

Taking now the broad resonance function, $R_{n,2}$, we 
calculate numerically the $n=0,\pm 1$ contributions of fast modes
to $D_{\mu\mu}$ --see Eqs.~(\ref{Dmumu_Fast_n0_LAZLAZ})
and~(\ref{Dmumu_Fast_n1_LAZLAZ}), for 20 combinations of $\eps$ and
$\delta$: $\eps \in \{10^{-1},10^{-2},10^{-3},10^{-4}\}$, $\delta
\mathcal{M}_{\rm A} \in
\{1,0.33,0.1,0.033,0.01\}$. 
In Fig.~\ref{FirstEigenvalue},
$(v/\Lambda_{1})/l$ is shown as a function of $\eps$ as a broad,
blue band in the right panel, which illustrates the range of values 
we find for $0.01 \leq \delta \mathcal{M}_{\rm A} \leq 1$. All values of
$(v/\Lambda_{1})/l$ are outside the area shaded in grey, and, therefore, 
viable in the sense that the boundary layers of the flux tube are thin,
so that the observed anisotropy should correspond to the predicted one. 

In this energy range, we find no noticeable dependence of $g(\mu)$ on
$\eps$. Results for $\nu$ are presented at $\eps=10^{-3}$, as an
example. We plot in Fig.~\ref{FastLAZ} (upper left panel) $\nu$ 
versus $\mu$ for all five tested values of $\delta\mathcal{M}_{\rm
  A}$. As in Sect.~\ref{Fast_Rn1}, the minimum corresponds to the
separation between the low-$\mu$ region, where the $n=0$ term dominates, 
and the high-$\mu$ region, where the $n=\pm 1$ term dominates. The
width of the peak around $\mu=0$ increases with increasing $\delta
\mathcal{M}_{\rm A}$. This is reminiscent of the behaviour of the
contribution of pseudo-Alfv\'en modes in GS turbulence (models C and~D). 
Results for $g(\mu)$ are shown in the upper centre
panel. The size of the hot/cold spot decreases with increasing
$\delta \mathcal{M}_{\rm A}$ in the range $0.01-0.33$, but
at $\delta \mathcal{M}_{\rm A}=1$, where $\nu$ is almost independent of 
$\mu$, the anisotropy moves 
closer to the result for a dipole --- see the 
magenta dotted line and red dashed
line in the upper centre panel. This behaviour with $\delta
\mathcal{M}_{\rm A}$ is also visible in the lower right panel of Fig.~\ref{Half_Width}. 
The minimum $\theta_{1/2}$ ($\simeq 47^{\circ}$) is reached at
$\delta \mathcal{M}_{\rm A}=0.33$, although the value of
$\theta_{1/2}$ at $\delta \mathcal{M}_{\rm A}=0.1$ is not much
larger. We plot our predictions for $\Delta N/\langle N \rangle$ and
compare them with the 2\,PeV data set in the upper right and lower
centre panels of Fig.~\ref{FastLAZ}. 
The cold spot is too large for $\delta \mathcal{M}_{\rm A} = 0.01$, $0.033$, and~$1$.
On the other hand, $\delta \mathcal{M}_{\rm A} = 0.1$ and
$\delta \mathcal{M}_{\rm A} = 0.33$ provide a satisfactory fit to the
data. Although $\theta_{1/2}$ is slightly smaller for 0.33 than for
0.1, the shape of the anisotropy appears to be slightly better at
$\delta\mathcal{M}_{\rm A}=0.1$ --- in the
latter case, $\Delta N/\langle N \rangle$ is flatter in the region around
$160^{\circ} \leq {\rm RA} \leq 360^{\circ}$, see lower centre
panel. In the lower left panel, we plot a sky map for $\delta
\mathcal{M}_{\rm A} = 0.1$, as would be observed within the field of
view of IceTop, at ${\rm dec} \leq -35^{\circ}$. The qualitative
agreement with the IceTop sky map at 2\,PeV is good. However, none of the
values of $\delta \mathcal{M}_{\rm A} \in [0.01,1]$ provide an acceptable fit to
the smaller cold spot in the 400\,TeV data. We show this in the lower right panel which 
compares the two cases with the smallest cold spots ($\delta
\mathcal{M}_{\rm A} = 0.1$ and 0.33) to IceTop 400\,TeV data set. If
such a small cold spot at 400\,TeV were to be confirmed in the future, 
this could have important implications, as we discuss below.

\section{Discussion and perspectives}
\label{Discussion}

The model developed in Sect.~\ref{Model} assumes that CRs propagate
in magnetic turbulence with a strong background field. Our approach
requires the CR Larmor radius and mean free path to remain smaller
than the coherence length of the interstellar magnetic field. We
enforce these conditions by keeping $\eps \ll 1$, and restricting
our study to the non-shaded area of
Fig.~\ref{FirstEigenvalue}. Also, we take the anisotropy to be
parallel to the regular field. In a 3D~model, the perpendicular
component of the CR density gradient would give rise to a
perpendicular component in the anisotropy, but we neglect it here,
in line with the observations of~\citet{Schwadron2014}.  We further
assume the turbulence to be homogeneous, i.e. to have properties that do
not vary significantly within the $\sim 10$\,pc flux tube in which
we solve Eq.~(\ref{TransportEqn}). In general, the coefficient
$D_{\mu\mu}$ could vary within the flux tube, and the pitch-angle
scattering rate could also depend on the gyrophase. However, these
effects would make the problem intractable analytically. Finally, 
we note that imbalanced turbulence would lead to an asymmetry in the
large-scale anisotropy within our model. Such an asymmetry 
is absent from the examples we have chosen to test in this paper, but it
would be straightforward to include it.

Within the confines of this simplified model, the
results of the previous section demonstrate explicitly that the
large-scale anisotropy can take on a variety of forms, depending on
the turbulence model and its parameters. In some cases (e.g., models
C, D and F from Table~\ref{tableoverview}, with an unrealistically
large $\delta\mathcal{M}_{\rm A}=1$), this can resemble a simple dipole
anisotropy. However, this is not the generic form and is also not
predicted for any reasonable scenario, including that of isotropic
fast-mode turbulence.  In fact, only the physically unfounded
assumption of {\em isotropic pitch-angle scattering}, $D_{\mu\mu}
\propto 1 - \mu^{2}$, leads to a pure dipole anisotropy.  It is,
therefore, not only unsurprising, but also reassuring that
measurements of the large-scale anisotropy
\citep[e.g.,][]{Aartsen:2012ma} cannot be reconciled with a pure
dipole anisotropy.

A generic feature of the models we study is that the scattering
frequency $\nu$ has a peak around $\mu=0$. This corresponds to the
\lq\lq transit-time damping\rq\rq\ mentioned, for example,
by~\citet{Yan:2007uc}.  We argue that this leads to a flattening of
the large-scale CR anisotropy in directions approximately perpendicular
to the local magnetic field, which is compatible with
IceCube~\citep{Abbasi:2011zk} and IceTop~\citep{Aartsen:2012ma} data
at energies $\gtrsim 100$\,TeV. This feature may also exist in the
low-energy ($\sim10$\,TeV) data of IceCube, see Fig.~9
in~\cite{Aartsen:2016ivj}: the measured relative intensity is rather
flat on ${\rm RA} \approx 0^{\circ} - 100^{\circ}$, i.e. in the
direction opposite to the minimum at ${\rm RA} \approx 200^{\circ} -
250^{\circ}$. It is worth noting that such a feature naturally appears
in all cases of turbulence we studied, without any fine-tuning of the
theory. In the case of GS turbulence, the peak of $\nu$ around $\mu =
0$ is due to pseudo-Alfv\'en waves. In the case of compressible
turbulence, the peak is due to the $n=0$ term for fast modes in
Eq.~(\ref{EqnDmumu}). In physical terms, this flattening can be
understood as follows. In Eq.~(\ref{Eqn_f}), the CR density is
$\propto x$. If $\nu$ is large on $- \Delta \mu \leq \mu \leq + \Delta
\mu$, the \lq\lq last scattering surface\rq\rq\ for CRs arriving from
these directions is relatively close to Earth. Thence, they come from
closeby regions of the local interstellar medium with small $x$
values, where the CR densities and distribution functions are
similar. This leads to an anisotropy, $g$, that is approximately
constant on $- \Delta \mu \leq \mu \leq + \Delta \mu$.

The peak at $\mu\approx0$ may appear either narrow or broad depending
on the type of resonance function and on the level of scattering at
larger values of $|\mu|$. The dominant contribution to CR scattering
at large $|\mu|$ is shear Alfv\'en modes for GS turbulence, and the
$n=\pm 1$ term for fast modes. The minimum in $\nu$ occurs around the
value of $\mu$ where the dominant contribution to scattering changes.

An important difference between the two resonance functions we used is
that $R_{\rm n,2}$ tends to produce broader peaks around $\mu=0$ than
$R_{\rm n,1}$, unless $\delta \mathcal{M}_{\rm A}$ is very small ($\ll
0.01$). As a consequence, the half-width of the anisotropy, which
characterises the size of the spots at $\mu=\pm1$, is usually smaller
for $R_{\rm n,2}$ than for $R_{\rm n,1}$. We note that this is not
systematically the case: for example, when the contribution to $\nu$
from shear Alfv\'en waves at large $|\mu|$ is at about the same level
as the contribution from pseudo-Alfv\'en waves at smaller $|\mu|$, the
anisotropy tends towards a dipole and $\theta_{1/2}\rightarrow
60^{\circ}$, see, for example, model~D for GS turbulence with
$\mathcal{I}_{\rm A,S}=\mathcal{I}_{\rm A,S,2}$, $R_{\rm n}=R_{\rm
  n,2}$, $\eps=10^{-1}$, and $\delta \mathcal{M}_{\rm A}=1$ in
Fig.~\ref{LAZLAZ_1} (upper row). Nonetheless, in most tested cases
that lie in the allowed region in Fig.~\ref{FirstEigenvalue}, i.e.,
where condition~(\ref{Eqnmfpcondition}) is satisfied,
$\theta_{1/2}\lesssim 60^{\circ}$ (resp. $> 60^{\circ}$) with $R_{\rm
  n,2}$ (resp. $R_{\rm n,1}$), as can be seen by comparing the values
of $\theta_{1/2}$ for the filled circles and triangles in the two
panels in Fig.~\ref{Half_Width}. Thus, $R_{\rm n,2}$ can produce cold
and hot spots in the direction of field lines ($\mu=\pm 1$) that are
narrower than those expected for a dipole anisotropy. In general,
$R_{\rm n,2}$ provides a better fit to IceTop data than $R_{\rm n,1}$,
and the existing TeV--PeV CR anisotropy data tends to favour
moderately broad resonance functions. 

In all the cases studied in which condition~(\ref{Eqnmfpcondition}) is
satisfied, $R_{\rm n,1}$ provides anisotropies whose $\theta_{1/2}$
are too large to be compatible with IceTop observations. The model of
isotropic fast modes with $R_{\rm n}=R_{\rm n,1}$ can be safely ruled
out: all cases lie in the allowed region of
Fig.~\ref{FirstEigenvalue}, but the shape of the anisotropy is clearly
incompatible with observations. GS turbulence with $R_{\rm n,1}$
(models~A and~B) can also be ruled out for high-energy CRs when
condition~(\ref{Eqnmfpcondition}) is satisfied.  Strictly speaking,
however, GS turbulence with $R_{\rm n,1}$ cannot be ruled out
altogether, because we cannot exclude the possibility that we live in
a region of space where $v/\Lambda_{1}$ is too large for
condition~(\ref{Eqnmfpcondition}) to be satisfied. In this case, the
anisotropy is not given by Eq.~(\ref{FormulaGmu}), but is, instead,
completely determined by the unknown conditions at the boundary of our
flux tube, which would make it essentially impossible ever to infer
constraints on the turbulence from measurements of the CR anisotropy.

Using $R_{\rm n,2}$, we find that both GS turbulence and isotropic
fast modes can fit IceTop 2\,PeV data. GS turbulence with
$\delta \mathcal{M}_{\rm A} = 0.1-0.33$, and 
power-spectrum $\mathcal{I}_{\rm A,S,1}$ ($\mathcal{I}_{\rm
  A,S,2}$), provides a good fit for $\eps=10^{-1}$ ($\eps=10^{-2}$). For isotropic
fast modes, the 2\,PeV data is fitted with the same range of $\delta
\mathcal{M}_{\rm A}$. Slight changes in the spectrum
$\mathcal{I}_{\rm A,S}$ for GS turbulence have a sizeable impact on the
CR anisotropy: $\mathcal{I}_{\rm A,S,2}$ leads to a larger value
of $\theta_{1/2}$ than $\mathcal{I}_{\rm A,S,1}$, at the same value of 
$\eps$ and other parameters. The function $\mathcal{I}_{\rm A,S}$ is not
explicitly specified by the theory
of~\cite{Goldreich:1994zz}. \cite{Cho:2001hf} claim that an
exponential function ($\mathcal{I}_{\rm A,S,2}$) for the shape of the
cutoff in $k_{\parallel}$ provides a better fit to their simulations
for Alfv\'en modes, than the Heavside function ($\mathcal{I}_{\rm
  A,S,1}$) used by~\cite{Chandran:2000hp}.  We note that a strong
dependence of $\nu$ and $g$ on the shape of $\mathcal{I}_{\rm A,S}$
implies that the CR anisotropy data can be used as a way to measure
$\mathcal{I}_{\rm A,S}$, if, locally, CRs scatter on GS turbulence.

It is interesting to note that $\theta_{1/2}$ increases with CR energy
for GS turbulence (for both resonance functions), at $\delta$ or
$\delta \mathcal{M}_{\rm A}$ fixed. An increase of the size of the
deficit in IceTop data was reported in~\cite{Aartsen:2012ma}, between
the 400\,TeV and 2\,PeV sets. We find that GS turbulence with
$\mathcal{I}_{\rm A,S}=\mathcal{I}_{\rm A,S,2}$, $R_{\rm n}=R_{\rm
  n,2}$, and $\delta \mathcal{M}_{\rm A}=0.33$ provides a good fit:
around $\eps \simeq 10^{-3}$, model~D fits the 400\,TeV data well, 
reproducing its deficit at $\simeq 30^{\circ}$. Increasing the CR
energy by a factor similar to that between the two data sets, 
model~D also fits the 2\,TeV data, with the same turbulence parameters. The
level of anisotropy in the fluctuations in a GS power-spectrum depends on $|{\bf k}|$, 
being more isotropic at longer wavelengths: low-energy CRs 
\lq\lq see\rq\rq\ a more anisotropic spectrum than those with higher
energy, and, therefore, $\theta_{1/2}$ varies with CR energy. On the
other hand, $\theta_{1/2}$ does not vary noticeably with energy for
isotropic fast modes. Here, CRs with
different energies \lq\lq see\rq\rq\ the same angular distribution of modes with
which they can resonate. Only the normalisation is different, with
more power being present at smaller $|{\bf k}|$. 

If one takes the energy-dependence in IceTop data at face value, the
absence of changes in the shape of the CR anisotropy with $\eps$ for
isotropic fast modes is an argument against them, i.e., against either
fast modes being isotropic, or fast modes providing
scattering. \citet{Aartsen:2016ivj} show only one data set for IceTop
(1.6\,PeV), and the angular size of the deficit in the $\sim 10$\,TeV
data from IceCube does not appear to be noticeably smaller than that
in IceTop 2\,PeV data.  Nevertheless, the observation of an
energy-dependence in the half-width of the anisotropy could suggest a
GS-like, $|{\bf k}|$-dependent anisotropy in the turbulence
power-spectrum.  Given that fast modes may suffer anisotropic damping
\citep[e.g.,][]{Yan:2007uc}, it is not entirely implausible that also
their spectrum is anisotropic, and we speculate that this would
produce a qualitatively similar effect.  However, more detailed
numerical simulations, taking into account effects such as the
potential energy dependence of the width of the resonance function,
would be needed to clarify such an interpretation of the data.

The applicability of our model, i.e., of the diffusion approximation
in the vicinity of the Earth, is determined by the magnitude of the
smallest positive eigenvalue $\Lambda_1$, and not, as commonly
assumed, the CR mean free path (MFP), $\lambda_{\parallel}$, defined by
Eq.~(\ref{MFP_Eqn}). Nevertheless, these quantities are related, and,
for completeness, we plot $\lambda_{\parallel}$ for the six
turbulence models in Appendix~\ref{appendixB}. From
Figs.~\ref{FirstEigenvalue} and \ref{MFP_All_Figure}, it is clear that
a small variation in the shape of $\mathcal{I}_{\rm A,S}$ for GS
turbulence has a substantial impact, with $\mathcal{I}_{\rm A,S,2}$
providing a larger $\Lambda_1$ and a smaller MFP than
$\mathcal{I}_{\rm A,S,1}$, in line with the findings
of~\cite{Yan:2002qm}. Most of this difference can be attributed to the
contribution from (shear) Alfv\'en modes.  A sharp cutoff in
$k_{\parallel}$ in the power-spectrum, such as that in
$\mathcal{I}_{\rm A,S,1}$, leads to a complete absence of these modes
at large $k_{\parallel}$ ($k_{\parallel}>k_{\perp}^{2/3}l^{-1/3}$),
and, consequently, to a relatively low scattering rate for CRs with
large $|\mu|$. In contrast, an exponential cutoff in $k_{\parallel}$,
such as that in $\mathcal{I}_{\rm A,S,2}$, leaves some power in modes
with $k_{\parallel}>k_{\perp}^{2/3}l^{-1/3}$, even though most of it
is concentrated in modes with wave vectors perpendicular to field lines. It
is this small increase in power at larger $k_{\parallel}$ that
produces the strong increase in $\Lambda_1$ and the corresponding
decrease in $\lambda_{\parallel}$.

Our results also show that 
$\Lambda_1$ increases (and $\lambda_{\parallel}$ decreases) 
when increasing $\delta$ or $\delta\mathcal{M}_{\rm A}$. Both for GS turbulence and
isotropic fast modes, and for a physically-relevant $\delta \sim 3
\times 10^{-5}$, scattering is more effective with the resonance function
$R_{\rm n,2}$ than with $R_{\rm n,1}$, unless $\delta\mathcal{M}_{\rm A} \ll 0.1$.

In the majority of cases studied, Figs.~\ref{FirstEigenvalue} and
\ref{MFP_All_Figure} confirm the claim of~\citet{Yan:2002qm} that the
MFP is significantly smaller for isotropic fast modes than for GS
turbulence. If fast modes are present to a non-negligible level in the
local interstellar medium, they should then provide the dominant
contribution to CR scattering.  Nevertheless, $\lambda_{\parallel}$ in
GS turbulence with $\mathcal{I}_{\rm A,S,2}$ and $R_{\rm n,2}$ is
smaller than in isotropic fast modes at $\eps \sim
10^{-2}-10^{-1}$. While the problem raised by~\citet{Chandran:2000hp}
concerning CR confinement in the Galaxy and the large value of
$\lambda_{\parallel}$ in GS turbulence remains in most cases, we note
that using $\mathcal{I}_{\rm A,S,2}$ and $R_{\rm n,2}$ with large
$\delta\mathcal{M}_{\rm A}$ decreases the tension. In this connection,
it is important to note that CR diffusion coefficients deduced from
the boron-to-carbon ratio reflect CR propagation on kiloparsec scales, and do
not rule out a much larger value of $\lambda_{\parallel}$ within our
local flux tube. Provided CRs are well confined in the halo, where the
physical conditions are different, $\lambda_{\parallel}$ could be
large in the entire disk without compromising their confinement in the
Galaxy on the correct time scale.

Finally, we point out that for anisotropic turbulence, scattering can
be more effective with increasing CR energy.  In GS turbulence, this
trend is seen in Fig.~\ref{FirstEigenvalue}, and is reflected in the
decrease of $\lambda_{\parallel}$ with $\eps$ in
Fig~\ref{MFP_All_Figure}, at relatively large $\eps$. This is due to
the fact that the power-spectrum of GS turbulence becomes more
isotropic at small $|{\bf k}|$, providing more scattering. In the
literature, the CR diffusion coefficient, and, therefore,
$\lambda_{\parallel}$, are usually assumed to grow with energy as a
power law. While the CR spectrum favours such a dependence on energy
on the scale of the Galaxy, this does not rule out the proposed local
energy dependence, a subtlety that is worth keeping in mind when
analysing observables such as the {\em amplitude} of the CR
anisotropy, that depend on the local value of $\lambda_{\parallel}$.


In the absence of a CR flux along $x$, the effects we describe cannot
be responsible for the observed anisotropy. However, this is an
unlikely scenario, because CR density gradients, which would drive a
flux, are expected to exist for three different reasons. Firstly, CRs
are injected into the Galaxy at different positions in the disk, so
that those from nearby, recent sources necessarily create a
gradient~\citep[e.g.][]{Erlykin:2006ri,Blasi:2011fm}. Secondly, the
dependence of CR density on galactocentric radius also induces a
gradient, and, thirdly, CR escape in the halo may do this too.  Which
of these effects dominates in the vicinity of Earth is not clear, but
is also unimportant for our work. Above $\approx 100$\,TeV, the
anisotropy flips by $\approx 180^{\circ}$ with respect to its
direction at $\sim 1-10$\,TeV~\citep{Aartsen:2016ivj}, most likely
because of a reversal of the local gradient when projected onto the
field lines in our local flux tube. This can be naturally explained by
a change in the dominant source of the
gradient~\citep[e.g.][]{Ahlers:2016njd}. Whether the anisotropy points
in one direction along the field lines, or the opposite, does not
affect our predictions for $g(\mu)$, and our theory can be applied
both above and below 100\,TeV.

\cite{Haverkorn:2008tb} measured the outer scale of the interstellar
magnetic turbulence in the disk between $\sim 1$ and 100\,pc depending
on location. Within $\sim10$\,pc from Earth, the direction of the 
field lies out of the Galactic plane, and does not coincide with that
of the kiloparsec-scale \lq\lq regular Galactic magnetic field\rq\rq, which
approximately follows the spiral
arms~\citep{Heiles1996,Han:2006ci}. This is
consistent with our assumption of local turbulence with an outer scale 
of a few tens of parsecs, whose orientation within a coherence length
determines the orientation of our \lq\lq flux tube\rq\rq.
This turbulence may
be driven by supernovae, which can stir the interstellar medium 
on scales up to
$\approx 100$\,pc. However, interstellar turbulence may also 
be driven on 
a much smaller scale, without affecting our study. 
The relevant $l$ corresponds to that of the
turbulence in the flux tube on which CR effectively scatter, 
which can also be driven, for example, by 
CR streaming. For
$400\,{\rm TeV}-2\,{\rm PeV}$ CR in a few $\mu$G field, the range
$\eps = 10^{-4}-10^{-1}$ used in our study corresponds to $l \sim 1 -
100$\,pc.

The polarization of optical light from nearby
stars~\citep{Frisch:2012zj} probes the magnetic field direction
within ten to a few tens of parsecs from Earth, while the direction
inferred from IBEX data \citep{Zirnstein2016} corresponds to that of
the magnetic field just outside the heliopause. (See also
\citet{Burlaga2016} for in-situ measurements from Voyager.) The
direction of the field within $\sim 10$\,pc from Earth is known with
an accuracy of $20^{\circ}$--$30^{\circ}$.  For example, the
direction deduced from the IBEX ribbon has varied by $\approx
10^{\circ}$ over the years
\citep[see][]{Funsten2009,Funsten2013,Schwadron2009}. In
Sect.~\ref{Results}, our predictions for $\Delta N/\langle N
\rangle$ are presented for a fixed direction of the magnetic
field. A more elaborate fitting procedure could also take into account
the uncertainty in the directions of the interstellar field and
of the gradient of CR density, at the expense of introducing
additional degrees of freedom. However, the available observational
database does not, as yet, require such an approach.



We point out that IceTop data~\citep{Aartsen:2012ma} already sets
interesting constraints on the properties of both the turbulence and
CR transport, even if the 2\,PeV data alone cannot, as yet,
distinguish between GS turbulence and isotropic fast modes. It is also
possible that a $|{\bf k}|$-dependent anisotropy in the power-spectrum
of fast modes would lead to an energy-dependence of the size of the
cold spot, similar to that in GS turbulence. We expect that more
stringent constraints could be set by conducting a systematic analysis
of all existing measurements of the large-scale anisotropy, including
those at $\sim (1-10)$\,TeV energies, and those in the Northern
hemisphere, provided that the distortions arising from
heliospheric fields can be reliably identified and removed.

The opposite approach is also possible: by constraining $g(\mu)$ from
observations, one can infer the {\em shape} of $D_{\mu\mu}$ versus
$\mu$ from Eq.~(\ref{FormulaGmu}), without knowing the turbulence
properties, or making any assumption about the resonance function.

IceCube and HAWC Collaborations have measured the CR anisotropy and
presented it in the form of an angular power spectrum
\citep{TheHAWC:2015vta,Aartsen:2016ivj}.  Assuming the large-scale
anisotropy has the form of a dipole, \citet{Ahlers:2013ima} and
\citet{Ahlers:2015dwa} argued that the $C_{\ell}$'s with $\ell \geq 2$
can be understood as arising from deviations of the anisotropy from
its ensemble-average, due to the given realisation of the turbulence
within a CR mean free path from Earth.  Our study raises the
possibility of an alternative interpretation. Since the large-scale
anisotropy predicted in our turbulence models is not a dipole, it can
{\em a priori} contribute to the $C_{\ell}$'s, at any $\ell$, and may
dominate the contribution from fluctuations about the ensemble
average. A directional analysis of the higher-order multipoles in the
data may help to disentangle these two contributions, since multipoles
arising from the large-scale anisotropy are directed along the local
magnetic field.

\section{Summary and conclusions}
\label{Conclusions}

Our results demonstrate that the 
large-scale CR anisotropy is {\em not} expected to be a dipole, even
for isotropic turbulence, but instead encodes valuable information
on the statistical properties of the local interstellar
magnetic turbulence.

We present predictions of this anisotropy for both Goldreich-Sridhar (GS)
turbulence (Sect.~\ref{AlfvenModes}), and for isotropic fast modes
(Sect.~\ref{FastModes}), using two types of resonance function,
and two parameterisations of the fluctuation spectrum.
Our main findings are:
\begin{itemize}

\item The angle-dependent CR scattering frequency exhibits a peak
  perpendicular to the local field lines ($\mu=0$). In the case of
  incompressible turbulence, this peak is due to pseudo-Alfv\'en
  modes. At larger values of $|\mu|$, scattering is provided
  by (shear) Alfv\'en modes. (See the blue and orange lines in the 
  upper left panels of Figs.~\ref{CHALAZ} and \ref{LAZCHA}.)

\item Under most conditions, a broad resonance function produces a
  broader peak, leading to cold/hot spots in the CR anisotropy in the
  direction of the field lines that are smaller than those expected
  for a dipole. The available data seem to favour moderately broad resonance
  functions. For example, the lower right panel of Fig.~\ref{FastCHA} 
  shows that IceTop data is not compatible with fast modes with a 
  narrow resonance function.

\item For GS turbulence, the half width of the anisotropy
  $\theta_{1/2}$ (size of the cold/hot spot) tends to increase with
  increasing CR energy (see the red and orange lines in the upper right and 
  lower left panels of Fig.~\ref{Half_Width}), and we identify parameters that simultaneously
  give a good fit both to the 400\,TeV data set and to the 2\,PeV
  data set of IceTop \citep{Aartsen:2012ma}, as shown in the upper 
  right (400\,TeV) and lower right (2\,PeV) panels of Fig.~\ref{LAZLAZ_2}.

\item In contrast, isotropic fast mode turbulence does not 
  produce the observed change in spot size with energy. The lower right 
  and lower middle panels of Fig.~\ref{FastLAZ} compare, respectively, the 
  400\,TeV and 2\,PeV data with our predictions for fast modes 
  with a moderately broad resonance function. 
  We speculate that introducing a $|{\bf k}|$-dependent anisotropy in 
  the power spectrum of fast modes might alleviate this problem.

\item We confirm that the CR mean free path is, in most cases,
  significantly smaller for isotropic fast modes than for GS
  turbulence (see Fig.~\ref{MFP_All_Figure}). 
  If fast modes are present at a non-negligible level in
  the local interstellar medium, CR should mostly scatter on
  them. For GS turbulence, a small change in the function used for the
  cutoff in $k_{\parallel}$ of the GS power-spectrum has important
  consequences.

\end{itemize}

Observations of the shape of the CR anisotropy can be used as a new way
  to probe the still poorly known statistical properties of the
  turbulence. Published IceTop data sets already place
  interesting constraints on the interstellar turbulence and CR
  transport properties, and it is to be expected that more stringent constraints
would follow from a more detailed and systematic study of all existing
  CR anisotropy data, using data from experiments in both the Northern
  and the Southern hemispheres, and at different median CR
  energies, down to TeV energies.

\acknowledgements

We thank Markus Ahlers, Tony Bell, and Huirong Yan for useful discussions, and the 
anonymous referee for a detailed and constructive report.

\appendix

\section{Appendix A: Formulae for $D_{\mu\mu}$}
\label{appendixA}

We calculate $D_{\mu\mu}$ numerically, using the following formulae
($\tilde{k}_{\perp}=k_{\perp}l$,
$\tilde{k}_{\parallel}=k_{\parallel}l$, $\tilde{k}=kl$, and $\xi$
denotes the cosine of the {\it wave} pitch-angle with respect to the
direction of local magnetic field lines):

\begin{itemize}

\item Model~A, Goldreich-Sridhar (GS) turbulence with
  $\mathcal{I}_{\rm A,S}=\mathcal{I}_{\rm A,S,1}$ and $R_{\rm
    n}=R_{\rm n,1}$ (Section~\ref{I1_Rn1}):

$n=\pm 1$ terms of the contribution of Alfv\'en modes:
\begin{equation}
  D_{\mu\mu}^{A,-1} + D_{\mu\mu}^{A,+1} = \frac{4 v^{2}(1-\mu^2)}{3 l \eps^{2}} \iint {\rm d}\tilde{k}_\parallel {\rm d}\tilde{k}_\perp  
 h \left( \frac{\tilde{k}_\parallel}{\tilde{k}_\perp^{2/3}} \right) \, \frac{J_{1}^{2}(z)}{z^{2}} \, \frac{v_{\rm A}\tilde{k}_\perp^{-5/3}}{v^{2}(\tilde{k}_\parallel \mu - \eps^{-1})^2 + v_{\rm A}^2 \tilde{k}_\perp^{4/3}} \;,
\label{Dmumu_Alf_n1_CHACHA}
\end{equation}
and $n=0$ term of the contribution of pseudo-Alfv\'en modes:
\begin{equation}
  D_{\mu\mu}^{S,0} = \frac{2 v^{2}(1-\mu^2)}{3 l \eps^{2}} \iint {\rm d}\tilde{k}_\parallel {\rm d}\tilde{k}_\perp h \left( \frac{\tilde{k}_\parallel}{\tilde{k}_\perp^{2/3}} \right) \, \frac{\tilde{k}_\parallel^2}{\tilde{k}_\parallel^2 + \tilde{k}_\perp^2} \, J_{1}^{2}(z) \, \frac{v_{\rm A}\tilde{k}_\perp^{-5/3}}{v^{2}(\tilde{k}_\parallel \mu)^2 + v_{\rm A}^2 \tilde{k}_\perp^{4/3}} \;.
\label{Dmumu_Slow_n0_CHACHA}
\end{equation}

  \item Model~B, GS turbulence with $\mathcal{I}_{\rm A,S}=\mathcal{I}_{\rm A,S,2}$ and $R_{\rm n}=R_{\rm n,1}$ (Section~\ref{I2_Rn1}):
\begin{equation}
  D_{\mu\mu}^{A,-1} + D_{\mu\mu}^{A,+1} = \frac{4 v^{2}(1-\mu^2)}{3 l \eps^{2}} 
\iint {\rm d}\tilde{k}_\parallel {\rm d}\tilde{k}_\perp \, \exp \left( - \frac{\tilde{k}_\parallel}{\tilde{k}_\perp^{2/3}} \right) \frac{J_{1}^{2}(z)}{z^{2}} \, \frac{v_{\rm A}\tilde{k}_\perp^{-5/3}}{v^{2}(\tilde{k}_\parallel \mu - \eps^{-1})^2 + v_{\rm A}^2 \tilde{k}_\perp^{4/3}} \;,
\label{Dmumu_Alf_n1_CHALAZ}
\end{equation}
and
\begin{equation}
  D_{\mu\mu}^{S,0} = \frac{2 v^{2}(1-\mu^2)}{3 l \eps^{2}} 
\iint {\rm d}\tilde{k}_\parallel {\rm d}\tilde{k}_\perp \, \exp \left( - \frac{\tilde{k}_\parallel}{\tilde{k}_\perp^{2/3}} \right) \, \frac{\tilde{k}_\parallel^2}{\tilde{k}_\parallel^2 + \tilde{k}_\perp^2} \, J_{1}^{2}(z) \, \frac{v_{\rm A}\tilde{k}_\perp^{-5/3}}{v^{2}(\tilde{k}_\parallel \mu)^2 + v_{\rm A}^2 \tilde{k}_\perp^{4/3}} \;.
\label{Dmumu_Slow_n0_CHALAZ}
\end{equation}
We also calculate the contribution of pseudo-Alfv\'en modes with $n=\pm 1$:
\begin{equation}
  D_{\mu\mu}^{S,-1}+  D_{\mu\mu}^{S,+1} = \frac{v^{2}(1-\mu^2)}{3 l \eps^{2}} 
\iint {\rm d}\tilde{k}_\parallel {\rm d}\tilde{k}_\perp \, \exp \left( - \frac{\tilde{k}_\parallel}{\tilde{k}_\perp^{2/3}} \right) \, \frac{\tilde{k}_\parallel^2}{\tilde{k}_\parallel^2 + \tilde{k}_\perp^2} \, 
{\left( J_{0}(z) - J_{2}(z) \right)^{2}}\, \frac{v_{\rm A}\tilde{k}_\perp^{-5/3}}{v^{2}(\tilde{k}_\parallel \mu - \eps^{-1})^2 + v_{\rm A}^2 \tilde{k}_\perp^{4/3}} \;.
\label{Dmumu_Slow_n1_CHALAZ}
\end{equation}

  \item Model~C, GS turbulence with $\mathcal{I}_{\rm A,S}=\mathcal{I}_{\rm A,S,1}$ and $R_{\rm n}=R_{\rm n,2}$ (Section~\ref{I1_Rn2}):
\begin{equation}
  D_{\mu\mu}^{A,-1}  +D_{\mu\mu}^{A,+1} = \frac{2\sqrt{\pi} v \sqrt{1-\mu^2}}{3 l \eps^{2} \sqrt{\delta \mathcal{M}_{\rm A}}} 
\iint {\rm d}\tilde{k}_\parallel {\rm d}\tilde{k}_\perp \frac{\tilde{k}_\perp^{-7/3} h \left( \tilde{k}_\parallel/\tilde{k}_\perp^{2/3} \right)}{\tilde{k}_\parallel} \, \frac{J_{1}^{2}(z)}{z^{2}} \, \exp \left( - \frac{\left( \mu - (\tilde{k}_\parallel \eps)^{-1} \right)^2}{(1-\mu^{2})\, \delta \mathcal{M}_{\rm A}} \right) \;,
\label{Dmumu_Alf_n1_LAZCHA}
\end{equation}
\begin{equation}
  D_{\mu\mu}^{S,0} = \frac{\sqrt{\pi} v \sqrt{1-\mu^2}}{3 l \eps^{2}\sqrt{\delta \mathcal{M}_{\rm A}}} 
\iint {\rm d}\tilde{k}_\parallel {\rm d}\tilde{k}_\perp \frac{\tilde{k}_\parallel \tilde{k}_\perp^{-7/3} h \left( \tilde{k}_\parallel/\tilde{k}_\perp^{2/3} \right)}{\tilde{k}_\perp^2 + \tilde{k}_\parallel^2 } \, J_{1}^{2}(z) \, \exp \left( - \frac{\left( \mu - v_{\rm A}/v \right)^2}{(1-\mu^{2})\, \delta \mathcal{M}_{\rm A}} \right) \;,
\label{Dmumu_Slow_n0_LAZCHA}
\end{equation}
and
\begin{equation}
  D_{\mu\mu}^{S,-1}+  D_{\mu\mu}^{S,+1} = \frac{\sqrt{\pi} v \sqrt{1-\mu^2}}{6 l \eps^{2}\sqrt{\delta \mathcal{M}_{\rm A}}} 
\iint {\rm d}\tilde{k}_\parallel {\rm d}\tilde{k}_\perp \frac{\tilde{k}_\parallel \tilde{k}_\perp^{-7/3} h \left( \tilde{k}_\parallel/\tilde{k}_\perp^{2/3} \right)}{\tilde{k}_\perp^2 + \tilde{k}_\parallel^2 } \, {\left( J_{0}(z) - J_{2}(z) \right)^{2}} \, \exp \left( - \frac{\left( \mu - (\tilde{k}_\parallel \eps)^{-1} \right)^2}{(1-\mu^{2})\, \delta \mathcal{M}_{\rm A}} \right) \;.
\label{Dmumu_Slow_n1_LAZCHA}
\end{equation}

  \item Model~D, GS turbulence with $\mathcal{I}_{\rm A,S}=\mathcal{I}_{\rm A,S,2}$ and $R_{\rm n}=R_{\rm n,2}$ (Section~\ref{I2_Rn2}):
\begin{equation}
  D_{\mu\mu}^{A,-1}+  D_{\mu\mu}^{A,+1} = \frac{2\sqrt{\pi} v \sqrt{1-\mu^2}}{3 l \eps^{2} \sqrt{\delta \mathcal{M}_{\rm A}}} 
\iint {\rm d}\tilde{k}_\parallel {\rm d}\tilde{k}_\perp \frac{\tilde{k}_\perp^{-7/3}}{\tilde{k}_\parallel} \, \frac{J_{1}^{2}(z)}{z^{2}} \, \exp \left(-\frac{\tilde{k}_\parallel}{\tilde{k}_\perp^{2/3}} - \frac{\left( \mu - (\tilde{k}_\parallel \eps)^{-1} \right)^2}{(1-\mu^{2})\, \delta \mathcal{M}_{\rm A}} \right) \;,
\label{Dmumu_Alf_n1_LAZLAZ}
\end{equation}
\begin{equation}
  D_{\mu\mu}^{S,0} = \frac{\sqrt{\pi} v \sqrt{1-\mu^2}}{3 l \eps^{2}\sqrt{\delta \mathcal{M}_{\rm A}}} 
\iint {\rm d}\tilde{k}_\parallel {\rm d}\tilde{k}_\perp \frac{\tilde{k}_\parallel \tilde{k}_\perp^{-7/3}}{\tilde{k}_\perp^2 + \tilde{k}_\parallel^2 } \, J_{1}^{2}(z) \, \exp \left(-\frac{\tilde{k}_\parallel}{\tilde{k}_\perp^{2/3}} - \frac{\left( \mu - v_{\rm A}/v \right)^2}{(1-\mu^{2})\, \delta \mathcal{M}_{\rm A}} \right) \;,
\label{Dmumu_Slow_n0_LAZLAZ}
\end{equation}
and
\begin{equation}
  D_{\mu\mu}^{S,-1} + D_{\mu\mu}^{S,+1} = \frac{\sqrt{\pi} v \sqrt{1-\mu^2}}{6 l \eps^{2}\sqrt{\delta \mathcal{M}_{\rm A}}} 
\iint {\rm d}\tilde{k}_\parallel {\rm d}\tilde{k}_\perp \frac{\tilde{k}_\parallel \tilde{k}_\perp^{-7/3}}{\tilde{k}_\perp^2 + \tilde{k}_\parallel^2 } \, {\left( J_{0}(z) - J_{2}(z) \right)^{2}} \, \exp \left(-\frac{\tilde{k}_\parallel}{\tilde{k}_\perp^{2/3}} - \frac{\left( \mu - (\tilde{k}_\parallel \eps)^{-1} \right)^2}{(1-\mu^{2})\, \delta \mathcal{M}_{\rm A}} \right) \;.
\label{Dmumu_Slow_n1_LAZLAZ}
\end{equation}

  \item Model~E, isotropic fast modes with $R_{\rm n}=R_{\rm n,1}$ (Section~\ref{Fast_Rn1}):

$n=0$ term for fast modes:
\begin{equation}
  D_{\mu\mu}^{F,0} = \frac{v^{2} (1-\mu^2)}{l \eps^{2}} 
\int {\rm d}\tilde{k} \tilde{k}^{-1} 
\int_{0}^{1} {\rm d}\xi \, \xi^{2} \, J_{1}^{2}(z) \, \frac{v_{\rm A}}{\left(v \xi \tilde{k} \mu - v_{\rm A} \tilde{k} \right)^{2} + v_{\rm A}^{2} \tilde{k}} \;,
\label{Dmumu_Fast_n0_CHALAZ}
\end{equation}
and $n=\pm 1$ term for fast modes:
\begin{equation}
  \begin{aligned}
  D_{\mu\mu}^{F,-1}+  D_{\mu\mu}^{F,+1} = \frac{v^{2} (1-\mu^2)}{4 l \eps^{2}} 
\int {\rm d}\tilde{k} \tilde{k}^{-1} 
\int_{0}^{1} {\rm d}\xi \, \xi^{2} \, {\left( J_{0}(z) - J_{2}(z) \right)^{2}} \, \frac{v_{\rm A} }{v^{2} \left( \xi \tilde{k} \mu - \eps^{-1} \right)^{2} + v_{\rm A}^{2} \tilde{k}} \;.
  \end{aligned}
\label{Dmumu_Fast_n1_CHALAZ}
\end{equation}

  \item Model~F, isotropic fast modes with $R_{\rm n}=R_{\rm n,2}$ (Section~\ref{Fast_Rn2}):
\begin{equation}
  D_{\mu\mu}^{F,0} = \frac{\sqrt{\pi} v \sqrt{1-\mu^2}}{2 l \eps^{2}\sqrt{\delta \mathcal{M}_{\rm A}}} 
\int {\rm d}\tilde{k} \int_{0}^{1} {\rm d}\xi \, {\tilde{k}^{-5/2} \xi} \, J_{1}^{2}(z) \, \exp \left( - \frac{(\mu - v_{\rm A}/(v \xi ))^{2}}{(1-\mu^{2}) \delta \mathcal{M}_{\rm A}} \right) \;,
\label{Dmumu_Fast_n0_LAZLAZ}
\end{equation}
and
\begin{equation}
  D_{\mu\mu}^{F,-1}+  D_{\mu\mu}^{F,+1} = \frac{\sqrt{\pi} v \sqrt{1-\mu^2}}{8 l \eps^{2}\sqrt{\delta \mathcal{M}_{\rm A}}} \int {\rm d}\tilde{k} \int_{0}^{1} {\rm d}\xi \, {\tilde{k}^{-5/2} \xi} \, {\left( J_{0}(z) - J_{2}(z) \right)^{2}} \, \exp \left( - \frac{(\mu - (\tilde{k} \xi \eps)^{-1} )^{2}}{(1-\mu^{2}) \delta \mathcal{M}_{\rm A}} \right) \;.
\label{Dmumu_Fast_n1_LAZLAZ}
\end{equation}

\end{itemize}

\section{Appendix B: CR mean free paths}
\label{appendixB}

\begin{figure}
  \centerline{\includegraphics[width=0.49\textwidth]{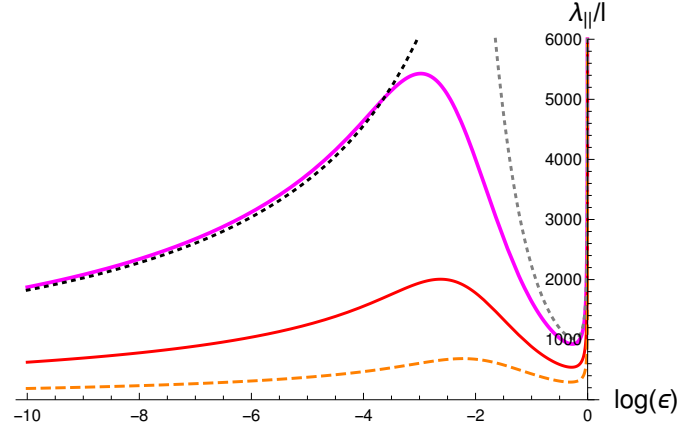}}
  \caption{$\lambda_{\parallel}/l$ as a function of $\log(\eps)$, for
    GS turbulence with $\mathcal{I}_{\rm A,S}=\mathcal{I}_{\rm
      A,S,1}$, $R_{\rm n}=R_{\rm n,1}$ (model~A), and for 3 values of $\delta$:
    $\delta=10^{-5}$ (thick magenta solid line), $3 \times 10^{-5}$
    (red solid line), and $10^{-4}$ (orange dashed line). Dotted lines
    for the behaviours of $\lambda_{\parallel}/l$ at $\delta=10^{-5}$
    in the limit of small or large $\eps$, according
    to~\cite{Chandran:2000hp}: Eq.~(\ref{mfp_limsmalleps}) for $\eps
    \ll - \delta \ln \eps$ (black), and Eq.~(\ref{mfp_limlargeeps})
    for $\eps \gg - \delta \ln \eps$ (grey).}
\label{Chandran_MFP}
\end{figure}

We calculate the scattering mean free path $\lambda_{\parallel}$ using
Eqs.~(\ref{MFP_Eqn}) and~(\ref{FormulaGmu_Dmumu_Sym}). We plot
$\lambda_{\parallel}/l$ versus $\log(\eps)$ in Fig.~\ref{Chandran_MFP}
for GS turbulence with $\mathcal{I}_{\rm A,S}=\mathcal{I}_{\rm A,S,1}$
and $R_{\rm n}=R_{\rm n,1}$. We use here the fitting formulae for
$D_{\mu\mu}$ provided by~\cite{Chandran:2000hp}. 
The thick magenta solid line corresponds to $\delta = 10^{-5}$,
the red solid one to $\delta = 3 \times 10^{-5}$, and the orange
dashed one to $\delta = 10^{-4}$. $\lambda_{\parallel}/l$ grows with
$\eps$ at small $\eps$. At such ``small'' values of $\delta$,
$\lambda_{\parallel}/l$ reaches a peak and then decreases with
$\eps$. \cite{Chandran:2000hp} provided two formulae for the limiting
behaviours of $\lambda_{\parallel}$ at small and large $\eps$:
\begin{itemize}
  \item For $\eps^{3/2} \ll - \delta \ln \eps$, i.e. at ``small'' $\eps$ (black dotted line in Fig.~\ref{Chandran_MFP}),
\begin{equation}
  \frac{\lambda_{\parallel}}{l} \approx \frac{3\,(5/2-3\pi/4)}{- \delta\ln \eps} \approx \frac{0.43}{- \delta \ln \eps}\;.
\label{mfp_limsmalleps}
\end{equation}
  \item For $\eps^{3/2} \gg - \delta \ln \eps$, i.e. at ``large'' $\eps$ (grey dotted line in Fig.~\ref{Chandran_MFP}),
\begin{equation}
  \frac{\lambda_{\parallel}}{l} \approx 2.64\,(- \delta \ln \eps)^{-5/11} \eps^{-9/11}\;.
\label{mfp_limlargeeps}
\end{equation}
\end{itemize}

\begin{figure*}
  \centerline{\includegraphics[width=0.49\textwidth]{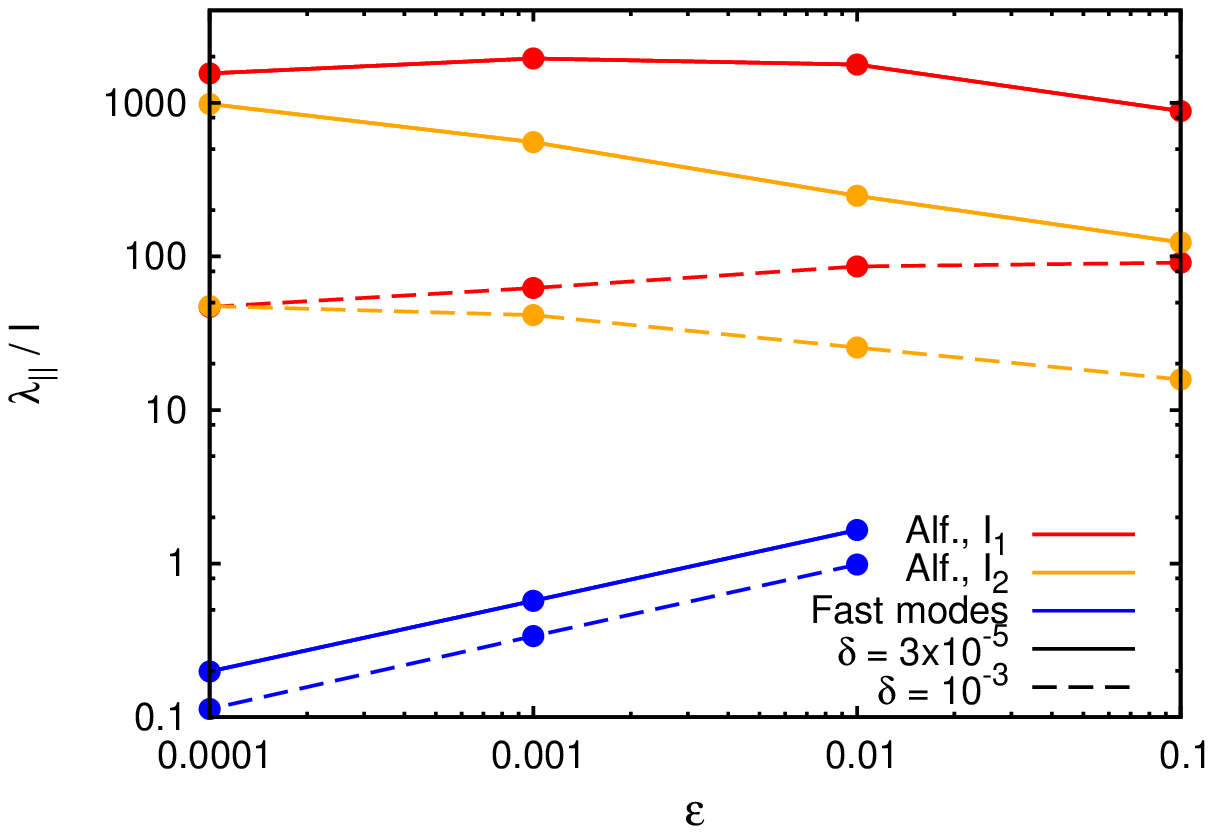}
              \hfil
              \includegraphics[width=0.49\textwidth]{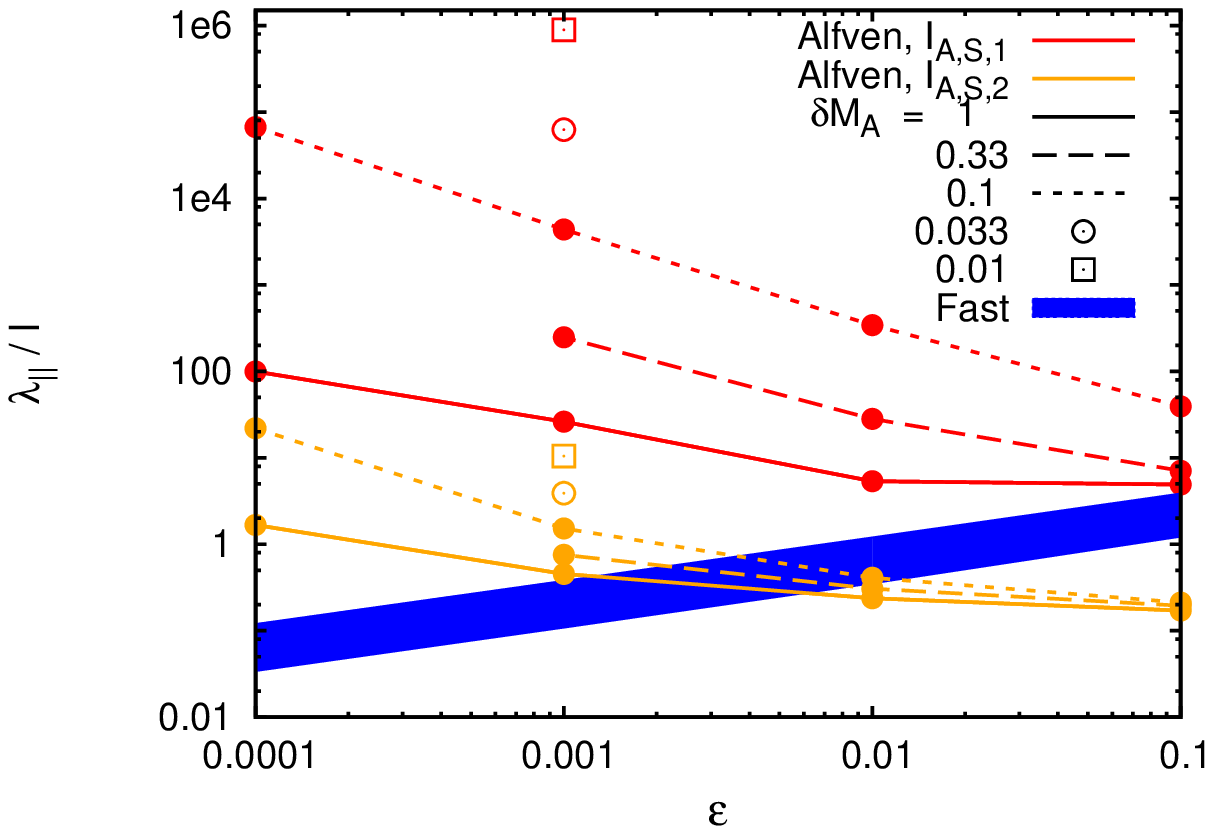}
              }
              \caption{$\lambda_{\parallel}/l$ as a function of
                $\eps$, for GS turbulence and for isotropic fast
                modes. {\it Left panel:} resonance function $R_{\rm
                  n}=R_{\rm n,1}$ (models~A, B \& E). {\it Right panel:} $R_{\rm
                  n}=R_{\rm n,2}$ (models~C, D \& F).}
\label{MFP_All_Figure}
\end{figure*}

In Fig.~\ref{MFP_All_Figure}, we plot $\lambda_{\parallel}/l$ versus
$\eps$, both for GS turbulence (red lines or symbols for
$\mathcal{I}_{\rm A,S,1}$, orange for $\mathcal{I}_{\rm A,S,2}$) and
for isotropic fast modes (blue lines or area). The left panel is for
calculations made with the resonance function $R_{\rm n}=R_{\rm n,1}$,
and the right panel is for $R_{\rm n}=R_{\rm n,2}$. Each line or
symbol type corresponds to a different value of $\delta$ or
$\delta\mathcal{M}_{\rm A}$, see keys.

\section{Appendix C: Additional calculations of the scattering rate and anisotropy}
\label{appendixC}

\begin{figure*}
  \centerline{
              \includegraphics[width=0.32\textwidth]{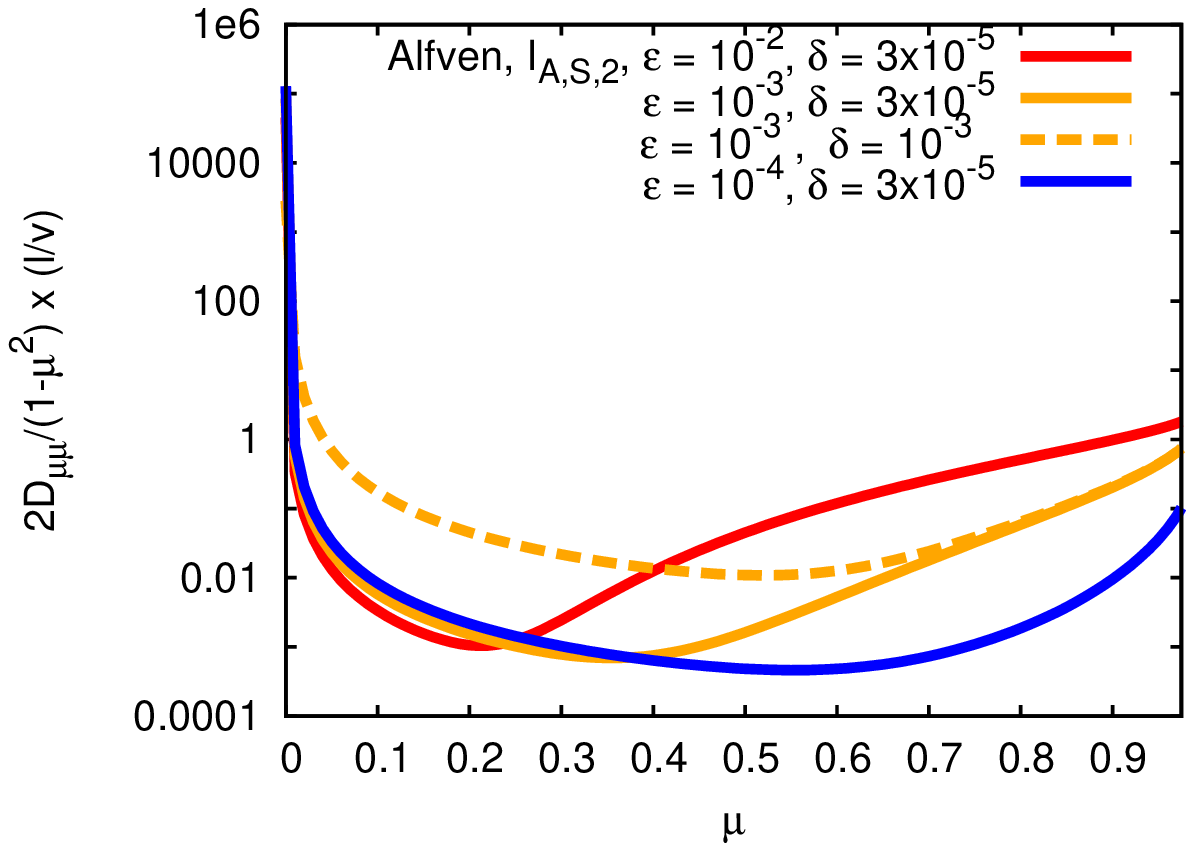}
              \hfil
              \includegraphics[width=0.32\textwidth]{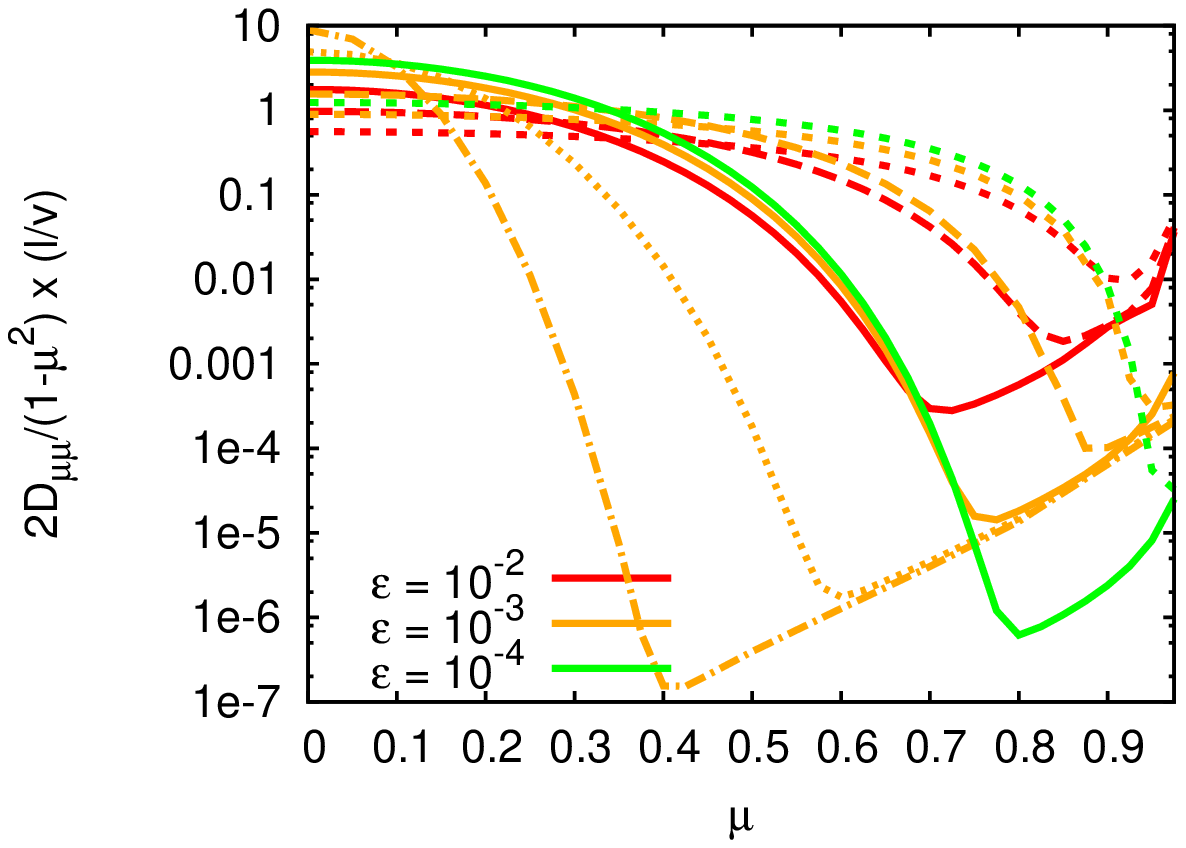}
              \hfil
              \includegraphics[width=0.32\textwidth]{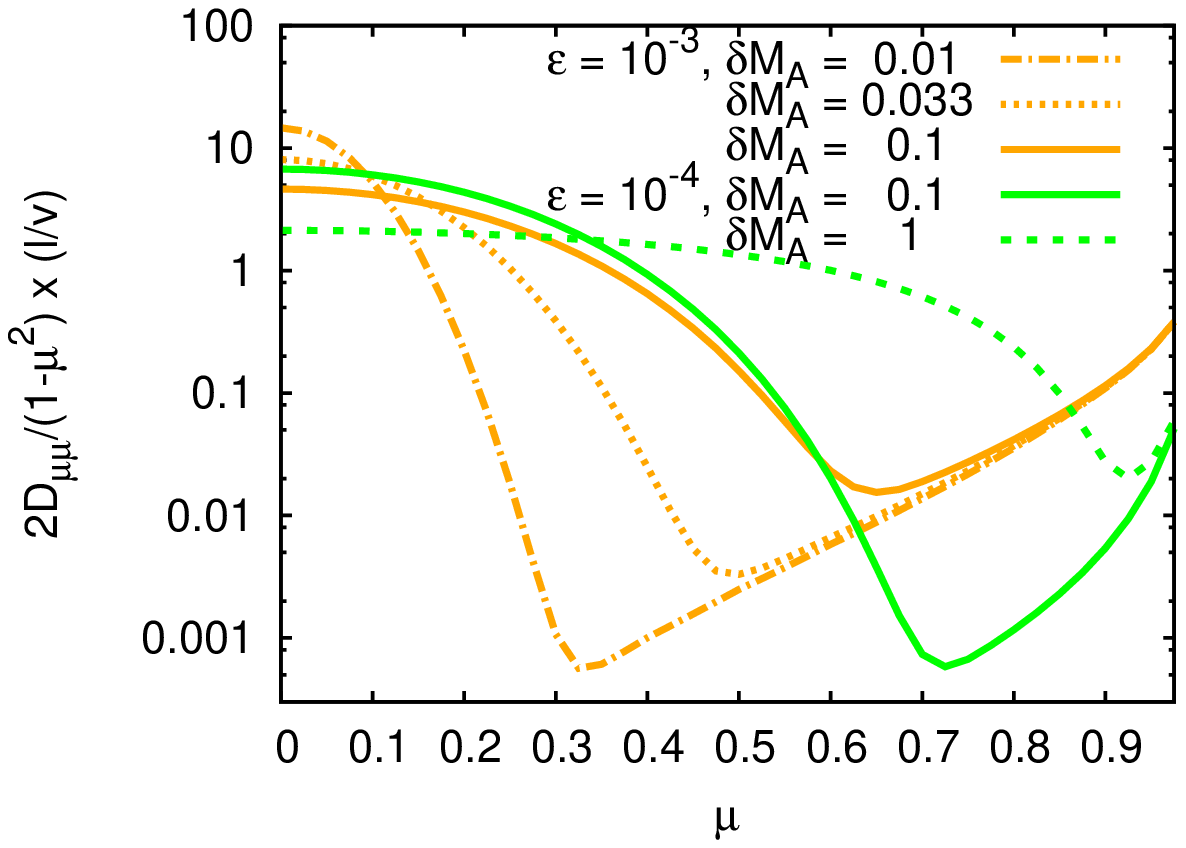}
              }
  \centerline{\includegraphics[width=0.32\textwidth]{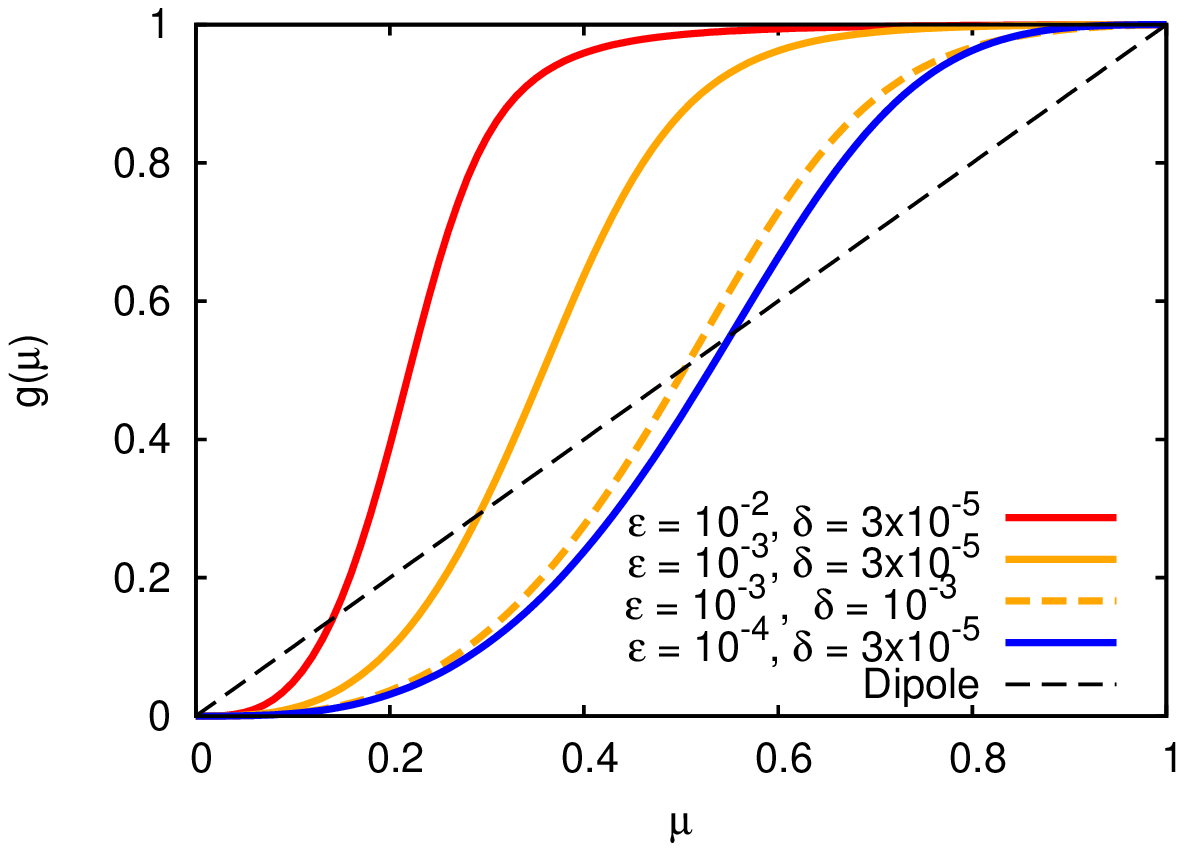}
              \hfil
              \includegraphics[width=0.32\textwidth]{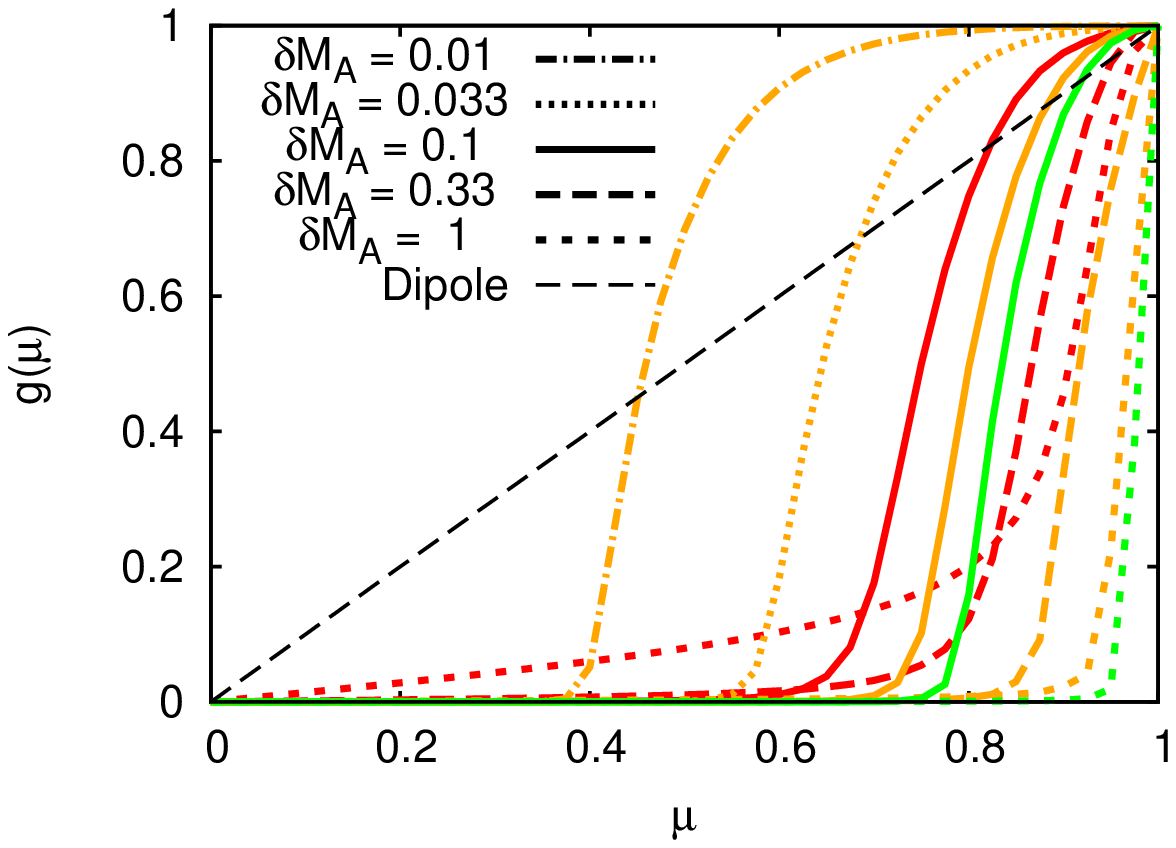}
              \hfil
              \includegraphics[width=0.32\textwidth]{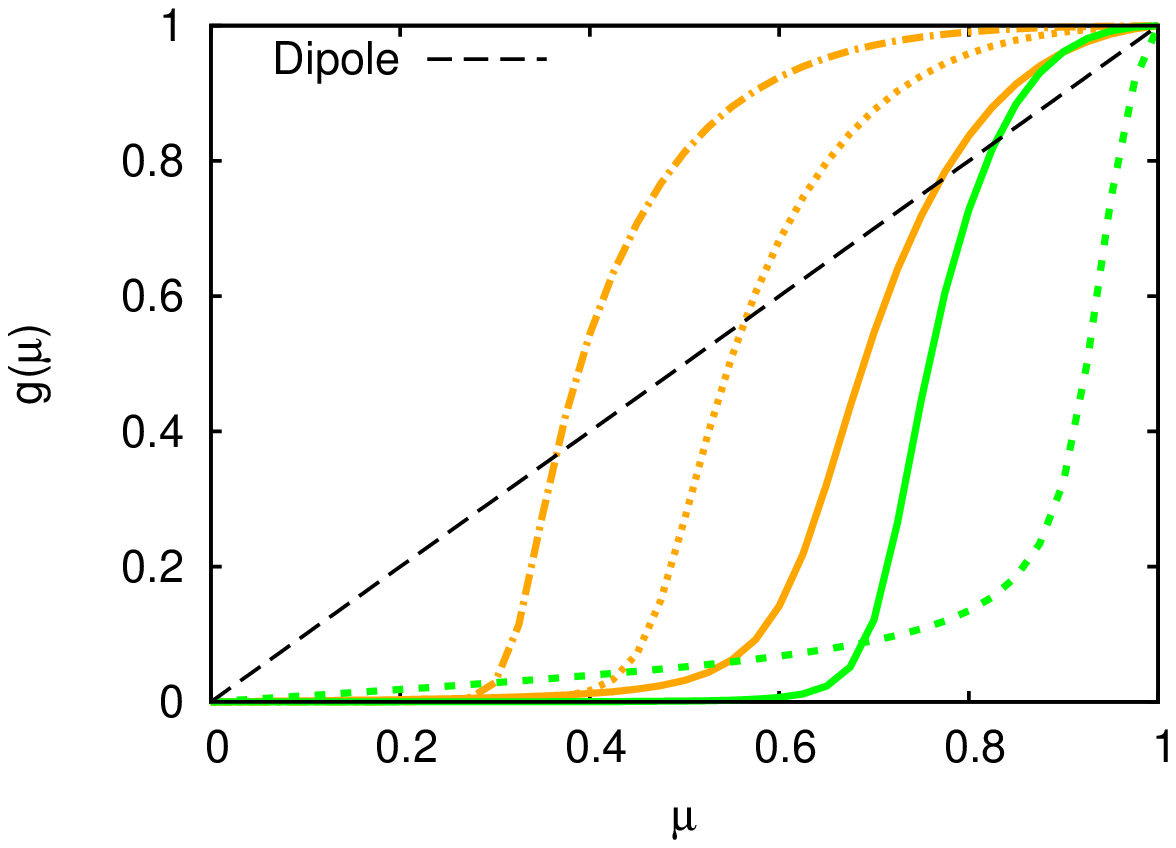}
              }
  \caption{$\nu$ ({\it upper row}) and $g$ ({\it lower row}) as functions of $\mu$, for the cases whose parameters are within the shaded areas in Fig.~\ref{FirstEigenvalue}. GS turbulence with: $\mathcal{I}_{\rm A,S}=\mathcal{I}_{\rm A,S,2}$ and $R_{\rm n}=R_{\rm n,1}$ ({\it left column}), $\mathcal{I}_{\rm A,S,1}$ and $R_{\rm n,2}$ ({\it middle column}), and $\mathcal{I}_{\rm A,S,2}$ and $R_{\rm n,2}$ ({\it right column}). See keys for parameters: each {\it column} has its own set of line types and colors.}
\label{Extra_CHALAZ_LAZCHA_LAZLAZ}
\end{figure*}

We present in Fig.~\ref{Extra_CHALAZ_LAZCHA_LAZLAZ} calculations of
$\nu = 2D_{\mu\mu}/(1-\mu^{2})\times (l/v)$ (upper row) and $g(\mu)$
(lower row), for the cases whose parameters are within the shaded
areas in Fig.~\ref{FirstEigenvalue} ---except two points for GS
turbulence with $\mathcal{I}_{\rm A,S}=\mathcal{I}_{\rm A,S,2}$ and
$R_{\rm n}=R_{\rm n,2}$, which were studied in Sect.~\ref{I2_Rn2}
($\{\eps,\delta\mathcal{M}_{\rm A}\}=\{10^{-3},0.33\}$ and
$\{10^{-3},1\}$).

The left column in Fig.~\ref{Extra_CHALAZ_LAZCHA_LAZLAZ} is for GS
turbulence with $\mathcal{I}_{\rm A,S}=\mathcal{I}_{\rm A,S,2}$ and
$R_{\rm n}=R_{\rm n,1}$, the middle one for $\mathcal{I}_{\rm A,S,1}$
and $R_{\rm n,2}$, and the right one for $\mathcal{I}_{\rm A,S,2}$ and
$R_{\rm n,2}$. Each column has its own set of line types and colors,
sey keys for values of $\eps$, $\delta$, or $\delta\mathcal{M}_{\rm
  A}$.

\bibliographystyle{apj}
\bibliography{references}


\end{document}